\documentstyle[11pt]{article}
\skip\footins 5mm plus 2mm minus 2mm
\begin{document}
\setlength{\parindent}{0 mm}

\vspace*{-0.5in}

\newcommand {\ds} {\displaystyle \sum}
\def\bd{\begin{displaymath}}
\def\ed{\end{displaymath}}
\def\be{\begin{equation}}
\def\ee{\end{equation}}
\def\p{ \partial}
\large

\newtheorem{theorem}{Theorem}
 \newtheorem{proposition}{Proposition}
 \newtheorem{definition}{Definition}

\renewcommand{\thesection}{\arabic{section}}

\smallskip

\centerline { {\bf   MULTIPLE HAMILTONIAN STRUCTURE }}
\centerline { {\bf  OF BOGOYAVLENSKY-TODA LATTICES}}

 \vskip 1  cm
\centerline { Pantelis A. Damianou}

\vskip 1 cm

 \centerline { Department of
Mathematics and Statistics } \centerline  { University of Cyprus}

\centerline { P. O. Box 20537, 1678 Nicosia, Cyprus } \centerline {
Email: damianou@ucy.ac.cy}

 \vskip 2 cm \centerline { \bf {ABSTRACT }}
\bigskip
{\it This paper is  mainly a  review of the multi--Hamiltonian nature of Toda and generalized Toda lattices
corresponding to the classical  simple Lie groups but it includes also  some new results.
   The areas  investigated include  master symmetries, recursion operators, higher    Poisson
brackets, invariants and group symmetries for the systems. In addition to the positive  hierarchy  we
also consider the negative hierarchy which is crucial in establishing the bi--Hamiltonian structure for each
particular simple Lie group. Finally, we include some results on point and Noether symmetries
  and an interesting connection
with the exponents of simple Lie groups. The case of exceptional simple Lie groups is still an open problem. }

\vskip 1cm {\bf Mathematics Subject Classification:} \  37J35, \
22E70 and 70H06.

\vskip 1cm
{\bf Key words} Toda lattice, Poisson brackets, master symmetries, bi-Hamiltonian systems, group symmetries, simple Lie
groups.

\newpage

\centerline{{\bf CONTENTS}}

\begin{enumerate}
\item
 {\bf Introduction}

\item
 {\bf Background}

\noindent
{\bf 2.1} Schouten bracket

\noindent
{\bf 2.2} Poisson manifolds

\noindent
{\bf 2.3} Symplectic and Lie--Poisson manifolds

\noindent
{\bf 2.4} Local theory

\noindent
{\bf 2.5} Cohomology

\noindent
{\bf 2.6} Bi--Hamiltonian systems

\noindent
{\bf 2.7} Master symmetries

\item
 {\bf $A_n$ Toda lattice}

\noindent
{\bf 3.1} Definition of the system

\noindent
{\bf 3.2} Multi--Hamiltonian structure

\noindent
{\bf 3.3} Properties of $\pi_n$ and $X_n$

\noindent
{\bf 3.4} The Faybusovich--Gekhtman approach

\noindent
{\bf 3.5}  A theorem of Petalidou

\noindent
{\bf 3.6}  A recursive process of Kosmann--Schwarzbach and Magri.

\item
 {\bf Lie group symmetries}

\item
 {\bf The Toda lattice in natural coordinates}

\noindent
{\bf 5.1} The Das--Okubo--Fernandes approach

\noindent
{\bf 5.2}  Negative Toda hierarchy

\noindent
{\bf 5.3} Master integrals and master symmetries

\noindent
{\bf 5.4} Noether symmetries

\noindent
{\bf 5.5} Rational Poisson brackets

\item
 {\bf Generalized Toda systems associated with simple Lie groups}

\item
 {\bf $B_n$ Toda systems}

\noindent
{\bf 7.1} A rational bracket for a central extension of $B_n$

\noindent
{\bf 7.2} A recursion operator for Bogoyavlensky--Toda systems of type $B_n$

\noindent
{\bf 7.3} Bi--Hamiltonian formulation of $B_n$ systems

\item
 {\bf $C_n$ Toda systems}

\noindent
{\bf 8.1} A recursion operator for Bogoyavlensky--Toda systems of type $C_n$

\noindent
{\bf 8.2} Bi--Hamiltonian formulation of $C_n$ systems

\item
 {\bf $D_n$ Toda systems}

\noindent
{\bf 9.1} A recursion operator for Bogoyavlensky--Toda systems of type $D_n$

\noindent
{\bf 9.2} Master symmetries

\noindent
{\bf 9.3}  A recursion operator for $D_n$ Toda systems in natural $(q,p)$ coordinates

\noindent
{\bf 9.4} Bi--Hamiltonian formulation of $D_n$ systems

\item
 {\bf Conclusion}

\noindent
{\bf 10.1} Summary of results

\noindent
{\bf 10.2} Open problems

\end{enumerate}

\newpage
\section   {    INTRODUCTION}

In this paper we review  the bi--Hamiltonian and multiple Hamiltonian nature of the Toda lattices
corresponding to simple Lie groups. These are systems that generalize the usual finite, non--periodic
Toda lattice (which corresponds to a root system of type $A_n$). This
generalization is due to Bogoyavlensky \cite{bogo}. These systems were studied extensively in
 \cite{kostant} where the solution of the system was connected intimately with the representation
theory of simple Lie groups. There are also studies by
 Olshanetsky and Perelomov
\cite{olshanetsky} and Adler, van Moerbeke \cite{avm}.
 We will call such systems
the Bogoyavlensky-Toda lattices.

We begin with the following more general definition which involves systems with exponential
interaction: Consider a Hamiltonian of the form
\be
H={1 \over 2} ({\bf p}, {\bf p})+\sum_{i=1}^m e^{ ({\bf v}_i\,,\, {\bf q}) }  \ , \label{a1}
\ee
where ${\bf q}=(q_1, \dots, q_N)$,  ${\bf p}=(p_1, \dots, p_N)$, ${\bf v}_1, \dots, {\bf v}_m$ are vectors in ${\bf R}^N$ and $(\ , \ )$ is the standard
inner product in ${\bf R}^N$. The set of vectors  $\Delta=\{ {\bf v}_1, \dots, {\bf v}_m \}$ is  called the spectrum of the system.

In this paper we limit our attention to the case where the spectrum is a system of simple roots for a simple Lie algebra
 ${\cal G}$. In this case $m=l={\rm rank} \,  {\cal G}$. It is worth mentioning that the case where $m, N$ are arbitrary is
an open and unexplored area of research. The main exception is the work of Kozlov and Treshchev \cite{kozlov}  where
a classification of system (\ref{a1}) is performed under the assumption that the system possesses $N$ polynomial (in the
momenta) integrals.  We also note the papers by Ranada  \cite{ranada2} and   Annamalai,  Tamizhmani  \cite{anna}.
 Such systems are called Birkhoff integrable. For each  Hamiltonian in (\ref{a1}) we
associate a Dynkin type diagram as follows: It is a graph whose vertices correspond to the elements of  $\Delta$. Each pair of vertices
$v_i$, $v_j$ are connected by
\bd
{ 4 (v_i, v_j)^2 \over (v_i,v_i) (v_j, v_j) }
\ed
edges.

\bigskip
\noindent
{\bf Example:}
The usual Toda lattice corresponds to a Lie algebra of type $A_{N-1}$. In other words $m=l=N-1$ and we choose $\Delta$ to
be the set:
\bd
{\bf v}_1=(1,-1,0, \dots, 0),   \dots \dots \dots , {\bf v}_{N-1}=(0,0, \dots,0, 1,-1) \ .
\ed
The graph  is the usual Dynkin diagram of a Lie algebra of type $A_{N-1}$.
The Hamiltonian becomes
\be
  H(q_1, \dots, q_N, \,  p_1, \dots, p_N) = \sum_{i=1}^N \,  { 1 \over 2} \, p_i^2 +
\sum _{i=1}^{N-1} \,  e^{ q_i-q_{i+1}}  \ , \label{a2}
\ee
which is the well-known classical, non--periodic Toda lattice.

It is  more convenient to work, instead in   the space of  the natural $(q,p)$ variables, with the
Flaschka variables $(a,b)$ which are defined by:

\be
\begin{array}{lcl}
a_i& =&{1\over 2}e^{ {1\over 2} ({\bf v}_i, {\bf q})} \ \ \ \ \ i=1,2, \dots, m    \\
 b_i & = &-{ 1 \over 2} p_i  \ \ \ \ \ \ \ \ \ \ \ \ \  i=1,2, \dots, N \ .     \label{a3}
\end{array}
\ee

We end--up with a new set of polynomial  equations in the variables $(a,b)$.  One can write the equations
in Lax pair form (at least this is well--known for spectrum corresponding to simple Lie algebras);
see for example \cite{perelomov}. The Lax pair ($L(t), B(t)$) in ${\cal G}$ can be described in terms
of the root system as follows:

\bd
L(t)=\sum_{i=1}^l b_i(t) h_{\alpha_i} + \sum_{i=1}^l a_i(t) (x_{\alpha_i}+x_{-\alpha_i}) \ ,
\ed

\bd
B(t)=\sum_{i=1}^l a_i(t) (x_{\alpha_i}-x_{-\alpha_i})  \ .
\ed
As usual $h_{\alpha_i}$ is an element of  a fixed Cartan subalgebra and $x_{\alpha_i}$ is a  root vector corresponding to
the simple root $\alpha_i$.
The Chevalley invariants of ${\cal G}$  provide for the constants of motion. We will describe them separately
for each case.

In this paper we begin with a  review of  the  $A_N$ Toda system. The theory
and bi--Hamiltonian structure for this case is
well--developed and the results are well--known. We   present  a  review of the results in sections 3--5
and then, for the remaining part of the paper, we deal exclusively with the other classical simple Lie algebras
of type $B_N$, $C_N$ and $D_N$. We will demonstrate  that these systems are bi--Hamiltonian and then illustrate
with some small dimensional examples, namely $B_2$, $C_3$ and $D_4$.

The multi--Hamiltonian structure of the Toda lattice was established in \cite{damianou1}. For the
remaining Bogoyavlensky--Toda  systems it was established recently  in several papers.
The  results for $B_N$ Bogoyavlensky-Toda  lattice were computed in \cite{damianou2} in Flaschka coordinates,
and in \cite{joana} in $(q,p)$ coordinates. The $C_N$ case  is in \cite{damianou3} in
Flaschka coordinates, and \cite{joana} in natural $(q,p)$ coordinates. The $D_N$ case was settled   in
\cite{damianou4}. The bi--Hamiltonian structure of these systems was established recently in \cite{damianou9}.
 The negative Toda hierarchy was constructed in \cite{damianou5} and it was crucial in  establishing the bi--Hamiltonian
formulation in \cite{damianou9}.

 The construction of the bi--Hamiltonian pair  may be summarized as follows:

 Define a recursion operator ${\cal R}$  in $(a,b)$ space by finding a second bracket, $\pi_3$,  and inverting
the initial Poisson bracket $\pi_1$.  Define the negative recursion operator ${\cal N}$  by inverting
 the second Poisson bracket $\pi_3$. This
recursion operator is the inverse of the operator ${\cal R}$. Finally, define a new rational bracket
$\pi_{-1}$ by $\pi_{-1}={\cal N} \pi_1= \pi_1 \pi_3^{-1}\pi_1$.  We obtain
a bi--Hamiltonian formulation of the system:
\bd
\pi_1 \nabla H_2 = \pi_{-1} \nabla H_4  \ ,
\ed
where $H_i={1\over i}{\rm tr}\, L^i$. The brackets $\pi_1$ and $\pi_{-1}$ are
compatible and Poisson.

There is also an interesting connection with the exponents of the corresponding Lie group.
For example, in the case of $D_N$
there is a sequence of invariants $H_2, H_4, \dots, $   of
even degree and an additional invariant of degree $N$. Let  $\chi_i$ denote the  Hamiltonian vector field
generated by  $H_i$ and let $Z_0$ denote
 a conformal symmetry. Then we have
\begin{displaymath}
[Z_0, \chi_j]=f(j)\chi_j  \ .
\end{displaymath}
The values of $f(j)$ corresponding to independent $\chi _j$
generate all the exponents except one. When $Z_0$ acts on the Hamiltonian
vector field $\chi_P$, where $P$ is the invariant corresponding to the
Pfaffian  of the Jacobi matrix, we obtain the last exponent $N-1$.
For example, in the case of $D_5$ the exponents are 1, 3, 5, 7,  and  4. The independent
invariants are $H_2$, $H_4$, $H_6$, $H_8$ and $P_5$ where
  $H_{2i} = { 1 \over 2i} \  { \rm Tr} \ L^{2i}  $ and $P_5=\sqrt{{\rm det}\, L}$.
We obtain

\bd
\begin{array}{lcl}

[Z_0, \chi_2 ]&=&  \chi_2  \\

[Z_0, \chi_4 ]&=& 3 \chi_4        \\

[Z_0, \chi_6 ] &=& 5 \chi_6  \\

[Z_0, \chi_8 ] &=& 7 \chi_8  \\

[Z_0, \chi_{P_5} ]&=& 4 \chi_{P_5}  \ .

\end{array}
\ed

In other words, the coefficients on the right hand side are precisely the exponents of a simple
Lie group of type $D_5$.

\section  {BACKGROUND}
In this section we review the necessary background from Poisson and symplectic geometry, bi--Hamiltonian
systems, master symmetries and recursion operators.

\subsection{Schouten bracket}
 We  list some properties of the Schouten bracket following  \cite{lich}, \cite{ratiu},  \cite{vaisman}.
Let $M$ be a $C^{\infty}$  manifold, $N=C^{\infty}(M)$ the algebra of $C^{\infty}$ real--valued
functions on $M$.
 A contravariant, antisymmetric tensor of order $p$ will be called a $p$-tensor for short. These
 tensors form a superspace endowed with a  Lie--superalgebra
 structure via the  Schouten bracket.

\smallskip
The Schouten bracket assigns to each $p$-tensor $A$, and $q$-tensor $B$, a
 $(p+q-1)$-tensor, denoted by $[A,B]$.
For $p=1$ we have $[A,B]=L_A B$ where $L_A$ is the Lie-derivative in  the
direction of the vector field $A$.
\bigskip
 The bracket satisfies

\smallskip
\noindent
{\it i)} \begin{equation}
  [A,B]= (-1)^{pq} [B,A] \ .
\end{equation}
\smallskip
\noindent
{\it ii)} If $C$ is a $r$-tensor
   \begin{equation}
(-1)^{pq} [[B,C], A]+ (-1)^{qr} [[C,A],B] +(-1)^{rp} [[A,B],C]=0 \ .  \label{f9}
\end{equation}
\smallskip
\noindent
{\it iii)} \begin{equation}
[A, B\wedge C]= [A,B]\wedge C + (-1)^{pq+q} B \wedge [A,C]  \ .
\end{equation}

\subsection{Poisson Manifolds}

We review the basic definitions and properties of Poisson  manifolds following  \cite{lich},  \cite{vaisman},
 \cite{weinstein}.
A  Poisson structure on $M$ is a bilinear form, called the  Poisson bracket  $\{\ , \ \} : \
   N \times N \to N$ such that

\smallskip
\noindent
{\it i)} \begin{equation}
     \{ f, g \} = - \{ g, f \}
\end{equation}
\smallskip
\noindent
{\it ii)} \begin{equation}
\{ f, \{ g, h\} \} + \{ g, \{h, f \}\} + \{ h, \{ f, g \} \}=0
\end{equation}
\smallskip
\noindent
{\it iii)}\begin{equation}
 \{ f, gh \} = \{ f, g \} h + \{f, h \} g  \ .
\end{equation}
Properties {\it i)} and {\it ii)} define a Lie algebra structure on $N$. {\it ii)} is called
the Jacobi identity and {\it iii)} is the analogue of Leibniz rule from calculus. A   Poisson
manifold is a manifold $M$ together with a Poisson bracket $\{ \ , \ \}$.

\smallskip
 To a  Poisson bracket  one can associate a 2-tensor $\pi$ such that
 \begin{equation}
\{ f, g \}= \langle \pi , df \wedge dg \rangle \ .
\end{equation}
Jacobi's identity is equivalent to the condition $[ \pi,\pi]=0$ where $[ \ , \ ]$ is the
Schouten bracket. Therefore, one could define a Poisson manifold by specifying  a pair
$(M,\pi)$  where $M$ is a manifold and $\pi$ a 2-tensor satisfying $[\pi, \pi]=0$. In
local coordinates $(x_1, x_2, \dots, x_n)$,   $\pi$ is given by
\begin{equation}
\pi=\sum_{i,j} \pi_{ij} { \partial \over \partial x_i} \wedge { \partial \over \partial x_j}
\end{equation}

and
\begin{equation}
\{ f, g \}= \langle \pi , df \wedge dg \rangle =\sum_{i,j} \pi_{ij} { \partial f \over \partial x_i}
 \wedge { \partial  g \over \partial x_j} \ .
\end{equation}
In particular $ \{ x_i, x_j \} =\pi_{ij}(x)$.  Knowledge of the Poisson matrix $(\pi_{ij})$ is sufficient to
 define the bracket of arbitrary functions. The rank of the matrix $( \pi_{ij})$
at a point $x \in M$ is called the  rank  of the Poisson structure at $x$.

\smallskip
A function $F : M_1 \to M_2 $ between two Poisson manifolds is called
 a  Poisson mapping if
\begin{equation}
\{ f \circ F, g \circ F \}_1 = \{ f, g \}_2 \circ F
\end{equation}
for all $f, g \in  C^{ \infty}(M_2)$. In terms of tensors,  $F_* \pi_1=\pi_2$.
 Two Poisson manifolds are called   isomorphic, if there exists a diffeomorphism
 between them which is a Poisson mapping.

\smallskip
The Poisson bracket  allows one to associate a vector field to each element
 $f \in N$.  The vector field $\chi_f$ is defined by the formula
\begin{equation}
\chi_f (g) = \{f, g\} \ .
\end{equation}
It is called the  Hamiltonian vector field generated by  $f$. In terms of the
Schouten bracket
\begin{equation}
\chi_f = [ \pi, f] \ .
\end{equation}
Hamiltonian vector fields are  infinitesimal automorphisms  of the Poisson
 structure. These are vector fields $X$ satisfying $L_X \pi=0$. In the
case of Hamiltonian vector fields we have
\begin{equation}
L_{\chi_f} \pi =[\pi, \chi_f]=[\pi, [\pi, f]]=- 2 [[\pi, \pi], f]=0 \ .
\end{equation}

\smallskip
The Hamiltonian vector fields form a Lie algebra and in fact
\begin{equation}
[\chi_f, \chi_g] = \chi_  { \{f,g\}} \ .
\end{equation}
So, the map $f \to \chi_f$ is a Lie algebra homomorphism.

\smallskip
The Poisson structure defines a bundle map
\be
\pi^* : T^*M \to TM \label{f10}
\ee
 such
that
\begin{equation}
\pi^* (df) = \chi_f \ .
\end{equation}
The rank of the Poisson structure at a point $x \in M$ is the rank of
$\pi_x^* : T_x^*  M \to T_x M$.
Throughout this paper we  use the symbol $\pi$ to denote a Poisson tensor but occasionally we use
the same symbol to denote the matrix of the components of the tensor (i.e. the Poisson matrix). The same
 convention applies for the recursion operators.

\smallskip
The functions in the center of $N$ are called   Casimirs. It is the set
of functions $f$ so that $\{f, g \}=0$  for all $g \in N$. These are functions
which are constant along the orbits of Hamiltonian vector fields. The
differentials of these functions are in the kernel of $\pi^*$. In terms of
 the Schouten bracket a Casimir satisfies $[\pi, f]=0$.

\smallskip
Given a function $f$, there is a reasonable algorithm for constructing
a Poisson bracket in which $f$ is a Casimir. One finds two vector fields
 $X_1$ and $X_2$ such that $L_{X_1}f=L_{X_2}f =0$.  If in addition $X_1$,
 $X_2$ and $[X_1, X_2]$ are linearly dependent, then $X_1 \wedge X_2$ is
a Poisson tensor and $f$ is a Casimir in this bracket. In fact
\begin{equation}
[f, X_1 \wedge X_2]=[f,X_1] \wedge X_2 - X_1 \wedge [f,X_2]=0 \ .
\end{equation}

\smallskip
More generally, there is  a formula due to  Flaschka and  Ratiu which gives
locally a Poisson bracket when the number of Casimirs is 2 less than the
 dimension of the space.
  Let $f_1,f_2,\dots, f_r$ be functions on ${\bf R}^{r+2} $.

Then the formula
\begin{equation}
 \omega \{ g, h \} =df_1 \wedge \dots \wedge df_r \wedge dg \wedge dh
\end{equation}
where $\omega$ is a non-vanishing $r+2$ form, defines a Poisson bracket on
${\bf R}^{r+2}$ and the functions $f_1, \dots, f_r $ are Casimirs.
For more details on this formula, see \cite{marmo}.

\smallskip
Multiplication of a Poisson bracket by a Casimir gives another Poisson
bracket. Suppose $[\pi, \pi]=0$ and $[\pi, f]=0$. Then
\begin{equation}
[f\pi, f\pi]=f \wedge [f, \pi] \wedge \pi+ f \wedge \pi \wedge [\pi,f]+
f^2 [\pi,\pi]=0 \  .
\end{equation}

\bigskip
\subsection{Symplectic and Lie Poisson manifolds}

 The most basic examples of Poisson brackets are the symplectic and Lie-Poisson
brackets.

\bigskip

\noindent

{\it i)}\ { \bf Symplectic manifolds:} A {\it symplectic manifold}
is a pair $(M^{2n},\  \omega )$ where $M^{2n}$ is an even
dimensional manifold and $\omega$ is a closed, non-degenerate
2--form. The associated isomorphism
\begin{equation}
 \mu : TM \to T^* M
\end{equation}
extends naturally to  a tensor bundle isomorphism still denoted by $\mu$. Let $\lambda= \mu^{-1}$,
 $f \in N$ and let $\chi_f= \lambda(df)$ be the corresponding Hamiltonian vector field.
  The symplectic bracket is given by

\begin{equation}
   \{ f, g\}= \omega ( \chi_f , \chi_g )  \ .
\end{equation}
\smallskip

\noindent

In the case of ${ \bf R}^{2n}$,  according to a Theorem of Darboux,  there are  coordinates $(x_1,\dots, x_n, y_1,\dots, y_n)$, so that

\begin{equation}
\omega = \sum_{i=1}^n dx_i \wedge dy_i
\end{equation}
\smallskip

\noindent

and the Poisson bracket is the standard symplectic bracket on ${\bf R}^{2n}$.

\bigskip

\noindent

{\it ii)}\ {\bf Lie--Poisson :}  Let $M=\cal{ G}^*$ where ${\cal G}$ is a Lie algebra. For
$a \in {\cal G}$,  define the function $\Phi_a $ on ${\cal G}^*$ by

\begin{equation}
 \Phi_a (\mu ) = \langle a , \mu \rangle
\end{equation}
\smallskip

\noindent

where $\mu \in {\cal G}^*$ and $\langle \ , \  \rangle $  is the pairing

between ${\cal G}$ and ${\cal G}^*$. Define a bracket on ${\cal G}^*$ by

\begin{equation}
 \{ \Phi_a, \Phi_b \} = \Phi_{[a,b]} \ .
\end{equation}
\smallskip

\noindent

This  bracket is easily extended to arbitrary $C^{\infty }$ functions
on ${\cal G}^*$.  The bracket of linear functions is linear and every linear
bracket is of this form,  i.e.,   it is associated with a Lie algebra. Therefore, the
classification of linear Poisson brackets is equivalent to the classification of Lie
algebras.

\bigskip
\subsection{Local theory}

In his paper \cite{weinstein}  A. Weinstein proved  the so-called
 ``splitting theorem'',  which describes the local behavior of Poisson
 manifolds.

\smallskip

\begin{theorem}
 Let $x_0$ be a point in a Poisson manifold $M$. Then

near $x_0$, $M$ is isomorphic to a product $S \times N$ where $S$ is symplectic,
  $N$ is a Poisson manifold, and the rank of $N$ at $x_0$ is zero.
\end{theorem}
\noindent
$S$ is called the  symplectic leaf through  $x_0$ and $N$ is called the
  transverse Poisson structure at  $x_0$. $N$ is unique up to isomorphism.
So, through each point $x_0$ passes a symplectic leaf $S_{x_0}$ whose dimension
equals the rank of the Poisson structure  on $M$ at $x_0$. The bracket
on the transverse manifold $N_{x_0}$ can be calculated using  Dirac's constraint
bracket formula.

\bigskip
\noindent
\begin{theorem}
  Let $x_0$ be a point in a Poisson manifold $M$

and let $U$ be a neighborhood of $x_0$ which is isomorphic to  a product

$S\times N$ as in Weinstein's splitting Theorem. Let $p_i$, $ i=1,
\dots, 2n$ be functions on $U$ such that

\begin{equation}
 N = \{ x\in U \, \vert \,  p_i (x) = {\rm constant} \} \ .
\end{equation}
\noindent

Denote by $P = P_{ij} = \{ p_i, p_j \}$ and by $P^{ij}$

the inverse matrix of $P$. Then the bracket formula for the

transverse Poisson structure on $N$ is given as follows:

\begin{equation}
\{ F, G \}_N (x) = \{ \hat{F}, \hat{G} \}_M (x) + \sum_{i,j}^{2n}
\{ \hat{F}, p_i \}_M (x) P^{ij } (x) \{ \hat{G}, p_j \}_M (x) \label{f25}
\end{equation}

\noindent

for all $x \in N$, where $F$, $G$ are functions on $N$ and $\hat{F}$, $\hat{G}$

are extensions of $F$ and $G$ to a neighborhood of $M$. Dirac's formula depends

only on $F$, $G$, but not on the extensions $\hat{F}$, $\hat {G}$.
\end{theorem}

When $\mu$ is an element of ${\cal G}^*$, where ${\cal G}$ is a semi--simple Lie algebra, Cushman and Roberts
proved that, in suitable coordinates, the transverse structure is polynomial; see \cite{cushman} and \cite{damianou10}.
\bigskip
\subsection{Cohomology}
Cohomology of Lie algebras was introduced by Chevalley and Eilenberg in \cite{chevalley}.
Let ${\cal G}$ be a Lie algebra and let $\rho$ be a representation of $\cal G$ with representation
space $V$. A $q$-linear skew-symmetric mapping of $\cal G$ into $V$ will be called a
$q$-dimensional $V$-cochain. The $q$-cochains form a space $C^q({\cal G}, V)$. By definition,
$C^0 ({\cal G},V)=V$.

\smallskip
 We define a coboundary operator
$ \delta = \delta _q   :  C^q ( {\cal G}, V)  \to  C^ {q+1}  ( {\cal  G} , V)$  by the formula

\begin{equation}
\begin{array}{rcl}
(\delta f)(x_0, \dots, x_q) &= &\sum_{i=0}^q (-1)^q \rho(x_i) f(x_0, \dots, \hat x_i, \dots, x_q)+ \\
                             && \\
  & &\sum_{i<j} (-1)^{i+j} f([x_i, x_j], x_0, \dots, \hat x_i, \dots, \hat x_j, \dots, x_q) \ ,
\end{array}
\end{equation}
where $f \in C^q ({\cal G},V)$ and $x_0, \dots, x_q \in {\cal G}$.
 As can be easily checked $\delta_{q+1} \circ \delta_q=0$ so that
$\{ C^q ({\cal G}, V), \delta_ q \}$ is an algebraic complex. Define $Z^q({\cal G}, V)$ the space
of $q$-cocycles as the kernel of $\delta : C^q \to C^{q+1} $ and the space $B^q({\cal G}, V)$ of
$q$-coboundaries as the image $\delta C^{q-1}$. Since $\delta \delta =0$ we can define
\begin{equation}
H^q ({\cal G}, V) = {  Z^{q} ({ \cal G},V) \over B^{q} ({\cal G},V) } \ .
\end{equation}

\smallskip

 Lichnerowicz \cite{lich} considers the following cohomology defined on the tensors of a Poisson manifold.
Let $(M, \pi)$ be a Poisson manifold. If we set $B=C=\pi$ in (\ref{f9}) we get
\begin{equation}
[\pi, [\pi, A]]=0
\end{equation}
for every tensor $A$. Define a coboundary operator $\partial_{ \pi}$  which assigns to each $p$-tensor
$A$,  a $(p+1)$-tensor $\partial_{\pi} A$ given by
\begin{equation}
\partial_{\pi}A=-[\pi, A] \ .
\end{equation}
We have $\partial_{\pi}^2 A=[\pi, [\pi,A]]=0$ and therefore $\partial_{\pi}$ defines a cohomology.
 An element $A$ is a $p$-cocycle if $[\pi, A]=0$. An element $B$ is a $p$-coboundary if $B=[\pi,C]$,
 for some $(p-1)$-tensor $C$. Let
\begin{equation}
Z^n(M, \pi) = \{ A \in T_n \, \vert \,  [\pi, A]=0  \}
\end{equation}
and
\begin{equation}
B^n(M, \pi)= \{ B \,  \vert \,  B=[\pi, C], \  \ C \in T_{n-1} \} \ .
\end{equation}
The quotient
\begin{equation}
H^n(M, \pi)= {  Z^n(M, \pi) \over B^n(M, \pi) }
\end{equation}
is the $n$th cohomology group.

\bigskip
Let ${ \cal G}$ be a Lie algebra and consider the Lie-Poisson manifold
  ${\cal G}^*$.  Define a
representation $\rho$ of ${\cal G}$ with values in $C^{\infty} ({\cal G}^*)$ by
\begin{equation}
\rho (x_i) f= \sum_{j,k} c_{ij}^k { \partial f \over \partial x_j}
\end{equation}
where $x_i$ denotes coordinates on ${\cal G}^*$ and at the same time elements of a basis for
${\cal G}$. In other words,  $\rho(x_i)f =\{ x_i, f \}$,  where the bracket is the Lie-Poisson
 bracket on ${\cal G}^*$. We denote the $n$th cohomology group of ${\cal G}$ with respect
to this representation by
\begin{equation}
H^n ({\cal G}, C^{\infty}( {\cal G}^* )) \ .
\end{equation}
We have the following result:

\smallskip
\noindent

\begin{theorem}
\begin{equation}
H^n ({\cal G}^*, \pi) \cong H^n ({\cal G}, C^{\infty}( {\cal G}^* )) \ .
\end{equation}
\end{theorem}
The proof can be found in \cite{koszul} or \cite{damianou2}.

\subsection{Bi--Hamiltonian systems}
\bigskip
\noindent

\begin{proposition}
\smallskip

Let $(M, \pi_1)$, $(M, \pi_2)$ two Poisson structures on $M$. The following are equivalent:

\smallskip
\noindent
{\it i)} $\pi_1 +\pi_2$ is Poisson.

\smallskip
\noindent
{\it ii)} $[\pi_1, \pi_2]=0$.

\smallskip
\noindent
{\it iii)} $\partial_{\pi_1} \, \partial_{\pi_2}= -  \partial_{\pi_2} \, \partial_{\pi_1}$.

\smallskip
\noindent
{\it iv)} $\pi_1 \in Z^2 (M, \pi_2)$,  $\pi_2 \in Z^2(M, \pi_1)$.

\end{proposition}

\bigskip
Two tensors which satisfy the equivalent conditions are said to form a  Poisson pair  on $M$. The
corresponding Poisson brackets are called  compatible.

\bigskip
\newtheorem{lemma}{Lemma}
\begin{lemma}
Suppose $\pi_1$ is Poisson and $\pi_2 =  L_{X}  \pi_1 =-\partial_{\pi_1} X $ for some vector field $X$.
 Then $\pi_1$ is compatible with $\pi_2$.
 \end{lemma}

\smallskip
{\bf Proof:}
\bd
[\pi_1, \pi_2]=[\pi_1, -[\pi_1 , X]]= -\partial_{\pi_1} \partial_{\pi_1} X=0  \ .
\ed
 \hfill \rule{5pt}{5pt}

\smallskip
\noindent
 If  $\pi_1$  is
symplectic, we call the Poisson pair non-degenerate.  If we assume a non-degenerate pair we make
the following definition: The  recursion operator associated with a non-degenerate pair is the
$(1,1)$-tensor ${\cal R}$ defined by
\begin{equation}
{\cal R}=\pi_2  \pi_1^{ -1}.
\end{equation}

\bigskip
A   bi-Hamiltonian system is defined by specifying two Hamiltonian functions $H_1$, $H_2$ satisfying
\begin{equation}
X=\pi_1  \,  \nabla H_2 = \pi_2 \, \nabla  H_1 \ .
\end{equation}
We have the following result due to Magri \cite{magri}:

\smallskip
\noindent

\begin{theorem}
 Suppose  that we have a non--degenerate bi-Hamiltonian system on a manifold $M$, whose first
cohomology group is trivial. Then,  there exists a hierarchy of mutually commuting
functions $H_1, H_2, \dots $,  all in involution with respect to both brackets. If we denote by
$\chi_i$ the Hamiltonian vector field generated by $H_i$  with respect to the initial bracket $\pi_1$ then
the $\chi_i$  generate mutually commuting bi-Hamiltonian flows, satisfying the
Lenard recursion relations
\begin{equation}
\chi_{i+j}= \pi_i \,   \nabla H_j  \ ,
\end{equation}
where $\pi_{i}={\cal R}^{i-1} \pi_1$ are the higher order Poisson tensors.

\end{theorem}

 For further  information  on bi-Hamiltonian systems relevant to
Toda type systems  see \cite{falqui},  \cite{gelfand}, \cite{smirnov}, \cite{suris}.

\bigskip
\subsection{Master Symmetries}
\bigskip
We recall the definition and basic properties of master symmetries following
 Fuchssteiner \cite{fuchssteiner}.
Consider a differential equation on a manifold $M$  defined by a vector field $\chi$.
 We are mostly interested in the case where $\chi$ is a Hamiltonian vector field. A
 vector field $Z$ is a   symmetry of the equation  if
\begin{equation}
[Z, \chi]=0 \ .
\end{equation}
If $Z$ is time dependent, then a more general condition is
\begin{equation}
{ \partial Z \over \partial t} + [ Z, \chi]=0  \ .
\end{equation}
 A vector field $Z$ is called a master symmetry if
\begin{equation}
[[Z, \chi], \chi]=0 \ ,
\end{equation}
but
\begin{equation}
[Z, \chi] \not= 0  \ .
\end{equation}

Master  symmetries were first introduced by Fokas and Fuchssteiner in \cite{fokas1}  in
connection with the Benjamin-Ono Equation.

Suppose that we have a bi-Hamiltonian system defined by the  Poisson tensors $\pi_1$, $\pi_2$ and
the Hamiltonians $H_1$, $H_2$.
 Assume that $\pi_1$ is symplectic.  We define
the recursion operator ${\cal R} = \pi_2 \pi_1^{-1}$,  the higher flows
\begin{equation}
\chi_{i} = {\cal R}^{i-1} \chi_1 \ ,
\end{equation}
and the higher order Poisson tensors
\bd
\pi_i = {\cal R}^{i-1} \pi_1 \ .
\ed

 For a non-degenerate
bi-Hamiltonian system, master symmetries can be generated using a
method due to W. Oevel \cite{oevel2}.

\begin{theorem}
 Suppose that   $X_0$ is a conformal symmetry for both  $\pi_1$, $\pi_2$
and $H_1$, i.e.,  for some scalars $\lambda$, $\mu$, and $\nu$ we
have
\bd
{\cal L}_{X_0} \pi_1= \lambda \pi_1,  \ \ \ \ \ {\cal L}_{X_0}
\pi_2 = \mu \pi_2, \ \ \ \ \ \ \ \ {\cal L}_{X_0} H_1 = \nu H_1 \ .
\ed
 Then the vector fields
\bd
X_i = {\cal R}^i X_0
\ed
are master symmetries  and  we have

\smallskip
\noindent (a)

\bd
{\cal L}_{X_i} H_j = (\nu +(j-1+i) (\mu -\lambda)) H_{i+j}
\ed

\smallskip
\noindent (b)

\bd
{\cal L}_{X_i} \pi_j = (\mu +(j-i-2) (\mu -\lambda)) \pi_{i+j}
\ed

\smallskip
\noindent (c)
\bd
[X_i, X_j]= (\mu - \lambda) (j-i) X_{i+j}  \ .
\ed

\end{theorem}

\section{$A_N$ TODA LATTICE}

\subsection{Definition of the system}
Equation (\ref{a2}) is    the classical, finite, nonperiodic Toda lattice. This system was investigated in
  \cite{flaschka1},  \cite{flaschka2},  \cite{henon},  \cite{manakov}, \cite{moser}, \cite{moser2}, \cite{toda}
and numerous of other papers that are impossible to list here.

This type of Hamiltonian was considered first by Morikazu Toda  \cite{toda}. The original Toda lattice can be
viewed as a discrete version of the Korteweg--de Vries equation. It is called a lattice as in atomic lattice since
interatomic interaction was studied. This  system also appears  in Cosmology. It appears also
in the work of Seiberg and Witten on supersymmetric Yang--Mills theories and it has applications
in analog computing and numerical computation of eigenvalues. But the Toda lattice is mainly a
theoretical mathematical model which is important due to the rich mathematical structure encoded in it.

Hamilton's equations become

\bd
\begin{array}{lcl}
\dot q_j =p_j    \\
\dot p_j=e^{ q_{j-1}-q_j }- e^{q_j- q_{j+1}}  \ .
\end{array}
\ed

\smallskip
\noindent
The system is  integrable. One can find a set of  independent functions
$\{  H_1, \dots,  H_N \} $  which are constants of motion for Hamilton's equations.
To  determine the constants of motion, one uses Flaschka's transformation:
\be
  a_i  = {1 \over 2} e^{ {1 \over 2} (q_i - q_{i+1} ) }  \ \ , \ \ \ \ \ \ \ \ \ \ \ \ \ \
             b_i  = -{ 1 \over 2} p_i  \ .     \label{f1}
\ee

\smallskip
\noindent
Then

\be
\begin{array}{lcl}
 \dot a _i& = & a_i \,  (b_{i+1} -b_i )    \\
   \dot b _i &= & 2 \, ( a_i^2 - a_{i-1}^2 ) \ .  \label{f22}
\end{array}
\ee

\smallskip
\noindent
These equations can be written as a Lax pair  $\dot L = [B, L] $, where $L$ is the Jacobi matrix

\bd
 L= \pmatrix { b_1 &  a_1 & 0 & \cdots & \cdots & 0 \cr
                   a_1 & b_2 & a_2 & \cdots &    & \vdots \cr
                   0 & a_2 & b_3 & \ddots &  &  \cr
                   \vdots & & \ddots & \ddots & & \vdots \cr
                   \vdots & & & \ddots & \ddots & a_{N-1} \cr
                   0 & \cdots & & \cdots & a_{N-1} & b_N   \cr } \ ,
\ed

\smallskip
\noindent
and

\bd
    B =  \pmatrix { 0 & a_1 & 0 & \cdots & \cdots &  0 \cr
                 -a_1 & 0 & a_2 & \cdots & & \vdots  \cr
                    0  & -a_2 & 0 & \ddots &  & \cr
                    \vdots &  & \ddots & \ddots & \ddots & \vdots \cr
                     \vdots & & &  \ddots & \ddots & a_{N-1} \cr
                     0 & \cdots &\cdots &  & -a_{N-1}  & 0 \cr } \ .
\ed

\bigskip
\noindent
This is an example of an isospectral deformation; the entries of $L$ vary over time but the eigenvalues  remain constant.
 It follows that the  functions $ H_i={1 \over i} {\rm tr} \, L^i$ are  constants of
motion.  We note that
\bd
H_1=\sum_{i=1}^N b_i=-{1 \over 2} (p_1+p_2+ \dots + p_N)  \ ,
\ed
corresponds to the total momentum and
\bd
H_2=  H(q_1, \dots, q_N, \,  p_1, \dots, p_N) = {1\over 2} \sum_{i=1}^N b_i^2+  \sum_{i=1}^{N-1} a_i^2
\ed
is the Hamiltonian.

Consider ${\bf R}^{2N} $  with  coordinates $(q_1, \dots , q_N, p_1, \dots, p_N)$, the
standard symplectic bracket
\be
\{ f, g \}_s = \sum_{i=1}^N \left( {\partial f \over \partial q_i} {\partial g
\over \partial p_i} - {\partial f \over \partial p_i} { \partial g \over
\partial q_i} \right)   \ ,  \label{f12}
\ee

and the mapping $F: {\bf R}^{2N} \to {\bf R}^{2N-1}$ defined by

\bd
 F:  (q_1, \dots, q_N, p_1, \dots, p_N) \to (a_1,  \dots, a_{N-1}, b_1, \dots, b_N) \ .
\ed

There exists a bracket on ${\bf R}^{2N-1} $ which satisfies
\bd
\{f, g \} \circ F  = \{ f \circ F, g \circ F \}_s \ .
\ed

It  is a bracket  which (up to a constant multiple) is given by
\be
\begin{array}{lcl}
\{a_i, b_i \}& =&-a_i  \\
\{a_i, b_{i+1} \} &=& a_i  \label{f4}   \ ;
\end{array}
\ee
all other brackets are zero.
$H_1=b_1+b_2 + \dots +b_N$ is the only Casimir.   The Hamiltonian in this bracket is
 $H_2 = { 1 \over 2}\  { \rm tr}\  L^2$.  We also have involution of invariants,  $ \{  H_i, H_j \}=0$.
 The Lie algebraic interpretation of this bracket can be found in \cite{kostant}.
  We   denote this bracket by $\pi_1$.

The quadratic Toda bracket appears in conjunction with isospectral deformations of
Jacobi matrices. First, let $\lambda $ be an eigenvalue of $L$ with normalized
eigenvector $v$. Standard perturbation theory shows that

\bd
\nabla \lambda = (2 v_1 v_2, \dots, 2  v_{N-1} v_N, v_1^2, \dots, v_N^2)^T :=U^{\lambda} \ ,
\ed

where $\nabla  \lambda $ denotes $( { \partial \lambda \over \partial a_1}, \dots, {\partial \lambda
 \over \partial b_N})$. Some manipulations show that $U^{\lambda}$ satisfies

\bd
\pi_2 \,  U^{\lambda} = \lambda \,  \pi_1 \, U^{\lambda} ,
\ed
where $\pi_1$ and $\pi_2$ are skew-symmetric matrices. It turns out that $\pi_1$ is the matrix of
coefficients of the Poisson tensor (\ref{f4}), and $\pi_2$, whose coefficients are quadratic functions
 of the $a$'s and $b$'s, can be used to define a new Poisson tensor.
The quadratic Toda bracket appeared in a paper of Adler  \cite{adler} in 1979.
 It is a Poisson bracket in which the Hamiltonian vector field generated by $H_1$ is the
same as the Hamiltonian vector field generated by $H_2$ with respect to the $\pi_1$ bracket.  The defining relations are

\be
\begin{array}{lcl}
\{a_i, a_{i+1} \}&=&{ 1 \over 2} a_i a_{i+1} \\
\{a_i, b_i \} &=& -a_i b_i                    \\
\{a_i, b_{i+1} \}&=& a_i b_{i+1}    \\
\{b_i, b_{i+1} \}&=& 2\, a_i^2   \ ;  \label{f15}
\end{array}
\ee
all other brackets are zero.  This bracket has ${\rm det} \, L$ as Casimir and $H_1 ={\rm tr}\, L$
 is the Hamiltonian. The eigenvalues of $L$ are still in involution.
Furthermore, $\pi_2$ is compatible with $\pi_1$.
 We also have
\be
\pi_2 \nabla H_l = \pi_1 \nabla H_{l+1}  \ . \label{a98}
\ee

These relations are similar to the Lenard relations for  the KdV equation; they are
generally called the Lenard relations. Taking $l=1$ in (\ref{a98}), we conclude  that the Toda
lattice is bi--Hamiltonian. In fact, using results from \cite{damianou5},  we can prove that the
Toda lattice is multi--Hamiltonian:
\be
\pi_2 \,  \nabla H_1=\pi_1 \,  \nabla H_2 =\pi_0 \,  \nabla H_3 =\pi_{-1} \,  \nabla H_4= \dots  \label{a6}
\ee

\smallskip
Finally, we remark that further manipulations with the Lenard relations for the infinite
 Toda lattice,  followed by setting all but  finitely many $a_i$, $b_i$ equal to zero,
yield another Poisson bracket, $\pi_3$, which is cubic in the coordinates; see \cite{kup}. The defining
relations for $\pi_3$ are
\be
\begin{array}{lcl}
\{a_i, a_{i+1} \}&=&a_i a_{i+1} b_{i+1}   \\
\{a_i, b_i \}&=&     -a_i b_i^2-a_i^3 \\
\{a_i, b_{i+1} &=&     a_i b_{i+1}^2 +a_i^3 \\
\{a_i, b_{i+2} \}&=&     a_i a_{i+1}^2 \\
\{a_{i+1}, b_i \}&=&     -a_i^2 a_{i+1} \\
\{b_i, b_{i+1} \} &=&     2\, a_i^2 \, (b_i+b_{i+1}) \ ;
\end{array}
\ee
all other brackets are zero. The bracket $\pi_3$ is compatible with both $\pi_1$ and
$\pi_2$ and the eigenvalues of $L$ are still in involution.  The Casimir for this
bracket is $ {\rm tr}\, L^{-1}$.

The multi-Hamiltonian structure of the Toda lattice is well-known. The results
are usually presented either in the natural $(q,p)$ coordinates or in the
more convenient Flaschka coordinates $(a,b)$. In the former case the
hierarchy of higher invariants are generated by the use of a recursion
operator \cite{das}, \cite{fernandes}. In the later case one uses master symmetries as in
\cite{damianou1}, \cite{damianou2}. We have to point out that chronologically every result obtained so
far was done first in Flaschka coordinates $(a,b)$ and then transferred
through the inverse of Flaschka's transformation to the original $(q,p)$
coordinates. This is to be expected  since it is always easier to work with sums of polynomials
than with sums of exponentials.

The sequence of Poisson tensors can be extended to form an infinite hierarchy.
In order to produce the hierarchy of Poisson tensors one uses master symmetries.  The first three Poisson brackets
are precisely the linear, quadratic and cubic brackets  we mentioned above.
If a system is bi-Hamiltonian and one of the brackets is symplectic,  one can find a recursion
operator by inverting the  symplectic  tensor.  The recursion operator is then applied to
the initial symplectic bracket  to produce an infinite sequence.  However, in the case of
Toda lattice (in  Flaschka variables $(a,b)$) both operators are non-invertible and therefore this method
fails.   The absence of a recursion operator for the finite Toda lattice is also mentioned in
Morosi and Tondo \cite{morosi} where a Nijenhuis tensor for the infinite Toda lattice is calculated.
 Recursion operators were introduced  by Olver \cite{olver1}.

\subsection{Multi--Hamiltonian structure}

   In the case of Toda equations,
 the master symmetries map invariant functions  to other invariant functions.
 Hamiltonian vector fields are also preserved. New Poisson brackets are generated
by using Lie derivatives in the direction of these vector fields and they satisfy
interesting deformation relations.  We give a summary of the results:

\begin{itemize}

\item
There exists  a sequence of invariants

\bd
H_1, H_2, H_3, \dots,
\ed

where   $ H_i={1 \over i} {\rm Tr} \, L^i$,
\item
a corresponding sequence of Hamiltonian vector fields
\bd
\chi_1, \ \  \chi_2, \ \  \chi_3, \dots,
\ed
where $\chi_i=\chi_{H_i}$,
\item
a hierarchy of Poisson tensors

\bd
\pi_1, \pi_2, \pi_3, \dots,
\ed
where $\pi_i$ is polynomial, homogeneous, of degree $i$.
\item
Finally, one can determine  a sequence of master symmetries
\bd
X_1, X_2, X_3, \dots,
\ed
which  are used to create the hierarchies through Lie derivatives.
\end{itemize}

 We quote the results from    refs. \cite{damianou1}, \cite{damianou2}.

\begin{theorem}
\hfill \\
\smallskip
\noindent
{\it i) } $\pi_j$, \  $j\ge 1$ are all Poisson.

\smallskip
\noindent
{\it ii) } The functions $H_i$,        $i\ge 1$ are in involution
 with respect to all of the $\pi_j$.

 \smallskip
 \noindent
 {\it iii)}  $X_i (H_j) =(i+j) H_{i+j} $ ,  $i\ge -1$, $j\ge 1$.

 \smallskip
 \noindent
{\it iv)} $L_{X_i} \pi_j =(j-i-2) \pi_{i+j} $,  $i\ge -1$, $j\ge 1$.

\smallskip
\noindent
{\it v)} $ [X_i, \ X_j]=(j-i)X_{i+j}$, $i\ge 0$, $j\ge 0$.

\smallskip
\noindent
{\it vi)} $\pi_j\  \nabla \  H_i =\pi_{j-1}\   \nabla \  H_{i+1} $,
where $\pi_j$ denotes  the Poisson matrix  of the tensor $\pi_j$.
\end{theorem}

To define the vector fields $X_n$  one considers  expressions  of the
form

\begin{equation}
\dot L = [B, L] +L^{n+1}  \ . \label{f44}
\end{equation}

This equation is similar to a Lax equation, but in this case  the
eigenvalues satisfy $\dot {\lambda}= {\lambda}^{n+1}$ instead of $\dot
{\lambda}= 0 $.

We give an outline of the construction of the vector fields $X_n$. We
define $X_{-1} $ to be

\begin{equation}
 \nabla \  H_1 = \nabla \ {\rm Tr}\ L = \sum_{i=1}^N {\partial \ \over \partial b_i}  \ , \label{f5}
\end{equation}

\smallskip
\noindent
and $X_0$ to be the Euler vector field

\begin{equation}
X_0= \sum _{i=1}^{N-1} a_i {\partial \ \over \partial a_i} +
\sum_{i=1}^N b_i {\partial \ \over \partial b_i} \ . \label{f2}
\end{equation}

\smallskip
\noindent
We want $X_1$ to satisfy

\begin{equation}
       X_1 ({\rm Tr}\  L^n) =n {\rm Tr}\  L^{n+1} \ .
\end{equation}

\smallskip
\noindent
One way to find such a vector field is by considering the equation

\begin{equation}
\dot L = [B, L] +L^2  \ . \label{f6}
\end{equation}

\smallskip
\noindent
Note that the left hand side of this equation is a tridiagonal matrix while the
right hand side is pentadiagonal. We look for $B$ as a tridiagonal matrix

\begin{equation}
    B=   \pmatrix { \gamma_1 & \beta_1 & 0 & \cdots & \cdots \cr
                \alpha_1 & \gamma_2 & \beta_2 & \cdots & \cdots \cr
                0 & \alpha_2 & \gamma _3 & \beta _3 & \cdots \cr
                \vdots & \vdots & \ddots & \ddots & \ddots \cr}  \ .
\end{equation}

\smallskip
\noindent
 We want to choose the $\alpha_i$, $\beta _i$ and $\gamma_i$ so that the
 right hand side of equation  (\ref{f6}) becomes tridiagonal. One  simple solution
 is \  $\alpha _n = -(n+1) a_n $, \ $\beta _n = (n+1) a_n $,\  $ \gamma_n =0$.
  The vector field $X_1$  is defined by the right hand side of (\ref{f6}):

\begin{equation}
 X_1 = \sum_{n=1}^{N-1} \dot a_n {\partial \ \over \partial a_n} +
   \sum _{n=1}^N \dot b_n {\partial \ \over \partial b_n} \ ,
 \end{equation}

  \smallskip
  \noindent
  where

  \smallskip
  \noindent

\begin{equation}
 \dot a_n =  - na_n b_n + (n+2) a_n b_{n+1 }
\end{equation}

\begin{equation}
\dot b_n = (2n+3) a_n^2 + (1-2n) a_{n-1}^2 + b_n^2  \ .
\end{equation}

  To construct the vector field $X_2$ we consider the equation
\bd
 \dot L =[B, L]+L^3 \ .
 \ed
 \smallskip
  \noindent
  The calculations are similar to those for $X_1$. The matrix $B$ is now
  pentadiagonal and the system of equations slightly more complicated. The
  result is a vector field
 \bd
  X_2 = \sum_{n=1}^{N-1} \dot a _n  {\partial \ \over \partial a_n }+
  \sum _{n=1}^N \dot b _n {\partial \ \over \partial b_n}
\ed

  \smallskip
  \noindent
  where

$$
 \dot a_n = (2-n)a_{n-1}^2 a_n + (1-n) a_n b_n^2 +a_n b_n b_{n+1} +
       (n+1) a_n a_{n+1}^2+ (n+1) a_n b_{n+1}^2 +a_n^3 +\sigma _n a_n (b_{n+1}-b_n) $$

 $$ \dot b _n = 2 \sigma _n a_n^2 -2 \sigma _{n-1}  a_{n-1}^2 + (2n+2)  a_n^2 b_n +(2n+1) a_n^2 b_{n+1} +
              + (3-2n) a_{n-1}^2 b_{n-1} + (4-2n) a_{n-1}^2 b_n +b_n^3 \ , $$

  \smallskip
  \noindent
  with
\bd
 \sigma _n = \sum_{i=1}^{n-1} b_i \ ,
 \ed
  \smallskip
  \noindent
and $\sigma_1 =0$.

 We continue the sequence of master symmetries for $n \ge 3$ by

\begin{equation}
  [X_1, X_{n-1} ] = (n-2) X_n \ .
\end{equation}

\bigskip
\subsection{Properties of $X_n$ and $\pi_n$.}

\bigskip

It is well known that $\pi_1$, $\pi_2$, $\pi_3$ satisfy Lenard
relations
\begin{equation}
\pi_n \nabla H_l = \pi_{n-1} \nabla H_{l+1} \ , \ \ \ \ n=2,3 \ \ \ \ \forall l \ .
\end{equation}
We want to show that these relations hold for all values of $n$. We denote
the Hamiltonian vector field of $H_l$ with respect to the $n$th bracket
by $\chi_l^n$. In other words,
\begin{equation}
\chi_l^n=[\pi_n, H_l] \ . \label{f11}
\end{equation}
We prove the Lenard relations in an equivalent form.

\bigskip
\begin{proposition}
$\chi_l^{n+1}=\chi_{l+1}^n \ . $
\end{proposition}

\smallskip
\noindent
{\bf Proof:}  To prove this we need the identity
\begin{equation}
[X_1, \chi_l^n]=(n-3) \chi_l^{n+1} + (l+1) \chi_{l+1}^n
\end{equation}
which follows easily from $X_1(H_l)=(l+1) H_{l+1}$ and (\ref{f11}).
Therefore,

\begin{equation}
\begin{array}{lcl}
(n-3) \chi_l^{n+1}& =& [X_1, \chi_l^n]- (l+1) \chi_{l+1}^n \\
                 &=& [X_1, \chi_{l+1}^{n-1} -(l+1) \chi_{l+1}^n  \\
                 &=& (n-4) \chi_{l+1}^n +(l+2) \chi_{l+2}^{n-1} -(l+1)
                       \chi_{l+1}^n \\
                &=&(n-4) \chi_{l+1}^n+(l+2)\chi_{l+1}^n - (l+1) \chi_{l+1}^n \\
                 &=& (n-3) \chi_{l+1}^n    \ \ \ \ \ \ \ \ \ \      \ .
\hfill \rule{5pt}{5pt}
\end{array}
\end{equation}

\smallskip
Using the Lenard relations we can show that the functions $H_n$ are
in involution with respect to all of the brackets $\pi_n$.

\smallskip

\begin{proposition}
$\{H_i, H_j \}_n =0$, where $\{ \ ,\ \}_n$ is the bracket corresponding to
 $\pi_n$.
\end{proposition}

\smallskip
\noindent
{\bf Proof:} First we consider the Lie-Poisson Toda bracket. We have

\begin{equation}
\{H_1, H_j \}=0  \ \ \ \ \ \  \forall j \ ,
\end{equation}
since $H_1$ is a Casimir for $\pi_1$. Suppose that $\{ H_{i-1} , H_j\}=0$
\  $\forall \ j$.

\begin{equation}
\begin{array} {lcl}
i\{ H_i, H_j \} &=& \{ X_1(H_{i-1}), H_j \}  \\
        &=& -[\chi_j^1, [X_1, H_{i-1}]]      \\
       &=&[X_1, \{ H_{i-1}, H_j \}]+[H_{i-1}, (j+1) \chi_{j+1}^1 ] \\
     &=& (j+1) \{ H_{i-1} , H_{j+1} \} \\
     &=& 0    \ .
\end{array}
\end{equation}
Now we use induction on $n$. Suppose
\begin{equation}
\{ H_i, H_j \}_n =0 \ \ \ \  \forall \  i,\, j \ .
\end{equation}

\begin{equation}
\begin{array} {lcl}
\{H_i, H_j \}_{n+1} &=& \chi_i^{n+1}(H_j)  \\
              &=& \chi_{i+1}^n (H_j)  \\
               &=& \{ H_{i+1}, H_j \}_n  \\
             &=& 0   \ . \ \ \ \ \ \
\end{array}
\end{equation}

Of course, one can prove the involution of integrals  without using master symmetries. We present the
classical proof:

In this proof the symbol $\pi_n$ stands for the bundle map $\pi_n^*$ defined by (\ref{f10}).
 We first prove the involution of constants in  the Lie-Poisson Toda bracket.
We use the basic Lenard relation
\bd
\pi_2 \,  dH_1=\pi_1 \,  dH_2 \ .
\ed
Since $H_1$ is a Casimir in the linear bracket, we have $\pi_1 dH_1=0$.
We calculate
\bd
\begin{array} {lcl}
\{H_i, H_j \}_1 &=& <dH_i,\  \pi_1 dH_j>  \\
                &=&-<dH_j,\  \pi_1 dH_i> \\
                &=&-<dH_j, \  \pi_2 dH_{i-1}> \\
                &=&<dH_{i-1}, \  \pi_2 dH_j > \\
                &=&<dH_{i-1}, \  \pi_1 dH_{j+1}> \\
                &=& \{H_1, \  H_{j+i-1} \}_1=0 \ .
\end{array}
\ed
It is also easy to show involution in the second quadratic bracket:
\bd
\begin{array} {lcl}
\{H_i, H_j \}_2 &=& <dH_i,\  \pi_2 dH_j > \\
                &=& <dH_i,\  \pi_1 dH_{j+1} > \\
                &=&\{H_i,\  H_{j+1}\}_1=0 \ .
\end{array}
\ed

The general result follows from the Lenard relations $\pi_n dH_j=\pi_{n-1} dH_{j+1}$ and induction:
\bd
\begin{array} {lcl}
\{H_i, H_j \}_n &=& <dH_i, \pi_n dH_j > \\
                &=& <dH_i, \pi_{n-1} dH_{j+1} > \\
                &=&\{H_i, H_{j+1}\}_{n-1}=0 \ .
\end{array}
\ed

\hfill \rule{5pt}{5pt}

\smallskip
It is straightforward to verify that the mapping
\begin{equation}
f(a_1, \dots, a_{N-1}, b_1, \dots, b_N) =(a_1, \dots, a_{N-1},
1+b_1, \dots, 1+b_N)
\end{equation}
is a Poisson map between $\pi_2$ and $\pi_1+\pi_2$. Since $f$ is a
diffeomorphism, we have the isomorphism
\begin{equation}
\pi_2 \cong \pi_1 + \pi_2 \ .
\end{equation}
In other words, the tensor $\pi_2$ encodes sufficient information for
both the linear and quadratic Toda brackets. An easy induction generalizes
 this result: i.e.,
\begin{proposition}
\begin{equation}
\pi_n \cong \sum_{j=0}^{n-1}  \pmatrix{ n-1 \cr
                                         j }  \pi_{n-j}  \ .
 \end{equation}
\end{proposition}

\smallskip
The function $\mbox{ tr } \, L^{2-n}$, which is well-defined on the open
set $ \mbox{ det}\ L \not= 0$, is a Casimir for $\pi_n$, for $n \ge 3$.
The proof uses the Lenard type relation

\begin{equation}
\pi_n \nabla \lambda = \lambda \pi_{n-1} \nabla \lambda  \label{f13}
\end{equation}
satisfied by the eigenvalues of $L$. To prove the last equation, one
uses the relation
\begin{equation}
\pi_n \sum \lambda_k^{l-1} \nabla \lambda_k = \pi_{n-1} \sum \lambda_k^l
\nabla \lambda_k \ .
\end{equation}
But
\begin{equation}
\sum \lambda_k^{l-1} (\pi_k \nabla \lambda_k -\lambda_k \pi_{n-1} \nabla
\lambda_k)=0 \ ,
\end{equation}
for $l=1,2, \dots, N+1$, has only the trivial solution because the
 (Vandermonde) coefficient determinant is nonzero.

\bigskip
\begin{proposition}
For $n >2$, $\mbox{tr} \, L^{2-n}$ is a Casimir for $\pi_n$ on the open dense
set $\mbox{ det} \, L \not= 0$.
\end{proposition}

\smallskip
\noindent
{\bf Proof:}  For $n=3$,

\begin{equation}
\begin{array}{lcl}
\pi_3 \nabla \mbox{tr}\, L^{-1} &=& \pi_3 \sum_k -{ 1 \over \lambda_k^2} \nabla \lambda_k \\
                    &=& \sum_k  -{ 1 \over \lambda_k^2} \lambda_k \pi_2 \nabla \lambda_k \\
                    &=& - \sum_k \pi_1 \nabla \lambda_k   \\
                     &=& - \pi_1 \nabla \mbox{ tr}\, L =\chi_1^1 =0 \ .
\end{array}
\end{equation}
For $n>3$ the induction step is as follows:

\begin{equation}
\begin{array}{lcl}
\pi_n \nabla \mbox{tr}\, L^{2-n} &=& \pi_n \nabla \sum_k { 1 \over \lambda_k^{n-2}}  \\
                                 &=& \sum_k (2-n) { 1 \lambda_k^{n-1}} \pi_n \nabla \lambda_k \\
                                 &=& \sum_k (2-n) { 1 \over \lambda_k^{n-1} } \lambda_k \pi_{n-1} \nabla \lambda_k \\
                                 &=&  { n-2 \over n-3} \pi_{n-1}  \nabla \mbox{tr}\, L^{3-n}  \\
                                 &=& 0 \ .
\end{array}
\end{equation}

\smallskip

\subsection{The Faybusovich--Gekhtman approach}

In \cite{fayb} Faybusovich and Gekhtman find  another method of generating the multi--Hamiltonian structure for the
Toda lattice. Their method is important and will certainly have applications to other integrable systems, both finite and
infinite dimensional, solvable by the inverse spectral transform.
Their  work shows that the Hamiltonian formalism is built into the spectral theory.

 In the case of Toda lattice,
 the key ingredient is the Moser
map which takes the $(a,b)$ phase space of tridiagonal Jacobi matrices to a new space of variables $(\lambda_i, r_i)$
 where $\lambda_i$ is an eigenvalue of the Jacobi matrix and $r_i$ is the residue of  rational functions that appear
in the solution of Toda equations.
 The
Poisson brackets of Theorem 6 project onto some rational brackets in the space of Weyl functions and in particular, the
Lie--Poisson bracket $\pi_1$ corresponds to the Atiyah--Hitchin bracket \cite{atiyah}. We briefly describe the construction:

Moser in \cite{moser3} introduced the resolvent
\bd
R(\lambda)=\left( \lambda I - L \right)^{-1} \ ,
\ed

 and defined the Weyl function
\bd
f(\lambda)=\left( R(\lambda)e_1, \  e_1 \right) \ ,
\ed

where $e_1=(1,0, \dots, 0)$.

The function $f(\lambda)$ has a simple pole at $\lambda= \lambda_i$ and positive residue at $\lambda_i$ equal to $r_i$:

\bd
f(\lambda)=\sum_{i=1}^n {r_i \over \lambda -\lambda_i} \ .
\ed

The variables $(a,b)$ may be expressed as rational functions of $\lambda_i$ and $r_i$ using a continued fraction
expansion of $f(\lambda)$ which dates back to Stieltjes. Since the computation  of the continued  fraction from the partial
fraction expansion is a rational process the solution is expressed as a rational function of the variables $(\lambda_i, \ r_i)$.
The idea of Faybusovich and Gehtman is to construct a sequence of Poisson brackets on the space $(\lambda_i, r_i)$ whose
image under the inverse spectral transform are the brackets $\pi_i$ defined in Theorem 6. The Lie--Poisson bracket
$\pi_1$  corresponds to the Atiyah--Hitchin bracket on Weyl functions which in coordinate free form is written as
\bd
\{ f(\lambda), \ f(\mu) \}={ \left( f(\lambda)-f(\mu) \right)^2 \over \lambda- \mu} \ .
\ed

A rational function of the form ${q(\lambda) \over p(\lambda)}$ is determined uniquely by the dinstinct eigenvalues
of $p(\lambda)$,  $\lambda_1, \dots, \lambda_n$ and values of $q$ at these roots. The residue $r_i$ is equal to
 ${ q(\lambda_i) \over p^{\prime}(\lambda_i)}$
and therefore we may choose

\bd
\lambda_1, \dots, \lambda_n, q(\lambda_1), \dots, q(\lambda_n)
\ed
  as global coordinates on the
space of rational functions (of the form ${q \over p}$ with $p$ having simple roots and
$q, p$ coprime). We have to remark that the image of the Moser map  is a much larger set.

 The  $k$th Poisson bracket is defined by

\bd
\begin{array}{lcl}
\{\lambda_i, \ q(\lambda_i) \}& =&-\lambda_i^k q(\lambda_i)    \\
\{q(\lambda_i),\  q(\lambda_j) \}& = & \{ \lambda_i,\  \lambda_j \}=0 \ .
\end{array}
\ed
Let us denote this bracket by $w_k$.

On the other hand in  \cite{damianou1}, page 108,  there is a definition of vector fields on the space
 of eigenvalues of the
Jacobi matrix which are projections of the master symmetries $X_i$.
They are defined by
\bd
e_i=\sum_{j=1}^N \lambda_j^{i+1} { \partial \over \partial \lambda_j} \ .
\ed
One verifies easily that these vector fields satisfy the usual Virasoro type relation
\bd
[e_i, \ e_j]=(j-i) e_{i+j} \ .
\ed
If we denote by $F$ the function which sends the Jacobi matrix to its eigenvalues then
\bd
dF(X_1)=e_1
\ed
\bd
dF(X_2)=e_2 \ .
\ed
Therefore, it follows  by induction that
\bd
dF(X_i)=e_i \ .
\ed

Faybusovich and Gekhtman used the brackets $w_i$ and the vector fields $e_j$ to obtain the analogue of
Theorem 6 in the space of rational functions. The relations obtained correspond to the relations of Theorem
6 under the inverse of the  Moser map.

The explicit formulas for for the brackets $w_k$ can be deduced easily from the formulas in \cite{fayb}. They are
\bd
\begin{array}{lcl}
\{ r_i, \  r_j \}_k& =&\frac {\lambda_i^k+\lambda_j^k}{\lambda_i-\lambda_j} r_i r_j     \\
\{ r_i,\  \lambda_i \}_k& = & \lambda_i  r_i \\
\{ \lambda_i,\  \lambda_j \}_k&=& 0 \ .
\end{array}
\ed

In a recent paper \cite{vaninsky} Vaninsky has also explicit formulas in $(\lambda_i,\, r_i)$ coordinates
for the initial bracket $w_1$.
\subsection{A Theorem of Petalidou }

Finally, we mention an interesting result of Petalidou \cite{petalidou}. She proves the  following Theorem:
Suppose that $(M, \Lambda_0, \Lambda_1)$ is a bi--Hamiltonian manifold of odd dimension and let $p$ be a point
in $M$ of corank 1. If there exists locally  an infinitesimal automorphism $Z_0$ of $\Lambda_0$  which is
transverse to the symplectic leaf through $p$ and a vector field $Z_1$ which depends on a parameter $t$ such that
\bd
[\Lambda_1, Z_1]+[Z_1, {\partial \over \partial t}]\wedge Z_1=0 \ ,
\ed
and
\bd
[\Lambda_0, Z_1]+[\Lambda_1, Z_0]+[Z_1, {\partial \over \partial t}]\wedge Z_0=0 \ ,
\ed
then one can find a symplectic realization of both $\Lambda_0$ and $\Lambda_1$ by a pair of symplectic brackets
$\hat{ \Lambda}_0$, $\hat{ \Lambda}_1$
given by
\bd
\hat{ \Lambda}_i= \Lambda_i+Z_i \wedge {\partial \over \partial t} \ \ \ \ \ i=1,\, 2  \  \ .
\ed

This result applies in the case of the Toda lattice by taking $Z_0=X_{-1}$ and $Z_1=X_0$.
 She obtains  symplectic realizations of $\pi_1$ and $\pi_2$. Furthermore, she constructs symplectic
realizations of the whole sequence

\bd
\pi_1, \pi_2, \pi_3, \dots
\ed

of Theorem 6. The corresponding symplectic sequence is given by
\bd
\hat{\pi}_k=\pi_k+X_{k-2} \wedge {\partial \over \partial t} \ .
\ed

The tensors $\hat{\pi}_k$ may be generated by a  recursion operator since the initial tensor,  which is a multiple of
$\hat{\Lambda}_0$, is invertible.

\subsection{A recursive process of Kosmann--Schwarzbach and Magri }

In \cite{magri2}  Y. Kosmann--Schwarzbach and F. Magri consider the relationship between Lax and bi--Hamiltonian
formulations of integrable  systems. They introduce an equation, called the   Lax-Nijenhuis equation,
 relating the Lax matrix with the  bi--Hamiltonian pair  and they show that
every operator that satisfies that equation  satisfies also the Lenard recursion relations. They derive the
multi--Hamiltonian structure of the Toda lattice by defining a matrix $M$ and a vector $\lambda_{0}$ which arise
by manipulating the Lax--Nijenhuis equation. They show that
\bd
\pi_2= M \pi_1 + X \otimes \lambda_{0} \ ,
\ed
where $X$ is the Hamiltonian vector field $\chi_2$. In the next step of the recursive process they show that
\bd
\pi_3=M \pi_2 + X \otimes  \lambda_{1} \ ,
\ed
where $\lambda_{1}=M \lambda_{0}$.
In general,
\bd
\pi_{i+1}= M \pi_i + X \otimes  M^{(i-1)} \lambda_{0} \ .
\ed

\section{LIE GROUP SYMMETRIES OF THE TODA LATTICE}

Sophus Lie introduced his theory of continuous groups in order to study symmetry properties of differential equations.
 His approach allowed a unification of existing methods for solving ordinary differential equations as well as
classifications of symmetry groups of partial and ordinary differential equations.
 A symmetry group of a system of differential equations
 is  a Lie group acting on the space of independent and dependent
variables in such a way that solutions are mapped into other solutions.
Knowing the symmetry  group allows one to determine some special types of solutions
 invariant under a subgroup of the full symmetry group, and in some cases
  one can  solve the equations completely.
 Lie's
methods have been developed into powerful tools for examining  differential equations through group analysis.
 In many cases, symmetry groups are the only known means for finding concrete solutions to complicated equations.
The method applies of course to the case of Hamiltonian or Lagrangian systems, both autonomous and time dependent.
 Recently, the immense amount of computations needed for
determining symmetry groups of concrete systems has been greatly reduced by the implementation of computer algebra packages for symmetry analysis of differential equations.
 The symmetry approach to solving differential equations can
be found, for example, in the books of Olver \cite{olver2},   Bluman and Kumei \cite{bluman},  Ovsiannikov \cite{ovs}
and Ibragimov \cite{ib}.

Some properties of master symmetries  are clear: They preserve constants  of
motion, Hamiltonian vector fields and they generate a hierarchy of Poisson
brackets. We are interested in the following problem : Can one find a symmetry group of the system whose
infinitesimal generator is a given master symmetry?
In the case of Toda equations the answer
is negative. However, in this section we find a sequence consisting of
time dependent evolution vector fields whose time independent part is
a master symmetry. Each master symmetry $X_n$ can be written in the
form $Y_n + t Z_n$ where $Y_n$ is a time dependent symmetry and $Z_n$ is
 a  time independent Hamiltonian symmetry (i.e. a Hamiltonian vector field).

 In other words,  we find an  infinite sequence of evolution vector
   fields that are symmetries of equations (\ref{f22}). We do not know if every symmetry
   of Toda equations is included in this sequence.

  We  begin by writing  equations (\ref{f22}) in the form

\bd
\Gamma _j = \dot a_j -a_j b_{j+1} +a_j b_j =0   \quad
\Delta _j = \dot b_j -2 a_j^2 +2 a_{j-1}^2  =0  \ .
\ed

\smallskip
\noindent
We look for symmetries of  Toda equations. i.e.  vector fields of the
form

\bd
{\bf v} = \tau {\partial \over \partial t} + \sum_{j=1}^{N-1} \phi_j
{\partial   \over \partial a_j} + \sum_{j=1}^N \psi_j {\partial \over \partial b_j } \
\ed

\smallskip
\noindent
that  generate the symmetry group of the Toda equations.
\smallskip
\noindent
The first prolongation of ${\bf v}$ is

\bd{\rm pr}^{(1)} {\bf v} = {\bf v} + \sum_{j=1}^{N-1} f_j { \partial \over
\partial \dot a_j } + \sum_{j=1}^N g_j {\partial \over \partial \dot b_j} \ ,
\ed

\smallskip
\noindent
where
\bd
\begin{array}{lcl}
f_j&= &\dot{\phi}_j -\dot {\tau} \dot a_j  \\
g_j& = &\dot{\psi}_j -\dot {\tau} \dot b_j  \ .
\end{array}
\ed

 The infinitesimal condition for a group to be a symmetry of the system is
\bd
{\rm pr}^{(1)} (\Gamma_j)   = 0   \ \ \ \ \ \ \ \ \ \ \
 {\rm pr}^{(1)} (\Delta_j) =0  \ .
\ed

\smallskip
\noindent
Therefore we obtain the equations
 \be
\dot{\phi}_j - \dot{\tau} a_j (b_{j+1} -b_j) + \phi_j (b_j-b_{j+1} ) +
a_j \psi_j -a_j \psi_{j+1} =0  \label{j1} \ ,
\ee

\be
 \dot {\psi}_j - 2 \dot {\tau} (a_j^2 -a_{j-1}^2) -4 a_j \phi_j + 4 a_{j-1} \phi_{j-1} =0 \ . \label{j2}
\ee

\smallskip
\noindent
We first give some obvious solutions :

\smallskip
\noindent
{\it i)} \ \ $\tau  =0$,\ \  $ \phi_j =0 $,\  $\psi _j =1 $.
\bigskip
This is the vector field $X_{-1}$.

\smallskip
\noindent
{\it ii)} \ \ $\tau =-1$, \ $\phi_j=0$, \ $ \psi_j=0$.
\bigskip
The resulting vector field  is the time translation
$-{\partial \over \partial t} $ whose evolutionary representative is
\bd
 \sum_{j=1}^{N-1} \dot a_j {\partial \over \partial a_j} + \sum_{j=1}^N
\dot b_j {\partial \over \partial b_j} \ .
\ed

\smallskip
\noindent
This is the Hamiltonian vector field $\chi_{H_2} $. It generates a Hamiltonian
symmetry group.

\smallskip
\noindent
{\it iii)} \ \ $\tau = -1$, \ $ \phi_j=a_j$, \ $ \psi_j= b_j$.
\bigskip
Then

\bd
{\bf v} = -{\partial \over \partial t} + \sum_{j=1}^{N-1} a_j {\partial \over \partial  a_j}
+ \sum_{j=1}^N b_j {\partial \over \partial b_j} =- {\partial \over \partial t} + X_0 \ .
\ed

\smallskip
\noindent
This vector field generates the same symmetry as the evolutionary vector field
\bd
X_0 + t \chi_{H_2}  \ .
\ed

\smallskip
 We next look for some non obvious solutions. The vector field $X_1$ is not
 a symmetry, so we add a term which depends on time. We try
\bd
\phi_j = -j a_j b_j +(j+2) a_j b_{j+1} +
                    t (a_j a_{j+1}^2 +a_j b_{j+1}^2-a_{j-1}^2 a_j -a_j b_j^2 )
\ed

\bd
 \psi _j = (2j+3) a_j^2 +(1-2j) a_{j-1}^2 +b_j^2 +
  t (2 a_j^2 b_{j+1} +2 a_j^2 -2a_{j-1}^2 a_j -2a_{j-1}^2 b_j )  \ ,
\ed
\smallskip
\noindent
and $\tau =0$.

 \smallskip
 \noindent
 A tedious but straightforward calculation shows that $\phi_j$, $\psi_j$
 satisfy (\ref{j1}) and (\ref{j2}). It is also straightforward to check that the vector field
\bd
\sum \phi_j {\partial \over \partial a_j} +\sum \psi_j {\partial \over \partial b_j}
\ed
  is precisely  equal to $X_1 + t \chi_{H_3}$. The pattern suggests that
  $X_n + t \chi_{H_{n+2} } $ is a symmetry of Toda equations.

 \bigskip
  \smallskip
  \noindent
  \begin{theorem} The vector fields $X_n +t \chi_{n+2} $ are symmetries of
  Toda equations for $n\ge -1$.
\end{theorem}

  \bigskip
  \noindent
  {\bf Proof :}  Note that $\chi_{H_1} =0$ because $H_1$ is a Casimir for the
    Lie-Poisson  bracket.  We use the formula
\begin{equation}
 [X_n, \chi_l] =(l-1) \chi_{n+l} \ .
\end{equation}

\smallskip
\noindent
In particular, for $l=2$, we have $[X_n, \chi_2] =\chi_{n+2} $.

\bigskip
Since the Toda flow is Hamiltonian, generated by $\chi_2$, to
  show that $Y_n =X_n +t \chi_{n+2} $ are symmetries of Toda equations we must
verify the equation
\begin{equation}
{\partial Y_n \over \partial t} + [\chi_2, Y_n] = 0 \ .
\end{equation}

\smallskip
\noindent
But
\begin{equation}
\begin{array}{lcl}
{\partial Y_n \over \partial t} + [\chi_2,\  Y_n] &= &  {\partial Y_n \over
 \partial t} +[ \chi_2, \  X_n +t \chi_{n+2}   ]  \\
                                  & = &   \chi_{n+2} -[X_n,\  \chi_2] \\
                                  & = &\chi_{n+2} -\chi_{n+2} =0  \ .
\end{array}
\end{equation}
\hfill \rule{5pt}{5pt}

\newpage
\section{THE TODA LATTICE IN NATURAL COORDINATES}

In this section we define the positive and negative Toda hierarchies for the Toda lattice  in $(q,p)$ variables.
We follow reference \cite{damianou5}.

\subsection{The Das--Okubo--Fernandes approach}

Another approach, which explains the relations of Theorem 6
 is adopted in  Das and Okubo \cite{das},  and Fernandes
 \cite{fernandes}.
 In principle, their  method is
general and may work for other finite dimensional systems as well.
 This approach was
also used in \cite{costa} by da Costa and Marle in the case of the
Relativistic Toda lattice. The procedure is the following:
One defines a second Poisson bracket in the space of canonical variables
$(q_1, \dots, q_N, p_1,\dots, p_N)$.
This gives rise to
a recursion operator.
 The presence of a conformal symmetry
 as defined in Oevel \cite{oevel2} allows one, by using the recursion
 operator, to generate an infinite sequence of master symmetries. These, in
turn, project to the  space of the new variables $(a,b)$ to
produce a sequence of master symmetries in the reduced space.

Let $\hat{J}_1$ be the symplectic bracket (\ref{f12}) with Poisson  matrix

\bd
\hat{J}_1 = \pmatrix { 0 &  I \cr
                      -I &   0}   \ ,
\ed

where $I$ is the $N \times N$ identity matrix. We use $J_1=4 \hat{J}_1$. With this convention the
bracket $J_1$ is mapped precisely onto the bracket $\pi_1$ under the Flaschka transformation (\ref{f1}).
We define $\hat{J}_2$ to be the tensor
\bd
\hat{J}_2 = \pmatrix { A &  B \cr
                      -B &  C}   \ ,
\ed
where $A$ is  the skew-symmetric matrix defined by $a_{ij}=1=-a_{ji}$ for $i<j$, $B$ is the diagonal matrix
$(-p_1, -p_2, \dots, -p_N)$ and $C$ is the skew-symmetric matrix whose non-zero terms are
$c_{i,i+1}=-c_{i+1,i}=e^{q_i-q_{i+1}}$ for $i=1,2, \dots, N-1$.
We define  $J_2=2 \hat{J}_2$.
 With this convention the
bracket $J_2$ is mapped precisely onto the bracket $\pi_2$ under the Flaschka transformation.

It is easy to see that we have a bi-Hamiltonian pair. We define
\bd
h_1=-2(p_1+p_2+\dots +p_N) \ ,
\ed
and $h_2$ to be the Hamiltonian:
\bd
h_2=\sum_{i=1}^N \,  { 1 \over 2} \, p_i^2 +
\sum _{i=1}^{N-1} \,  e^{ q_i-q_{i+1}}  \ .
\ed
Under Flaschka's transformation  (\ref{f1}),  $h_1$ is mapped onto
$4(b_1+b_2+ \dots+b_N)=4 \  {\rm tr} L= 4 H_1$ and
$h_2$ is mapped onto $2 \  {\rm tr} L^2=4 H_2$.
Using  the relationship
\bd
\pi_2 \nabla \  H_1=\pi_1 \nabla \  H_2 \ ,
\ed
which follows from  Proposition 2,   we obtain, after  multiplication by 4, the
following pair:
\bd
J_1 \nabla \  h_2= J_2 \nabla \  h_1 \ .
\ed

We define the recursion operator as follows:

\bd
{\cal R}=J_2 J_1^{-1} \ .
\ed

The matrix form of ${\cal R}$ is quite simple:

\be
{\cal R} ={ 1 \over 2} \pmatrix{B &-A \cr
                       C& B }\ .     \label{c1}
\ee

This operator raises degrees and we therefore   call it the  positive Toda operator.
In $(q,p)$ coordinates, the  symbol  $\chi_i$ is a shorthand for  $\chi_{h_i}$. It is generated  as
usual by

\bd
\chi_i = {\cal R}^{i-1} \chi_1 \ .
\ed
In a similar fashion we obtain the  higher order Poisson tensors
\bd
J_i = {\cal R}^{i-1} J_1 \ .
\ed

We finally  define the conformal symmetry
\bd
Z_0=\sum_{i=1}^N  (N-2i+1) {\partial \over \partial q_i} +\sum_{i=1}^N p_i {\partial \over \partial p_i} \ .
\ed

It is straightforward to verify that
\bd
{\cal L}_{Z_0} J_1=- J_1 \ ,
\ed

\bd
{\cal L}_{Z_0} J_2=0  \ .
\ed

In fact, $Z_0$ is Hamiltonian  in the $J_2$ bracket with Hamiltonian function
${ 1 \over 2} \sum_{i=1}^N q_i$; see \cite{fernandes}. This observation will be generalized in
5.3.

In addition,
\bd
Z_0(h_1)=h_1
\ed

\bd
Z_0(h_2)=2h_2  \ .
\ed

Consequently, $Z_0$ is a conformal symmetry for $J_1$, $J_2$ and $h_1$. The constants appearing in Oevel's Theorem are
$\lambda=-1$, $\mu=0$ and $\nu=1$. Therefore,
 we end--up with the following deformation relations:

\bd
[Z_i, h_j]= (i+j)h_{i+j}
\ed

\bd
L_{Z_i}  J_j = (j-i-2) J_{i+j}
\ed

\bd
 [ Z_i, Z_j ]  = (j-i) Z_{i+j}  \ .
\ed

Switching to Flaschka coordinates, we obtain  relations  iii)- v) of Theorem 6.

\subsection{The negative Toda hierarchy}
To define the negative Toda hierarchy we use the inverse of
the positive recursion operator ${\cal R}$. We define
\bd
{\cal N}={\cal R}^{-1}=J_1 J_2^{-1} \ .
\ed

Obviously we can use the same conformal symmetry $Z_0=K_0$ and
  take  $\lambda=0$, $\mu=-1$ and $\nu=2$. In other words the role
of $\lambda$ and $\mu $ is reversed.
We  define the vector fields
\bd
K_i = {\cal N}^i K_0 ={\cal N}^i Z_0\  \ \ \ \ \ \ i=1,2, \dots
\ed

 which are master symmetries.   We  use the convention $Y_{-i}=K_{i}$ for $i=0,1,2, \dots$.
For example, $Y_{-1}=K_1 ={\cal N} Z_{0}=-2 \sum_{i=1}^N { \partial \over \partial p_i}$.
This vector field, in $(a,b)$ coordinates, is given by

\bd
X_{-1}=  \nabla \  H_1 = \nabla  \ {\rm tr}\ L = \sum_{i=1}^N {\partial \ \over \partial b_i}   \ .
\ed

This is precisely the same vector field (\ref{f5}). In that section,  $X_{-1}$
 was constructed through a different method.
Similarly, the vector field $Z_0$ corresponds  to the
 Euler vector field (\ref{f2}):

\bd
X_0= \sum _{i=1}^{N-1} a_i {\partial \ \over \partial a_i} +
\sum_{i=1}^N b_i {\partial \ \over \partial b_i} \ .
\ed

{\it Note: We use the  symbol $Y_{i}$ for a vector field in  $(q,p)$ coordinates
and $X_{i}$ for the same vector field in  $(a,b)$ coordinates.  Similarly, we
denote by $J_i$ a Poisson tensor in $(p,q)$ coordinates and $\pi_i$ the corresponding
Poisson tensor in $(a,b)$ coordinates. The index $i$ ranges over all integers. }

We now calculate, using Oevel's Theorem:
\bd
[Y_{-i}, Y_{-j}] =[K_i,K_j]=(\mu-\lambda)(j-i)K_{i+j} \\
                 =(-1)(j-i)K_{i+j}
                 =(i-j)Y_{-(i+j)} \ .
\ed
Letting $m=-i$ and $n=-j$ we obtain the relationship

\be
[Y_m, Y_n]=(n-m) Y_{m+n} \ , \label{d1}
\ee
for all $m$, $n$ negative.
The same relation holds in Flaschka coordinates.
In other words
\bd
[X_m, X_n]=(n-m)X_{m+n}  \ \ \ \ \ \ \forall\,  m,n \in {\bf Z}^{-} \ . \label{d11}
\ed

This last relation may be modified to hold for any two arbitrary  integers  $m$, $n$.
We suppose, without loss of generality, that $j >i$ and consider the bracket
of two master symmetries $K_i=Y_{-i}$ and $Z_j=Y_j$, one in the negative
hierarchy and the second in the positive hierarchy. i.e.
\bd
K_i={\cal N}^i Z_0=R^{-i} Z_0 \ ,
\ed
and
\bd
 Z_j={\cal R}^j Z_0 \ .
\ed
We proceed as in the proof  of Oevel's Theorem (see \cite{fernandes}):
First we note that
\bd
{\cal L}_{Z_0} {\cal R} = \left( {\cal L}_{Z_0} J_2\right) J_1^{-1}- J_2 J_1^{-1}
{\cal L}_{Z_0} J_1  J_1^{-1}=(\mu- \lambda) {\cal R} \ .
\ed
On the other hand
\bd
{\cal L}_{Z_0}{\cal N} = {\cal L}_{Z_0} \left( J_1 J_2^{-1} \right)= (\lambda- \mu){\cal N} \ .
\ed

Finally,
\begin{eqnarray*}
[Y_{-i}, \ Y_j ]=[K_i, Z_j]=[{\cal N}^i Z_0, {\cal R}^j Z_0] \\
                ={\cal N}^i{\cal L}_{Z_0} \left( {\cal R}^j \right) Z_0 - {\cal R}^j {\cal L}_{Z_0}
\left( {\cal N}^i \right) Z_0  \\
         = {\cal N}^i j(\mu- \lambda) {\cal R}^j Z_0 - {\cal R}^j i (\lambda -\mu) {\cal N}^i Z_0 \\
           = j(\mu- \lambda) {\cal R}^{j-i} Z_0 - i(\lambda -\mu) {\cal R}^{j-i} Z_0 \\
= (i+j) (\mu- \lambda){\cal R}^{j-i} Z_0 \\
= (i+j) (\mu- \lambda)Y_{j-i}   \ .
\end{eqnarray*}

In the case of Toda lattice
 $\mu=0$ and $\lambda=-1$, therefore
\bd
[Y_{-i}, Y_j]= (i+j) Y_{j-i} \ .
\ed
We deduce that (\ref{d1}) holds for any integer value of the index.

We define  $W_{i}=J_{3-i}$. This is necessary since the conclusions of Oevel's Theorem assume that
the index begins at $i=1$ and is positive. We compute
\bd
{\cal L}_{Y_{-i}} J_{-j}={\cal L}_{K_i} W_{j+3} \\
                   =(\mu +(j+3-2-i)(\mu-\lambda)) W_{i+j+3}\\
                  =(i-j-2)W_{i+j+3}
                 =(i-j-2)J_{-(i+j)} \ .
\ed

Letting $m=-i$ and $n=-j$ we obtain
\bd
L_{Y_m}J_n=(n-m-2) J_{n+m}  \ ,
\ed

for $n$, $m$ negative integers.
Switching  to Flaschka coordinates we deduce that the relation iv) of Theorem 6  holds also for negative
values of the index. In other words
\bd
L_{X_i} \pi_j =(j-i-2) \pi_{i+j} ,  \ \ \  i\le 0,\ \ \ \  j\le 0 \ .
\ed

Again, a straightforward  modification of the proof of Oevel's Theorem shows that the last relationship holds
for any integer value of $m$, $n$.   We have shown that conclusions  iv) and v) of Theorem 6 hold for
integer values of the index. In fact, it is not difficult to demonstrate all the other parts of Theorem 6.

\begin{theorem}
The conclusions of Theorem 6 hold for any integer value of the index.
\end{theorem}
\noindent
{\it Proof}:
\noindent

We need to prove parts {\it i)}, {\it ii)}, {\it iii)} and {\it vi)} of the Theorem.

\medskip
\noindent
{\it i)}
The fact that $J_n$ are Poisson for $n \in {\bf Z}$ follows from properties of the recursion operator.
 The similar result in $(a,b)$ coordinates  follows easily from properties of the Schouten bracket, and
the fact that $J_n$ and  $\pi_n$  are $F-$related.
We have $\pi_n=F_* J_n$,  therefore
\bd
[\pi_n, \pi_n]=[F_*(J_n), F_*(J_n)]=F_*[J_n, J_n]=F_*(0)=0 \ .
\ed
The vanishing of the Schouten bracket is equivalent to the Poisson property.

\noindent
{\it iii)}
The case where $i$ and $j$ are both of the same sign was already proved.
We next note that  $X_n(\lambda)=\lambda^{n+1}$ if $\lambda$ is an eigenvalue of $L$.
This follows from equation (\ref{f44}) which is used to define the vector fields $X_n$ for
$n\ge 0$.
We would like to extend the formula $X_n(\lambda)=\lambda^{n+1}$ for $n<0$. Since $X_{-1}(\lambda)=1$ we consider $X_{-2}$.
We look at the equation
\bd
[X_{-2}, X_n]=(n+2)X_{n-2} \ .
\ed
We act on $\lambda$ with  both sides of the equation and let $X_{-2}(\lambda)=f(\lambda)$. We obtain the equation
\bd
(n+1) \lambda f( \lambda)-f^{\prime}( \lambda)  \lambda^2 =(n+2) \ .
\ed
This is a linear first order ordinary differential equation with general solution
\bd
f( \lambda)={ 1 \over  \lambda} + c  \lambda^{n+1} \ .
\ed
Since $n$ is arbitrary,  we obtain $f( \lambda)= { 1 \over \lambda}$.
In order to  calculate $X_{-3} (\lambda)$ we use
\bd
X_{-3}=-[X_{-1}, X_{-2}] \ .
\ed
We  obtain
\bd
X_{-3} (\lambda)=X_{-2}X_{-1} (\lambda)-X_{-1}X_{-2} (\lambda)= \\
-X_{-1}( { 1 \over \lambda}) = \\
{ 1 \over \lambda^2}  \ .
\ed
The result follows by induction.

Finally we calculate
\bd
X_i(H_j)={ 1 \over j} X_i \left( \sum \lambda_k^j \right) \\
={ 1 \over j}  \left( \sum X_i \lambda_k^j \right) \\
=\sum \lambda_k^{j-1} X_i (\lambda_k) \\
= \sum \lambda_k^{j-1} \lambda_k^{i+1} \\
=\sum \lambda_k^{i+j} \\
=(i+j)H_{i+j} \ .
\ed

\noindent
{\it vi)} First we note that
 $\pi_j\  \nabla \,  H_i =\pi_{j-1}\   \nabla \,  H_{i+1} $,   holds for $i$, $j$ of the same sign. More
generally, in the positive (or the negative) hierarchy we have the Lenard  relations (\ref{f13}) for the eigenvalues, i.e.
\be
\pi_j \nabla \, \lambda_i= \lambda_i \pi_{j-1} \nabla \, \lambda_i  \ .  \ \ \ \ \ \ \ \ \  \  \label{d12}
\ee
Assume now that $i<0$,  $j>0$.  The calculation is straightforward:

\bd
\begin{array}{rcl}
\pi_j \nabla { 1 \over i} \sum \lambda_k^i& =& \sum \lambda_k^{i-1} \pi_j \nabla \lambda_i  \\
 &&  \\
                                      &=& \sum \lambda_i^l \pi_{j-1} \nabla \lambda_k  \\
 & &     \\
                                      &=& \pi_{j-1} \nabla { 1 \over i+1} \sum \lambda_k^{i+1} \ .
\end{array}
\ed
Therefore,
\be
\pi_j \nabla H_i = \pi_{j-1} \nabla H_{i+1}  \ . \label{d13}
\ee

In the case $i>0$ and $j<0$ we use exactly the same calculation but use (\ref{d12}) for the negative
hierarchy.

\noindent
{\it ii)}
 It is clearly enough to show
the involution of the eigenvalues of $L$ since  $H_i$ are  functions of the eigenvalues.
We prove   involution of eigenvalues  by  using the Lenard relations (\ref{d13}).
We give the proof for the case of the bracket $\pi_j$ with $j>0$ but if $j<0$ the proof is
identical.
First we show that the eigenvalues are in involution with respect to the  bracket $\pi_1$.
 Let $\lambda$ and $\mu$ be two distinct
eigenvalues and let $U$, $V$ be the gradients of $\lambda$ and $\mu$ respectively. We use
the notation $\{\ , \ \}$ to denote the bracket $\pi_1$ and $\left<\ , \ \right>$ the standard inner product.
 The Lenard relations
(\ref{d12})  translate into $\pi_2\,  U= \lambda \,  \pi_1\, U$ and $\pi_2\, V= \mu \, \pi_1 \, V$. Therefore,

\begin{eqnarray*}
\{\lambda, \mu \}&=& \left< \pi_1 U, V \right>  = { 1 \over \lambda} \left< \pi_2 U , V \right> \\
                 &=& -{ 1 \over \lambda} \left< U, \pi_2 V \right> = -{ 1 \over \lambda} \left< U,  \mu \pi_1 V \right> \\
                 &=& -{ \mu \over \lambda} \left< U, \pi_1 V \right>  = { \mu \over \lambda} \left< \pi_1 U,  V \right> \\
                 &=& { \mu \over \lambda} \{\lambda, \mu \}  \ .
\end{eqnarray*}

Therefore, $\{\lambda, \mu \}=0$.
To show the involution with respect to all brackets $\pi_j$,  and in view of part {\it iv)} of Theorem 6, it is
enough to show the following: Let $f_1$, $f_2$ be  two functions  in involution with respect to the Poisson bracket
 $\pi$, let $X$ be a vector field such that $X(f_i)=f_i^2$ for $i=1,2$. Define a Poisson bracket $w$ by
$w={\cal L}_X \pi $.  Then the functions $f_1$, $f_2$ remain in involution with respect to the bracket
$w$. The proof follows trivially if we write  $w={\cal L}_X \pi $
in Poisson form:
\bd
\{f_1, f_2 \}_w=X \{ f_1, f_2\}_{ \pi} - \{ f_1, X(f_2) \}_{\pi} - \{ X(f_1), f_2 \}_{ \pi}  \ .
\ed

\hfill \rule{5pt}{5pt}

{\it Remark:
We should point out that
\bd
H_n ={ 1 \over n} \, {\rm tr} L^{n} \ ,
\ed
makes sense for  $n\not=0$  but it is undefined for $n=0$. The reader should interpret the formulas
involving $H_0$ as a degenerate case, i.e.  $H_0={{\rm tr} L^0 \over 0}={N \over 0} = \infty$.  Therefore the result of $X_{-n}(H_n)=N$ where $N$ is
the size of $L$. It makes sense to define
\bd
X_m(H_0)=\lim_{n \to 0} { 1 \over n} X_m\left( { \rm tr} \, L^n \right) \ .
\ed
For example, $X_{-1} (H_0)$ is calculated by $X_{-1}\left( { 1 \over n}\, {\rm tr}\, L^n \right)= {\rm tr}\, L^{n-1}$.
Taking the limit as $n \to 0$ gives $X_{-1}(H_0)={\rm tr}\, L^{-1}=-H_{-1}$ which is the correct answer. }

\subsection{Master integrals and master symmetries}
In this  section we prove some further results and give some specific examples.

In subsection 5.1 we noticed that $Z_0$ is Hamiltonian with respect to the $J_2$ bracket with Hamiltonian
function $f= {1 \over 2} \sum_{i=1}^N q_i$.  This observation is due to  Fernandes \cite{fernandes}.
  This type of function is called a master integral. It is not a
constant of motion, but its derivative is.
We generalize the result as follows:

\begin{theorem}
The master symmetry  $Y_n$, $n \in {\bf Z}$  is the Hamiltonian vector field of $f$   with respect to the
$J_{n+2}$ bracket.
\end{theorem}

\noindent
{\it Proof}:
We will prove the result for the positive hierarchy $Z_n=Y_n$ but the proof for $Y_{-n}=K_n$ is similar.
As a first step we show that
\bd
Z_n(f)=0 \ \ \ \ \forall \ \  n\ge 0 \  .
\ed
We recall that

\bd
Z_0=\sum_{i=1}^N  (N-2i+1) {\partial \over \partial q_i} +\sum_{i=1}^N p_i {\partial \over \partial p_i} \ .
\ed

Since
\bd
\sum_{i=1}^N (N+1-2i)=0  \ ,
\ed
we obtain
\bd
Z_0(f)={1 \over 2} Z_0 ( \sum_{i=1}^N q_i)={1 \over 2}  ( \sum_{i=1}^N Z_0 (q_i))=0 \ .
\ed
By examining   the form  (\ref{c1}) of the   recursion operator ${\cal R}$ we deduce easily that the
 $q_i$ component of $Z_1$ is
\bd
Z_1(q_i)=-{ 1\over 2} \left[ (N-2i+1) p_i + \sum_{j>i} p_j -\sum_{j<i} p_j \right] \ .
\ed
In other words, the vector
\bd
\left( Z_1(q_1), \dots, Z_1(q_N) \right)
\ed
is the product $AP$ where

\bd
A=-{1 \over 2} { \pmatrix { Z_0(q_1) &  1 & 1 & \cdots & \cdots & 1 \cr
                   -1 & Z_0(q_2) & 1 & \cdots &  \cdots   & \vdots \cr
                   -1 & -1 & Z_0(q_3) & \ddots &  &  \cr
                   \vdots & & \ddots & \ddots & & \vdots \cr
                   \vdots & & & \ddots & \ddots & 1 \cr
                   -1 & \cdots & & \cdots & -1 & Z_0(q_N)   \cr } } \ ,
\ed

and $P$ is  the column vector $(p_1, p_2, \dots, p_N)^t$.
Note that $\sum_{i=1}^N a_{ij}=0$ and $\sum_{j=1}^N a_{ij}=-Z_0(q_i)$.
Therefore,
\begin{eqnarray*}
Z_1(f)={1 \over 2} (Z_1(q_1) + \dots + Z_1 (q_N)) ={ 1 \over 2}
\sum_{i,j} a_{ij} p_j= { 1\over 2} \left( \sum_j \left( \sum_i a_{ij} \right) p_j\right)=0 \ .
\end{eqnarray*}
In the same fashion  one proves that $Z_2(f)=0$.

For $n>2$, we proceed by induction.
\bd
Z_n= {1 \over n-2} \left[ Z_1, Z_{n-1} \right] \ .
\ed
Therefore,
\bd
Z_n(f)= {1 \over n-2} \left[ Z_1, Z_{n-1} \right](f) ={ 1 \over n-2} \left(  Z_1 Z_{n-1}f-Z_{n-1} Z_1 f \right)=0 \ ,
\ed
by the induction hypothesis.

To complete the proof of the Theorem, it is enough to show
\bd
Z_n=[J_{n+2}, f] \ ,
\ed
where $[\ , \ ]$ denotes the Schouten bracket.

First we note that
\bd
[ [ J_{n+1}, f], Z_1]+[[f,Z_1], J_{n+1}]+ [[Z_1, J_{n+1}], f]=0
\ed
due to the super Jacobi identity for the Schouten bracket.  Since
\bd
[Z_1,f]=Z_1(f)=0 \ ,
\ed
the middle term in the last identity is zero. We obtain
\bd
[Z_1, [J_{n+1}, f]]= [[ Z_1, J_{n+1}],f] \ .
\ed
Finally, we calculate using induction:
\begin{eqnarray*}
Z_n= { 1 \over n-2} [Z_1, Z_{n-1}]  \\
     = { 1 \over n-2} [ Z_1, [J_{n+1}, f]] \\
     = { 1 \over n-2} [[ Z_1, J_{n+1}],f] \\
      ={ 1 \over n-2}  \left( [ (n-2) J_{n+2}, f] \right) \\
        = [J_{n+2}, f] \ .
\end{eqnarray*}

\hfill \rule{5pt}{5pt}

 The result of the  Theorem is striking. It shows that the master symmetries are determined once the
Poisson hierarchy is constructed. Of course  one  requires  knowledge of the function $f$.  The function $f$
may be constructed by using Noether's Theorem: The  symmetries of the Toda lattice in $(q,p)$ coordinates
 have been constructed in
 \cite{damianou7}, at least for two degrees of freedom.
The Lie algebra  for the potential of the Toda lattice with $N$ degrees of freedom
is
five  dimensional with generators

\be
\begin{array}{rcl}
X_1 &=&  { \p \over \p t}  \\
X_2&=&  t { \p \over \p t} -2 \ds_{i=2}^N (i-1)  { \p \over \p q_i} \\
X_3&=&     \left( \ds_{i=1}^N q_i \right)     \ds_{i=1}^N { \p \over \p q_i}  \\
X_4&=&  \ds_{i=1}^N { \p \over \p q_i }      \\
X_5 &=& t  \ds_{i=1}^N { \p \over \p q_i }   \ . \label{d3}
\end{array}
\ee
The non--zero bracket relations satisfied by the generators are

\begin{displaymath}
\begin{array}{rcl}

[X_1, X_2] &=& X_1   \\

[ X_1, X_5] &= &  X_4 \\

[ X_2, X_3] &= &-2 X_4 \\

[ X_2, X_5] &= &  X_5 \\

[ X_3, X_4] &= & -2 X_4 \\

[ X_3, X_5] &= & -2 X_5 \ .
\end{array}
\end{displaymath}

 This Lie algebra $L$ is solvable with $L^{(1)}= [L, L]=\{ X_1, X_4, X_5 \}$,
   $L^{(2)}=\{ X_4  \}$ and $L^{(3)}=\{ 0 \}$.

We examine the symmetry $X_5$.

A corresponding time dependent integral produced from   Noether's  Theorem is
\bd
I={ 1 \over 2} \sum_{i=1}^N q_i - { 1 \over 2} t \sum_{i=1}^N p_i = f+ { 1 \over 4} t h_1 \ .
\ed
Motivated by the results of \cite{damianou6}, \cite{ranada}, it makes sense to consider the time independent part of
$I$ which is precisely the function $f$. It is an interesting question whether this procedure  works
 for other integrable systems as well. We also remark that the integrals are also determined from the
knowledge of  the Poisson brackets
and the function $f$. For example, it follows easily from Theorem 9 that
\bd
h_{i+1}= { 1 \over i+1} \{ h_i, f \}_3 \ ,
\ed
where $\{\ , \ \}_3$ denotes the cubic Toda bracket.

\subsection{Noether Symmetries}
We recall that Noether's Theorem   states that for a first order Lagrangian,
the action integral
$\int_{t_1}^{t_2} L dt $
is invariant under the infinitesimal transformation generated by the differential operator, known as a Noether
symmetry,
\be
X = T { \partial~ \over \partial t } +  \sum_{i=1}^N
Q_i { \partial ~\over \partial q_i}   \label{e2}
\ee
if there exists a function $F$, known as a  gauge term, such that
\be
\dot{F} = T {\p L \over \p t}+ \sum_{i=1}^N  Q_i {\p L \over \p q_i}+ \sum_{i=1}^N \left( \dot{Q}_i-
\dot{q}_i\dot{T}\right) {\p L \over \p \dot{q}_i}+\dot{T}L \ .   \label{e3}
\ee
When the corresponding Euler--Lagrange equation is taken into account, equation (\ref{e3}) can be manipulated
to yield the first integral
\be
I = F-\left[ TL+ \sum_{i=1}^N \left( Q_i-\dot{q}_i T\right){\p L \over  \p  \dot{q}_i} \right] \ .  \label{e4}
\ee
Thus to every Noether symmetry there is an associated  first integral.
Consider the Lagrangian to be of the form

\begin{equation}
L={ 1\over 2} \sum_{i=1}^N \dot{q_i}^2 - V(q_1,q_2, \dots, q_N) \label{e1} \ .
\end{equation}
 We summarize the  results of \cite{sopho} in the following Theorem:

\begin{theorem}
Let  ${\bf X}$ be the $N\times 1$ vector with entries $Q_i$, ${\bf x}$ the
vector with entries $q_i$ and
${\bf b}$ the vector with entries $b_i(t)$. Let ${\bf A}$ be an $N\times N$
skew-symmetric matrix with
constant entries. Finally we denote by ${\bf I_N}$ the $N \times N$ identity
matrix.
If $X$, given by (\ref{e2}), is a Noether symmetry
then the infinitesimals must be of the form
\begin{eqnarray}
T& = & T(t)  \nonumber  \\
 {\bf X} &= &  \left( {\bf A} + {1\over 2}   {d T \over dt}
{\bf I_N}\right) {\bf x}+{\bf b} \label{e5}
\end{eqnarray}
and the gauge term is restricted to
\be
F= {1\over 4} {d ^2 T \over dt^2}
\sum_{i=1}^N q_i^2 + \sum_{i=1}^N {
d b_i(t) \over d t} q_i+ d(t) .  \label{e6}
\ee
\end{theorem}

The associated first integral $I$ is equal to
\bd
F+TH-\sum_{i=1}^N Q_i p_i \ ,
\ed
where $H$ is the Hamiltonian.

By examining the form (\ref{d3}) of the generators for the Toda lattice we conclude, using Theorem 10, that only
 $X_1$, $X_4$ and $X_5$ are Noether symmetries. The corresponding integrals provided by Noether's Theorem are
the Hamiltonian $H=h_2$, the total momentum $h_1=p_1+ \dots + p_N$ and
$f+th_1$.

In order to obtain more integrals we
 consider generalized Noether  symmetries.  That is, the infinitesimals in  (\ref{e2}) do not
just depend on $t, q_1, \dots, q_N$ but also on $\dot{q}_1, \dots, \dot{q}_N$. For Lagrangians of the form (\ref{e1})
for one, two and three degrees of freedom, all the possible  point Noether symmetries are classified in \cite{sopho}.
The following results are from \cite{damianou8}.

 In the case of generalized symmetries,
we can take without loss of generality $T=0$. Hence, we consider operators of the form
\be
G=\sum_{i=1}^N  \eta_i { \p \over \p q_i} \label{e10}
\ee
where the infinitesimals of  (\ref{e2}) and (\ref{e10}) are related by
\bd
\eta_i=Q_i-\dot{q}_i T \ .
\ed
Using (\ref{e1}) and (\ref{e10}), Noether's condition (\ref{e3}) becomes

\be
f_t+\sum_{i=1}^N\dot q_if_{q_i} +\sum_{i=1}^N \ddot q_i f_{\dot q_i}=
\sum_{i=1}^N \eta_i V_{q_i}+\sum_{i=1}^N \dot q_i \left ( \eta_{it}+
\sum_{j=1}^N \dot q_j \eta_{iq_j} +\sum_{j=1}^N\ddot q_j \eta_{i\dot q_j}\right ) \ . \label{e12}
\ee

We consider equation (\ref{e12}) in the case of the Toda lattice with two degrees of freedom.
 By assuming various forms of the $\eta_i$ (i.e. linear, quadratic or arbitrary) we can
 solve this  equation and produce the following integrals
one of which ($I_3$) is new:

\bd
I_1=p_1+p_2  \ , \qquad I_2=(p_1-p_2)^2+4 e^{q_1-q_2}  \ ,
\ed

\bd
I_3={ p_1-p_2 + \sqrt{ I_2} \over p_1-p_2-\sqrt{ I_2}} \exp \left( \sqrt{I_2} {q_1+q_2 \over p_1+p_2} \right) \ .
\ed
Note that $H={1\over 4} \left( I_1^2+I_2 \right)$ and
 that the function $ G={ q_1 +q_2 \over p_1+p_2}$ which appears in the exponent of $I_3$
satisfies $\{G,H \}=1$.

The existence of the integral $I_3$  shows that the two degrees of freedom Toda lattice is super--integrable with three integrals of motion
$\{I_1, I_2, I_3 \}$. A Hamiltonian system with $N$ degrees of freedom is called super-integrable if it possesses
$2N-1$ independent  integrals of motion. Of course these integrals cannot be all in involution.
 Based on this computation for 2 degrees of freedom we make the following conjecture:

\noindent
{\bf Conjecture:}
The Toda lattice is super--integrable.

\subsection{Rational Poisson brackets}
\bigskip
The rational brackets in $(q,p)$ coordinates are given by complicated expressions
that are quite hard  to write in explicit form. When projected in the space of $(a,b)$ variables
they give rational brackets whose numerator is polynomial and the denominator is the determinant of the
Jacobi matrix. We  give  examples of these brackets and master symmetries for
 $N=3$.

For example, the tensor $J_0$ is a homogeneous rational bracket of degree 0. It is
defined by
\bd
J_0={\cal N} J_1 = J_1 J_2^{-1} J_1 \ .
\ed
In the case of three particles  the corresponding  bracket $\pi_0$ is given  as
follows:
First define the skew-symmetric matrix $A$
by

\begin{eqnarray*}
a_{12}&= &  -{1 \over 2}a_1 a_2(b_3+b_1-b_2) \\
a_{13}&=&   a_1(a_2^2-b_2 b_3)  \\
a_{14}&=&  -a_1(a_2^2-b_1 b_3)  \\
a_{15}&=& a_1 a_2^2  \\
a_{23}&=& -a_1^2 a_2 \\
a_{24}&=&a_2 (a_1^2-b_1 b_3) \\
a_{25}&=& -a_2(a_1^2-b_1 b_2) \\
a_{34}&=&-2 a_1^2 b_3 \\
a_{35}&=& 0  \\
a_{45}&=&  -2 a_2^2 b_1  \ .
\end{eqnarray*}

The matrix of the tensor $\pi_0$ is defined by
\be
\pi_0={1  \over {\rm det}\, L} A  \label{d14}
\ee
where $ {\rm det} L= b_1 b_2 b_3- a_2^2 b_1 -a_1^2 b_3 $. This formula defines a Poisson bracket with one single
Casimir $H_2={1 \over 2} {\rm tr} \, L^2$.  The bracket is defined on the open dense set ${\rm det}\, L \not= 0$.
 Taking $H_3={1 \over 3} {\rm tr} \, L^3$ as the Hamiltonian we have another bi--Hamiltonian formulation of the
system:
\bd
\pi_1 \,  dH_2=\pi_0 \,  dH_3 \ .
\ed
In fact we have infinite pairs of such formulations since
\bd
\pi_2 \,  dH_1=\pi_1 \,  dH_2 =\pi_0 \,  dH_3 =\pi_{-1} \,  dH_4= \dots
\ed

The explicit formulas for the vector fields $X_1$ and $X_2$ are given in 3.1  therefore we  give
an example for the vector field $X_{-2}$.  In the case $N=3$ it is given by

 \bd
X_{-2} = { 1 \over {\rm det} L} \left( \sum_{i=1}^2 r_i  {\partial \ \over \partial a_i }+
  \sum _{i=1}^3 s_i {\partial \ \over \partial b_i} \right)
  \ed

  where

  \bd
  \begin{array}{rcl}

r_1 & =& {1\over 2}a_1(b_1-b_2-2 b_3)  \\
r_2 &= & { 1\over 2}a_2(b_3-2b_1-b_2)    \\
    s_1&=& b_2b_3-a_1^2-a_2^2 \\
s_2&=&b_1 b_3+a_1^2+a_2^2 \\
s_3&=& b_1b_2-a_1^2-a_2^2 \ .
\end{array}
\ed

Finally, we consider  the Casimirs of these new Poisson brackets.

\begin{theorem}
 The Casimir of $\pi_n$ in the open dense set ${\rm det}\, L \not= 0$  is Tr $L^{2-n}$ for all $n \not=2$. The Casimir of
$\pi_2$ is det $L$.

\end{theorem}

\noindent
\underline{Proof:}

For $n\ge 1$ the result  was proved in Proposition 5.
Therefore, we only  have to show that the Casimir of $\pi_{-m}$    is tr $L^{m+2}$  ($m\ge 0$).
This follows from (\ref{d13}) and the fact that
 $H_1= {\rm Tr}\, L$ is the Casimir for the Lie-Poisson
bracket $\pi_1$:
\bd
0=\pi_1 \nabla H_1=\pi_0 \nabla H_2 = \pi_{-1} \nabla H_3= \dots    \ .
\ed

\smallskip
\section{ GENERALIZED TODA SYSTEMS   ASSOCIATED WITH SIMPLE LIE GROUPS}
\vskip 2cm

In this section we consider mechanical systems which generalize the
finite, nonperiodic Toda lattice. These systems correspond to Dynkin
diagrams. They are special cases of (\ref{a1}) where the spectrum corresponds to a
 system of simple roots for a simple
Lie algebra. It is well known that irreducible root systems classify
simple Lie groups. So, in this generalization for each simple Lie algebra
 there  exists a mechanical system of Toda type.

\smallskip
The generalization is obtained from the following simple observation: In
terms of the natural basis $q_i$ of weights, the simple roots of $A_{n-1}$
are
\begin{equation}
q_1-q_2,\,  q_2-q_3, \dots, q_{n-1}-q_n \ .
\end{equation}
On the other hand, the potential for the Toda lattice is of the form

\begin{equation}
e^{q_1-q_2} +e^{q_2-q_3} + \dots +e^{q_{n-1}- q_n}  \ .
\end{equation}
We note that the angle between $q_{i-1}-q_i$ and $q_i -q_{i+1}$ is
$ { 2 \pi \over 3}$ and the lengths of $ q_i -q_{i+1} $ are all
equal. The Toda lattice corresponds to a Dynkin diagram of type
 $A_{n-1}$.

\smallskip
More generally, we consider potentials of the form
\begin{equation}
U=c_1 \,  e^{f_1(q)} + \dots + c_l \,  e^{f_l(q)}
\end{equation}
where $c_1, \dots, c_l$ are constants, $f_i (q)$ is
linear  and $l$ is the
rank of the simple Lie algebra.  For each Dynkin diagram we
construct a Hamiltonian system of Toda type. These systems
are interesting not only because they are integrable, but also for their
fundamental importance in  the theory of semisimple Lie groups. For example Kostant in
\cite{kostant}  shows that the integration of these systems and the theory
of the finite dimensional representations of semisimple Lie
groups are equivalent.

\smallskip
For reference, we give a complete list of the Hamiltonians for
 each simple Lie algebra.

\bigskip
\noindent
$\underline { A_{n-1}}$

\medskip
$$ H= { 1 \over 2} \sum_1^n p_j^2 + e^{ q_1- q_2} + \cdots + e^{ q_{n-1}-q_n} $$

\medskip
\noindent
$\underline{ B_n}$
\bigskip
 $$ H= { 1 \over 2} \sum_1^n p_j^2 + e^{ q_1- q_2} + \cdots + e^{ q_{n-1}-q_n}
   + e^{q_n} $$
\medskip
\noindent
$\underline{ C_n}$
\medskip
 $$ H= { 1 \over 2} \sum_1^n p_j^2 + e^{ q_1- q_2} + \cdots + e^{ q_{n-1}-q_n}
   + e^{ 2 q_n} $$
\medskip
\noindent
$\underline{ D_n}$
\medskip
 $$ H= { 1 \over 2} \sum_1^n p_j^2 + e^{ q_1- q_2} + \cdots + e^{ q_{n-1}-q_n}
   + e^{q_{n-1} +q_n} $$
\medskip
\noindent
$\underline{ G_2}$
\medskip
 $$ H= { 1 \over 2} \sum_1^3 p_j^2 + e^{ q_1- q_2} +  e^{  -2 q_1 +q_2 +q_3}
    $$
\medskip
\noindent
$\underline{ F_4}$
\medskip
 $$ H= { 1 \over 2} \sum_1^4 p_j^2 + e^{ q_1- q_2} +e^{ q_2 -q_3} +e^{ q_3}
 + e^{ { 1 \over 2} (q_4 -q_1 -q_2 -q_3)} $$

\medskip
\noindent
$\underline{ E_6}$
\medskip
 $$ H= { 1 \over 2} \sum_1^8 p_j^2 +  \sum_1^4 e^{ q_j- q_{j+1} } +e^{ -(q_1+q_2)}
+ e^ { { 1 \over 2} ( -q_1 +q_2+ \dots + q_7-q_8)} $$

\medskip
\noindent
$\underline{ E_7}$
\medskip
 $$ H= { 1 \over 2} \sum_1^8 p_j^2 +  \sum_1^5 e^{ q_j- q_{j+1} } +e^{ -(q_1+q_2)}
+ e^ { { 1 \over 2} ( -q_1 +q_2+ \dots + q_7-q_8)} $$

\medskip
\noindent
$\underline{ E_8}$
\medskip
 $$ H= { 1 \over 2} \sum_1^8 p_j^2 +  \sum_1^6 e^{ q_j- q_{j+1} } +e^{ -(q_1+q_2)}
+ e^ { { 1 \over 2} ( -q_1 +q_2+ \dots + q_7-q_8)} $$
\bigskip

We should note that the Hamiltonians in the  list are not unique. For example, the
$A_2$ Hamiltonian is

\begin{equation}
H={ 1 \over 2} p_1^2 + { 1 \over 2} p_2^2 + { 1 \over 2} p_3^2 + e^{q_1-q_2} + e^{q_2-q_3} \ .
\end{equation}
An equivalent system is

\begin{equation}
H(Q_i, P_i) = { 1 \over 2} P_1^2 + { 1 \over 2} P_2^2 + e^ { \sqrt { { 2 \over 3}} ( \sqrt{3} Q_1 +Q_2 )}
+ e^{ -2 \sqrt{ { 2 \over 3}} Q_2} \ .
\end{equation}
 The second Hamiltonian is obtained from the first by using the
 transformation

\begin{eqnarray}
 Q_1 & =& { \sqrt2 \over 4} (q_1+q_2 -2 q_3) \\
             Q_2&=& { \sqrt 6 \over 4} (q_2-q_1)   \\
             P_1&=& { 2 \over \sqrt2}  (p_1+p_2)  \\
             P_2&=& { 2 \over \sqrt 6} (p_2-p_1)   \ .
\end{eqnarray}

\bigskip
Another example  is the following two systems, both corresponding
to a Lie algebra of type $D_4$:
\bd
\sum_{i=1}^4 { p_i^2 \over 2} + e^{q_1} +e^{q_2} +e^{q_3} +e^{ { 1 \over 2} (q_4 -q_1 -q_2-q_3)}
\ed

\bd
\sum_{i=1}^4 { p_i^2 \over 2} + e^{q_1-q_2} +e^{q_2-q_3} +e^{ q_3-q_4} +e^{q_3+q_4} \ .
\ed

 Finally, let us recall the definition of exponents for a semi-simple group
$G$. An excellent reference is the book by Collingwood and McGovern  \cite{collin}.  Let $G$ be a connected, complex,
 simple Lie Group $G$.  We
form the de Rham cohomology groups $H^i (G, {\bf C})$ and the
corresponding Poincar\'e polynomial of $G$:
\begin{displaymath}
p_G(t)=\sum d_i t^i \ ,
\end{displaymath}
 where $d_i=  $dim  $H^i (G, {\bf C})$. A Theorem of Hopf shows that the
cohomology algebra is   a finite product of $l$
 spheres of odd dimension  where $l$ is the rank of $G$.. This Theorem implies that
\bd
p_G(t)= \prod_{i=1}^l (1+t^{2 e_i +1} )\ .
\ed
The  positive integers $\{ e_1, e_2, .... , e_l \}$ are called the
{\it exponents} of $G$. One can also extract the exponents from
the root space decomposition of $ G$.  The connection
with the invariant polynomials is the following: Let $H_1, H_2,
..., H_l$ be  independent, homogeneous, invariant polynomials of
degrees $m_1, m_2, .. , m_l$. Then $ e_i=m_i- 1$.
 The exponents of a simple Lie group  are given in the following list:

\bigskip
\noindent
$\underline { A_{n-1}}$

\medskip
$$ 1,2,3,...., n-1 $$

\medskip
\noindent
$\underline{ B_n}$, $\underline{ C_n}$
\bigskip
  $$ 1,3,5, ...., 2 n-1 $$
\medskip
\noindent
$\underline{ D_n}$
\medskip
$$ 1,3,5,...., 2 n-3, n-1 $$

\medskip
\noindent
$\underline{ G_2}$
\medskip
 $$1,5 $$
\medskip
\noindent
$\underline{ F_4}$
\medskip
$$1,5,7,11 $$

\medskip
\noindent
$\underline{ E_6}$
\medskip
$$1,4,5,7,8,11  $$

\medskip
\noindent
$\underline{ E_7}$
\medskip
$$1,5,7,9,11,13,17 $$

\medskip
\noindent
$\underline{ E_8}$
\medskip
$$1,7,11,13,17,19,23,29$$
\bigskip

\bigskip
\section {$B_n$ TODA SYSTEMS}

\subsection{ A rational bracket for a central extension of  $B_n$-Toda }

In this subsection we show that the $B_n$ Toda system is bi--Hamiltonian by considering a central extension
of the the corresponding Lie algebra in analogy with  $gl(n,\,  {\bf C})$ which is  a
central extension of $sl(n,\, {\bf C})$ in the case of $A_n$ Toda.

\bigskip
Another way to describe these generalized Toda systems, is to give a Lax pair
representation in each case. It can be shown that the equation $\dot L =[B,L]$
is  equivalent to the equations of motion generated by the Hamiltonian
 $H_2 = { 1 \over 2} {\rm tr} \, L^2 $ on the orbit through $L$ of the coadjoint
action of  $B_{-}$ (lower triangular group) on the dual of its Lie algebra, ${\cal B}_{-}^*$. The space ${\cal B}_{-}^*$ can
be identified with the set of symmetric matrices. This situation, which corresponds
 to $sl(n, {\bf C})=A_{n-1}$ can be generalized to other semisimple Lie algebras. We
use notation and definitions from Humphreys \cite{humphreys}.

\smallskip
Let ${\cal G}$ be a semisimple Lie algebra, $\Phi$ a root system for ${\cal G}$,
$\Delta = \{ \alpha_1, \dots , \alpha_l \} $ the simple roots, $h$ a Cartan
subalgebra and ${\cal G}_{\alpha}$ the root space of $\alpha$. We denote by $x_{\alpha}$ a
generator of ${\cal G}_{\alpha}$.  Define
\bd
{\cal B}_{-}=h \oplus \sum_{\alpha <0} {\cal G}_{\alpha} \ .
\ed
There is an automorphism $\sigma$ of ${\cal G}$, of order 2, satisfying
$\sigma (x_{\alpha})=x_{- \alpha}$ and $\sigma ( x_{-\alpha})= x_{\alpha}$. Let ${\cal K}=
\{ x \in \, {\cal G} \, \vert \, \sigma (x)=-x  \}$. Then we have a direct sum
decomposition ${\cal G}={\cal B}_{-} \oplus {\cal K}$.  The Toda flow is a coadjoint flow on
${\cal B}_{-}^*$ and the coadjoint invariant functions on ${\cal G}^*$, when restricted to
${\cal B}_{-}^*$ are still in involution.
\smallskip
The Jacobi elements are of the form
\bd
L=\sum_{i=1}^l \, b_i h_i + \sum_{i=1}^l a_i \, (x_{\alpha_i} + x_{-\alpha_i}) \ .
\ed
We define
\bd
B=\sum_{i=1}^l a_i \,  (x_{\alpha_i }- x_{-\alpha_i}) \ .
\ed
The generalized Toda flow takes the Lax pair form:
\bd
\dot L=[B,L]  \ .
\ed

\smallskip
The $B_n$ Toda systems were shown to be Bi-Hamiltonian. The second bracket
 can be found in \cite{damianou2}. It turned out to be a rational bracket and it was obtained by
using Dirac's constrained bracket formula (\ref{f25}).  The idea is to use the inclusion of $B_n$ into $A_{2n}$  and to
restrict the hierarchy of brackets from $A_{2n}$ to $B_n$ via Dirac's bracket.
Straightforward restriction does not work. We briefly describe the procedure in the
case of $B_2$.

\bigskip
The Jacobi matrices  for $A_4$ and $B_2$ are given by

\begin{equation}
 L_{A_4} =  \pmatrix { b_1 & a_1 & 0 & 0& 0 \cr
              a_1 & b_2 & a_2 & 0& 0 \cr
              0& a_2 & b_3 & a_3 & 0 \cr
              0 & 0& a_3 & b_4 & a_4  \cr
               0& 0&0& a_4 & b_5 }  \ ,
 \end{equation}

and
\begin{equation}
 L_{B_2}=  \pmatrix { b_1 & a_1 & 0 & 0& 0 \cr
              a_1 & b_2 & a_2 & 0& 0 \cr
              0& a_2 & b_3 & -a_2 & 0 \cr
              0 & 0& -a_2 &  2 b_3 -b_2 & -a_1  \cr
               0& 0&0& -a_1 & 2 b_3 -b_1  }  \ .
\end{equation}
 Note that $L_{A_4}$ lies in ${\rm gl} (4, {\bf C})$  instead of $ {\rm sl} (4, {\bf C})$. Therefore we have
added an additional variable in $L_{B_2}$. We define

\begin{eqnarray}
p_1 &=& a_1 +a_4  \nonumber \\
p_2& = & a_2 +a_3\nonumber \\
p_3&= & b_1 +b_5 - 2 b_3\nonumber \\
p_4 &=& b_2 +b_4 -2 b_3 \nonumber \ .
\end{eqnarray}
It is clear that we obtain $B_2$ from $A_4$ by setting $p_i=0$ for
$i=1,2,3,4$. We calculate the matrix $P=\{ p_i, p_j \}$.  The bracket
  used  is the quadratic Toda  (\ref{f15}) on $A_4$.

\begin{eqnarray}
\{p_1, p_2 \}&=& { 1 \over 2} (a_1 a_2 -a_3 a_4)\nonumber \\
\{p_1, p_3 \}&=& a_4 b_5 -a_1 b_1 \nonumber \\
\{p_1 ,p_4 \}&=& a_1 b_2 -a_4 b_4 \nonumber \\
\{p_2, p_3 \}&=& 2(a_3 b_3 -2 a_2 b_3 )\nonumber \\
\{p_2, p_4 \}&=& a_3 b_4 + 2 a_3 b_3 -2 a_2 b_3 -a_2 b_2\nonumber \\
\{p_3, p_4 \}&=& -2 a_4^2 -4 a_3^2 + 4 a_2^2 +2 a_1^2 \nonumber \ .
\end{eqnarray}

If we evaluate at a point in $B_2$ we get

\begin{eqnarray}
\{p_1, p_2 \}&=& 0   \nonumber \\
\{p_1, p_3 \}&=& -2 a_1 b_3    \nonumber \\
\{p_1 ,p_4 \}&=& 2 a_1 b_3 \nonumber \\
\{p_2, p_3 \}&=&  -4 a_2 b_3   \nonumber \\
\{p_2, p_4 \}&=& -6 a_2 b_3    \nonumber \\
\{p_3, p_4 \}&=&  0  \nonumber \ .
\end{eqnarray}

Therefore the matrix $P$ is given by

\bd
P=\pmatrix { 0 &0& -2 a_1 b_3 & 2 a_1 b_3  \cr
             0 &0& -4 a_2 b_3 & -6 a_2 b_3 \cr
             2 a_1 b_3 & 4 a_2 b_3 & 0 & 0 \cr
             -2 a_1 b_3 & 6 a_2 b_3 & 0 & 0    } \ ,
\ed
and $P^{-1}$ is the matrix

\bd
P^{-1}= \pmatrix  { 0 & 0 & { 3 \over 10 a_1 b_3 } & -{ 1 \over 5 a_1 b_3} \cr
          0&0 & { 1 \over 10 a_2 b_3 } & { 1 \over 10 a_2 b_3}  \cr
          -{  3 \over 10 a_1 b_3 } & { 1 \over 5 a_1 b_3} & 0 &0 \cr
            -{ 1 \over 10 a_2 b_3} & -{ 1 \over 10 a_2 b_3 }& 0 & 0}  \ .
\ed

Using Dirac's formula (\ref{f25}) we obtain a homogeneous quadratic bracket on $B_2$ given by

\begin{eqnarray}
\{a_1, a_2 \} &=& { a_1 a_2 ( 3 b_3 -b_2-2 b_1) \over 10 b_3}\nonumber \\
\{a_1, b_1 \} &=& { -a_1 (10 b_1 b_3 -2 b_1 b_2 -3 b_1^2 -a_1^2) \over 10 b_3} \nonumber \\
\{ a_1, b_2 \}&=& { a_1 (10 b_2 b_3 -3 b_2^2 -2 b_1 b_2 -4 a_2^2 -a_1^2) \over 10 b_3}\nonumber \\
\{a_1 , b_3 \}&=& { a_1 (b_2-b_1) \over 5} \nonumber \\
\{a_2, b_1 \}&=& { a_2 ( 2 b_1 b_3 -2 b_1 b_2 +a_1^2 ) \over 10 b_3} \nonumber \\
\{a_2, b_2 \}&=& { -a_2 ( 8 b_2 b_3 -3 b_2^2 -6 a_2^2 -4 a_1^2 ) \over 10 b_3} \nonumber \\
\{a_2, b_3 \}&=& {a_2 (b_3-b_2) \over 5 } \nonumber \\
\{ b_1,b_2 \} &=& { 10 a_1^2 b_3 -3 a_1^2 b_2 -2 a_2^2 b_1 -3 a_1^2 b_1  \over 5 b_3}\nonumber \\
\{ b_1,b_3 \} &=& { 2 a_1^2 \over 5} \nonumber \\
\{ b_2,b_3 \}&=& { 2 \over 5} (a_2^2 -a_1^2)  \nonumber    \ .
\end{eqnarray}
The bracket satisfies the following properties which are analogous to the quadratic $A_n$ Toda (\ref{f15}).

\smallskip
\noindent
{\it i)} It is a homogeneous quadratic Poisson bracket.

\smallskip
\noindent
{\it ii)} It is compatible with the $B_2$ Lie-Poisson bracket.

\smallskip
\noindent
{\it iii)} The functions $H_n = { 1 \over n} \, {\rm tr} \, L^n$ are in involution in this
bracket.

\smallskip
\noindent
{\it iv)} We have Lenard type relations $\pi_2 \nabla H_i = \pi_1 \nabla H_{i+1} $ where $\pi_1$,
$\pi_2$ are the Poisson  matrices of the linear and quadratic $B_2$ Toda brackets respectively.

\smallskip
\noindent
{\it v)} The function ${\rm det}\, L$ is the Casimir.

\bigskip

\subsection{A recursion operator for Bogoyavlensky--Toda systems of type $B_n$}

 In this section, we  show that higher polynomial brackets exist also in the case of $B_n$ Toda  systems.
We will prove  that these systems  possess a recursion operator and we will
construct an infinite sequence of compatible Poisson brackets in which the constants of motion are
in involution.

\smallskip
 The Hamiltonian for $B_n$ is

\begin{equation}
 H= { 1 \over 2} \sum_1^n p_j^2 + e^{ q_1- q_2} + \cdots + e^{ q_{n-1}-q_n}
   + e^{q_n}  \ .
\end{equation}

 We make a  Flaschka-type  transformation

\begin{equation}
  a_i  = {1 \over 2} \,  e^{ {1 \over 2} (q_i - q_{i+1} ) } \ ,   \ \ \ \ \ \ \ \ \ \  a_n= { 1 \over 2} e^{ { 1 \over 2} q_n }
\label{a99}
\end{equation}

\begin{displaymath}
             b_i  = -{ 1 \over 2} p_i    \ .
\end{displaymath}

\smallskip
\noindent
Then

\begin{equation}
\begin{array}{lcl}
 \dot a _i& = & a_i \,  (b_{i+1} -b_i )    \ \ \ i=1, \dots, n \\
   \dot b _i &= & 2 \, ( a_i^2 - a_{i-1}^2 ) \ \ \ i=1, \dots, n  \ ,
\end{array}
\end{equation}
with the convention that $a_0=b_{n+1}=0$.

\smallskip
\noindent
These equations can be written as a Lax pair  $\dot L = [B, L] $, where $L$ is the  symmetric  matrix

\begin{equation}
  \pmatrix { b_1 &  a_1 &  & &  &    &    & \cr
                   a_1 &  \ddots  & \ddots  &   & &&   &  \cr
                    & \ddots & \ddots& a_{n-1} &  & &   &  \cr
                   &  & a_{n-1} & b_n & a_n &  & &  \cr
                    & &  & a_n & 0 &  -a_n     & & \cr
                    & & & & -a_n & -b_n & \ddots &  \cr
                    &  &&&& \ddots & \ddots & -a_1  \cr
                       &&&&& & -a_1 & -b_1     \cr }   \ ,
\end{equation}

and $B$ is the skew-symmetric part of $L$ (In the decomposition, lower Borel plus skew-symmetric).
We note that the determinant of $L$ is zero.

The mapping  $F: {\bf R}^{2n} \to {\bf R}^{2n}$, $(q_i ,p_i ) \to
(a_i ,b_i )$, defined by (\ref{a99}), transforms the standard
symplectic bracket   into  another symplectic bracket $\pi_1 $
given (up to a constant multiple) by

\bigskip
\noindent
\begin{equation}
\begin{array}{lcl}
\{ a_i, b_i \} & = &-a_i  \\
\{ a_i, b_{i+1} \}& =&  a_i \ .
\end{array}
\end{equation}

It is easy to show by induction that
\bd
{\rm det} \,  \pi_1= a_1^2 a_2^2 \dots  a_n^2 \ .
\ed

\smallskip

The invariant polynomials for $B_n$, which we denote by
\begin{displaymath}
  H_2, \  H_4, \  \dots \ H_{2n}
\end{displaymath}

are defined by $H_{2i} = { 1 \over 2i} \  { \rm Tr} \ L^{2i}  $.
 The degrees of the first $n$ (independent) polynomials are $2,4, \dots, 2n$   and
 the exponents of the corresponding Lie group are $1,2, \dots, 2n-1$.

\smallskip

We look for a bracket $\pi_3$ which satisfies
\begin{equation}
 \pi_3 \ \nabla \ H_2 = \pi_1 \ \nabla  \ H_4  \ .
\end{equation}

\smallskip
 Using trial and error,  we end up with the following homogeneous cubic bracket  $\pi_3$.

\begin{equation}
\begin{array}{lcl}
\{ a_i, a_{i+1} \}& = &a_i a_{i+1} b_{i+1}  \\
\{ a_i, b_i \}& = &-a_i b_i^2 -a_i^3   \ \ \ \ \ \ \ \ \ \ i=1,2, \dots , n-1 \\
\{ a_n,b_n \}& = &-a_n b_n^2 -2 a_n^3  \\
\{a_i, b_{i+2} \}& = &a_i a_{i+1}^2   \\
\{ a_i, b_{i+1} \}& =& a_i b_{i+1}^2 + a_i^3  \\
 \{ a_i, b_{i-1} \} &=& -a_{i-1}^2 a_i  \\
\{ b_i, b_{i+1} \}&= &2 a_i^2 (b_i +b_{i+1})  \ .
\end{array}
\end{equation}
\smallskip
We summarize the properties of this new bracket in the following:

\begin{theorem}  The bracket $\pi_3$ satisfies

\smallskip
\noindent
1. $\pi_3 $ is Poisson

\smallskip
\noindent
2. $\pi_3$ is compatible with $\pi_1$.

\smallskip
\noindent
3. $H_{2i}$ are in involution.

\smallskip
Define ${\cal R}=\pi_3 \pi_1^{-1}$. Then ${\cal R}$ is a recursion operator.
 We obtain a hierarchy
$$\pi_1, \pi_3, \pi_5, \dots $$
consisting of compatible Poisson brackets of odd degree in which   the constants of
motion are in involution.

\smallskip
\noindent
4. $\pi_{j+2} \  \nabla \  H_{2i} = \pi_j \  \nabla  \  H_{2i+2} \ \ \ \ \ \forall \ i, \ j \   \ . $

\end{theorem}

\smallskip
\noindent
The proof  of this result is in \cite{damianou2}.

It is interesting  to compute the cubic Poisson bracket in $(p,q)$ coordinates. We will see that
 in the expression for the master symmetry the exponents of the corresponding Lie-group appear explicitly.
 We reproduce the formula for the  cubic Poisson bracket   in $(p,q)$-coordinates from \cite{joana}.
\bigskip

\bd
\begin{array}{lcl}
\{ q_i, q_{i-1} \} = \{q_i, q_{i-2} \} = \dots = \{q_i, q_1 \} & =
& 2p_i \ \ \ \  i=2, \dots ,n  \\
 \{ p_i, q_{i-2} \} = \{p_i, q_{i-3} \} = \dots = \{ p_i, q_1 \} &
 = & 2 (e^{q_{i-1}-q_i}- e^{q_i -q_{i+1}}) \ \ \ \ i = 3, \dots,
n-1 \\ \{ p_n, q_{n-2} \} = \{p_n, q_{n-3} \} = \dots = \{ p_n,
q_1 \} &
 = & 2(e^{q_{n-1}-q_n}-e^{q_n})
\end{array}
\ed

\bd
\begin{array}{lcl}
\{ q_i, p_i \} & = & p_{i}^2 + 2 e^{q_i -q_{i+1}} \ \ \ \ i= 1,
\dots, n-1 \\ \{ q_n, p_n \} & = & p_{n}^2 + 2 e^{q_n} \\ \{
q_{i+1}, p_i \} & = & e^{q_i - q_{i+1}} \\ \{ q_i, p_{i+1} \} & =
& 2(e^{q_{i+1}-q_{i+2}}- e^{q_i -q_{i+1}}) \ \ \ \ i=1, \dots ,n-2
\\ \{ q_{n-1}, p_n \} & = & 2 e^{q_n}- e^{q_{n-1}-q_n}
\\ \{ p_i , p_{i+1} \} & = & -e^{q_i -q_{i+1}} (p_i + p_{i+1})
\end{array}
\ed

\bigskip

 In $(p,q)$-coordinates, $J_1$ is the (symplectic)
canonical Poisson tensor, $h_2$ is the Hamiltonian , $J_3$ is the cubic
Poisson tensor for $B_n$ and $Z_0$ is the conformal symmetry for
both $J_1$, $J_3$ and $h_2$.

So, with
 $$Z_0= \sum_{i=1}^{n}p_i \frac{\partial}{\partial p_i}+
 \sum_{i=1}^{n}2(n-i+1) \frac {\partial}{\partial q_i},$$

 we have

 $$ {\cal L}_{Z_0} J_1 =-J_1, \quad {\cal L}_{Z_0} J_3 = J_3, \quad
 {\cal L}_{Z_0} h_2 =2h_2 \ .$$
We obtain a hierarchy of Poisson tensors, master symmetries and invariants
which are obtained using Oevel's Theorem.
For example, we have
\be
 [Z_i, \chi_j]  =  (2j+1) \chi_{i+j}
 \ee
and the coefficients of the first $n$ independent Hamiltonian vector fields correspond
to the exponents of a Lie group of type $B_n$.

\subsection{Bi--Hamiltonian formulation of $B_n$ systems}

Following the procedure outlined in the introduction we obtain a bi--Hamiltonian
formulation of the system. In other words,  we define $\pi_{-1}={\cal N} \pi_1=\pi_1 \pi_3^{-1} \pi_1$ and
we use it to obtain the desired formulation.

 We illustrate with the $B_2$ Toda system. In this case ${\rm det}\, \pi_1= a_1^2 a_2^2$ and
 ${\rm det}\, \pi_3= a_1^2 a_2^2\,  \Delta^2 ={\rm det}\, \pi_1 \,  \Delta^2 $
 where

\bd
\Delta=a_1^4+2a_2^2a_1^2+2a_2^2b_1^2+b_1^2b_2^2-2a_1^2b_1b_2 \ .
\ed

The explicit formula for $\pi_{-1}$ is
\bd
\pi_{-1}={1 \over \Delta} A  \ ,
\ed
 where
\bd
A=\pmatrix{ 0 & -a_1 a_2 b_2 & -a_1(b_2^2+a_1^2+2 a_2^2) & a_1 (b_1^2+a_1^2+2 a_2^2)   \cr
           a_1 a_2 b_2&  0 &    a_1^2 a_2      & -a_2 (b_1^2+2 a_1^2)         \cr
            a_1(b_2^2+a_1^2+2 a_2^2)  &-a_1^2 a_2 & 0 & -2 a_1^2(b_1+b_2)   \cr
             -a_1 (b_1^2+a_1^2+2 a_2^2)  & a_2 (b_1^2+2 a_1^2)  & 2 a_1^2(b_1+b_2)   &0} \ .
\ed
This bracket is Poisson by construction. We will prove later that it is compatible with $\pi_1$.
We note that  $\Delta=\sqrt{ {\rm det}\, {\cal R}}$ and it is also equal to
 the product of the non--zero eigenvalues of $L$.
Using the rational bracket $\pi_{-1}$ we  establish  the
bi-Hamiltonian nature  of the system, i.e.
\bd
\pi_1 \nabla H_2 = \pi_{-1} \nabla H_4  \ .
\ed

\section{$C_n$ TODA SYSTEMS}

We now consider    $C_n$ Toda  systems.
We will prove  that these systems  also possess a recursion operator and we will
construct an infinite sequence of compatible Poisson brackets as in the $B_n$ case.
We also show that the systems are  bi--Hamiltonian.

\subsection{A recursion operator for Bogoyavlensky--Toda systems of type $C_n$}

\smallskip
 The Hamiltonian for $C_n$ is

\begin{equation}
 H= { 1 \over 2} \sum_1^n p_j^2 + e^{ q_1- q_2} + \cdots + e^{ q_{n-1}-q_n}
   + e^{ 2 q_n}  \ .
\end{equation}

 We make a  Flaschka-type  transformation

\begin{equation}
  a_i  = {1 \over 2} \,  e^{ {1 \over 2} (q_i - q_{i+1} ) } \ ,   \ \ \ \ \ \ \ \ \ \  a_n= { 1 \over \sqrt{2}} e^{  q_n }
\nonumber
\end{equation}

\begin{displaymath}
             b_i  = -{ 1 \over 2} p_i    \ .
\end{displaymath}

The equations in $(a,b)$ coordinates are
\begin{equation}
\begin{array}{lcl}
 \dot a _i& = & a_i \,  (b_{i+1} -b_i )    \ \ \ i=1, \dots, n-1 \cr
\dot  a_n&= & -2 a_n b_n \cr
   \dot b _i &= & 2 \, ( a_i^2 - a_{i-1}^2 ) \ \ \ i=1, \dots, n  \ ,
\end{array}
\end{equation}
with the convention that $a_0=0$.

These equations can be written as a Lax pair  $\dot L = [B, L] $, where $L$ is the  matrix
\begin{equation}
L=  \pmatrix { b_1 &  a_1 &  & &  &    &    & \cr
                   a_1 &  \ddots  & \ddots  &   & &&   &  \cr
                    & \ddots & \ddots& a_{n-1} &  & &   &  \cr
                   &  & a_{n-1} & b_n & a_n &  & &  \cr
                    & &  & a_n & -b_n &  -a_{n-1}     & & \cr
                    & & & & -a_{n-1} & \ddots & \ddots &  \cr
                    &  &&&& \ddots & \ddots & -a_1  \cr
                       &&&&& & -a_1 & -b_1     \cr }   \ ,
\end{equation}
and $B$ is the skew-symmetric part of $L$.
\smallskip

In the new variables $a_i$, $b_i$, the canonical  bracket  on ${\bf R}^{2n} $  is transformed into
a bracket  $\pi_1 $  which is given  by

\bigskip
\noindent
\begin{equation}
\begin{array}{lcl}
\{ a_i, b_i \} & = &-a_i  \ \ \ \ i=1,2, \dots,  n-1\cr
\{ a_i, b_{i+1} \}& =& a_i \ \ \ \ i=1,2, \dots,  n-1  \cr
\{a_n, b_n\}&=& -2 a_n  \ .
\end{array}
\end{equation}

\smallskip

The invariant polynomials for $C_n$, which we denote by
\begin{displaymath}
  H_2, \  H_4, \  \dots \ H_{2n},
\end{displaymath}

are defined by $H_{2i} = { 1 \over 2i} \  { \rm Tr} \ L^{2i}  $.

\smallskip

We look for a bracket $\pi_3$ which satisfies
\begin{equation}
 \pi_3 \ \nabla \ H_2 = \pi_1 \ \nabla  \ H_4  \ .
\end{equation}

\smallskip
The  bracket  $\pi_3$ was obtained in \cite{damianou3}:

\begin{equation}
\begin{array}{lcl}
\{ a_i, a_{i+1} \}& = &a_i a_{i+1} b_{i+1}   \ \ \ \ i=1,2,..,n-2 \\
\{a_{n-1}, a_n \}& = & 2 a_{n-1}a_n b_n   \\
\{ a_i, b_i \}& = &-a_i b_i^2 -a_i^3   \ \ \ \ \ \ \ \ \ \ i=1,2, \dots , n-1 \\
\{ a_n,b_n \}& = &-2a_n b_n^2 -2 a_n^3  \\
\{a_i, b_{i+2} \}& = &a_i a_{i+1}^2   \\
\{ a_i, b_{i+1} \}& =& a_i b_{i+1}^2 + a_i^3  \\
\{ a_{n-1}, b_n \}& =& a_{n-1}^3+a_{n-1} b_n^2-a_{n-1} a_n^2 \\
 \{ a_i, b_{i-1} \} &=& -a_{i-1}^2 a_i  \\
\{ a_n, b_{n-1} \} &=& -2a_{n-1}^2  a_n \\
\{ b_i, b_{i+1} \}&= &2 a_i^2 (b_i +b_{i+1})  \ .
\end{array}
\end{equation}

\smallskip
We summarize the properties of this  bracket in the following:

\begin{theorem}  The bracket $\pi_3$ satisfies

\smallskip
\noindent
1. $\pi_3 $ is Poisson

\smallskip
\noindent
2. $\pi_3$ is compatible with $\pi_1$.

\smallskip
\noindent
3. $H_{2i}$ are in involution.

\smallskip
Define ${\cal R}=\pi_3 \pi_1^{-1}$. Then ${\cal R}$ is a recursion operator.
 We obtain a hierarchy
$$\pi_1, \pi_3, \pi_5, \dots $$
consisting of compatible Poisson brackets of odd degree in which   the constants of
motion are in involution.

\smallskip
\noindent
4. $\pi_{j+2} \  \nabla\  H_{2i} = \pi_j \  \nabla \  H_{2i+2} \ \ \ \ \ \forall \ i, \ j \   \ . $
\end{theorem}

\smallskip
\noindent
The proofs  are precisely the same as in the case of $B_n$.

Even though it is not necessary to work in  $(p,q)$-coordinates, we reproduce the formulas
from \cite{joana} for completeness.
\bigskip

\begin{equation}
\begin{array}{lcl}
\{ q_i, q_{i-1} \} = \{q_i, q_{i-2} \} = \dots = \{q_i, q_1 \} & =
& 2p_i \ \ \ \  i=2, \dots ,n \\
 \{ p_i, q_{i-2} \} = \{p_i, q_{i-3} \} = \dots = \{ p_i, q_1 \} &
 = & 2( e^{q_{i-1}-q_i}- e^{q_i -q_{i+1}}) \ \ \ \ i = 3, \dots,
n-1\\ \{ p_n, q_{n-2} \} = \{p_n, q_{n-3} \} = \dots = \{ p_n, q_1
\} &
 = & 2 e^{q_{n-1}-q_n}- 4 e^{2 q_n}
\end{array}
\end{equation}

\begin{equation}
\begin{array}{lcl}
\{ q_i, p_i \} & = & p_{i}^2 + 2 e^{q_i -q_{i+1}} \ \ \ \ i= 1,
\dots, n-1 \\ \{ q_n, p_n \} & = & p_{n}^2 + 2 e^{2 q_n} \\ \{
q_{i+1}, p_i \} & = & e^{q_i - q_{i+1}} \\ \{ q_i, p_{i+1} \} & =
& 2e^{q_{i+1}-q_{i+2}}- e^{q_i -q_{i+1}} \ \ \ \ i=1, \dots ,n-2
\\ \{ q_{n-1}, p_n \} & = & 4 e^{2 q_n}- e^{q_{n-1}-q_n} \\
\{ p_i , p_{i+1} \} & = & -e^{q_i -q_{i+1}} (p_i + p_{i+1}) \ .
\end{array}
\end{equation}

For $C_n$, the conformal symmetry is

 $$Z_0= \sum_{i=1}^{n}p_i \frac{\partial}{\partial p_i}+
 \sum_{i=1}^{n}(2n-2i+1) \frac {\partial}{\partial q_i},$$

  and we have the same constants as in the case of $B_n$:

 $$ {\cal L}_{Z_0} J_0 =-J_0, \quad {\cal L}_{Z_0} J_1 = J_1, \quad
 {\cal L}_{Z_0} H_0 =2H_0.$$

 The relations of Oevel's Theorem are the same of the $B_n$ Toda
\begin{equation}
 [Z_i, \chi_j]  =  (2j+1) \chi_{i+j} \label{f55}
 \end{equation}

 \begin{equation}
 [Z_{i}, Z_{j}]  =  2(j-i) Z_{i+j}
 \end{equation}

 \begin{equation}
{\cal L}_{Z_{i}} J_{j}  =  (2(j-i)-1)J_{i+j}.
\end{equation}

Note that (\ref{f55}) gives a method of  generating the exponents.

\subsection{Bi--Hamiltonian formulation of $C_n$ systems}

In order to show that the $C_n$ Toda systems are bi--Hamiltonian we define $\pi_{-1}=\pi_1 \pi_3^{-1} \pi_1$.  This
is the second bracket required to obtain a bi--Hamiltonian pair. We illustrate with a small dimensional
example, namely $C_3$. The explicit formula for $\pi_{-1}$ is the following:
Let $A$ be  the  skew--symmetric $6\times 6$  matrix   defined by the following terms:
\bd
\begin{array}{lcl}
a_{12}&=&a_1 a_2 (b_2a_3^2+b_3^2b_2-a_2^2b_3-b_3b_2^2+a_2^2b_2+b_1^2b_3) \cr
a_{13}&=&-2a_1 a_3(a_2^2b_2+b_1^2b_3-b_3b_2^2)\cr
a_{14}&=&a_1(b_2^2a_3^2+a_2^4+b_2^2b_3^2-2a_2^2b_2b_3+a_1^2a_3^2+a_1^2b_3^2)\cr
a_{15}&=&-a_1(2a_2^4-2a_2^2b_2b_3+a_1^2a_3^2+a_1^2b_3^2+a_3^2b_1^2+b_3^2b_1^2)\cr
a_{16}&=&a_1(2a_2^4+2a_3^2b_1^2+a_2^2a_1^2-2a_2^2 b_2b_3-a_2^2b_1^2-2b_2^2a_3^2)\cr
a_{23}&=&2 a_2(b_1^2+a_1^2)b_3a_3\cr
a_{24}&=&-a_1^2a_2(a_3^2+b_3^2-2b_2b_3+a_2^2-2b_1b_3)\cr
a_{25}&=&a_2(2a_1^2a_3^2+2a_1^2b_3^2-2a_1^2b_2b_3+2a_2^2a_1^2-2a_1^2b_1b_3+a_3^2b_1^2+b_3^2b_1^2+
a_2^2b_1^2)\cr
a_{26}&=&-a_2(2a_1^2a_3^2+2a_3^2b_1^2+a_2^2b_1^2+2a_2^2a_1^2+a_1^4-2a_1^2 b_1b_2+b_2^2 b_1^2)\cr
a_{34}&=&2a_1^2 a_3(a_2^2-2b_1b_3-2b_2b_3)\cr
a_{35}&=&-2a_3(2a_2^2a_1^2-2a_1^2b_1b_3+a_2^2b_1^2-2a_1^2b_2b_3)\cr
a_{36}&=&2a_3(2a_2^2b_1^2+2a_2^2a_1^2+a_1^4-2a_1^2b_1b_2+b_2^2b_1^2)\cr
a_{45}&=&2a_1^2(a_3^2b_1+b_2a_3^2+b_1b_3^2+b_3^2b_2)\cr
a_{46}&=&2a_1^2(a_2^2b_1-2a_3^2b_1-2b_2a_3^2-a_2^2b_3)\cr
a_{56}&=&2(2a_1^2a_3^2b_1+2a_1^2b_2a_3^2+2a_1^2a_2^2b_3-2a_1^2a_2^2b_1+a_2^2b_1^2b_3+a_2^2b_1^2 b_2) \ .
\end{array}
\ed
 The  Poisson tensor  $\pi_{-1}$ is  of the form
\bd
\pi_{-1}= { 1 \over {\rm det}\, L} A \ ,
\ed
where

\bd
{\rm det}\, L= \sqrt {{\rm det}\, {\cal R} }=2a_2^2\, b_1^2\,b_2\, b_3-2 a_1^2 a_2^2\, b_1\, b_3-a_3^2\, b_1^2\, b_2^2+2a_3^2\, b_1\,b_2\,a_1^2
- a_1^4 a_3^2-b_1^2\,b_2^2\, b_3^2+2 a_1^2 b_1\, b_2\, b_3^2-a_1^4\, b_3^2-a_2^4 b_1^2 \ .
\ed

As in the case of $B_2$ we have

\bd
\pi_1 \nabla H_2 = \pi_{-1} \nabla H_4  \ .
\ed

\section{ $D_n$ TODA SYSTEMS}

 In this section, we  show that higher polynomial brackets exist also in
the case of
  $D_n$ Bogoyavlensky-Toda  systems. Using Flaschka coordinates, we will prove  that
these systems  possess a recursion operator and we will construct
an infinite sequence of compatible Poisson brackets in which the
constants of motion are in involution. We also show that the system is bi--Hamiltonian.

\subsection{  A recursion operator for  $D_n$ Bogoyavlensky-Toda systems in Flaschka
coordinates}

\smallskip
 The Hamiltonian for $D_n$ is

\begin{equation}
 H=   { 1 \over 2} \sum_1^n p_j^2 + e^{ q_1- q_2} + \cdots + e^{ q_{n-1}-q_n}
   + e^{q_{n-1} +q_n} \ \ \ \ \ \ \ \ \ \ \ n\ge 4 \ .
  \label{29}
\end{equation}

 We make a  Flaschka-type  transformation, $F: {\bf R}^{2n} \to {\bf
R}^{2n}$ defined by

\begin{displaymath}
 F:  (q_1, \dots, q_n, p_1, \dots, p_n) \to (a_1,  \dots, a_{n}, b_1,
\dots, b_n) \ ,
\end{displaymath}
with

\begin{equation}
  a_i  = {1 \over 2} \,  e^{ {1 \over 2} (q_i - q_{i+1} ) } \ ,   \ \ \ \ \
\ \ \ \ \  i=1,2, \dots, n-1, \ \ \ \ \ \  a_n= { 1 \over 2} e^{ { 1 \over 2}(q_{n-1}+ q_n) } \ , \label{30}
\end{equation}

\begin{displaymath}
             b_i  = -{ 1 \over 2} p_i,   \ \ \ \ \ \ \ \ \ \ i=1,2,\dots, n \ .
\end{displaymath}

\smallskip
\noindent Then

\begin{equation}
\begin{array}{lcl}
 \dot a _i& = & a_i \,  (b_{i+1} -b_i )  \ \ \ \ \ \   \ \ \ i=1,2,  \dots, n-1 \\
  \dot a_n& =& -a_n(b_{n-1}+ b_n)   \\
   \dot b _i &= & 2 \, ( a_i^2 - a_{i-1}^2 ) \ \ \ \ \ \ \ \ \ \ i=1,2,  \dots, n-2 \  {\rm and}\  i=n  \\
    \dot b_{n-1} &=& 2 (a_n^2+a_{n-1}^2-a_{n-2}^2)  \ .
\end{array} \label{31}
\end{equation}

\smallskip
\noindent These equations can be written as a Lax pair  $\dot L =
[B, L] $, where $L$ is the  symmetric  matrix

\begin{equation}
  \pmatrix { b_1 &  a_1 &  & &  &    &    & \cr
                   a_1 &  \ddots  & \ddots  &   & &&   &  \cr
                    & \ddots & \ddots& a_{n-1} & -a_n &0 &   &  \cr
                   &  & a_{n-1} & b_n & 0 & a_n & &  \cr
                     &  &- a_n & 0 &  -b_n &-a_{n-1} & &\cr
                    & &0 &a_n & -a_{n-1} & \ddots & \ddots &  \cr
                    &  &&&& \ddots & \ddots & -a_1  \cr
                       &&&&& & -a_1 & -b_1     \cr }   \ , \label{32}
\end{equation}

and $B$ is the skew-symmetric part of $L$ (In the decomposition,
lower Borel plus skew-symmetric).
\smallskip

The mapping  $F: {\bf R}^{2n} \to {\bf R}^{2n}$, $(q_i ,p_i ) \to
(a_i ,b_i )$, defined by (\ref{30}), transforms the standard
symplectic bracket $J_0$  into  another symplectic bracket $\pi_1 $
given (up to a constant multiple) by

\bigskip
\noindent
\begin{equation}
\begin{array}{lcl}
\{ a_i, b_i \} & = &-{ 1 \over 2}a_i \ \ \ \ \ \ i=1,2,  \dots ,n \\
 \{ a_i,b_{i+1} \}& =& {1 \over 2} a_i \ \ \ \ \ \ \ \ i=1,2,  \dots , n-1 \\
  \{a_n, b_{n-1} \}&=& -{ 1 \over 2} a_n.    \label{a12}
\end{array}
\end{equation}

\smallskip

We obtain a hierarchy of  invariant polynomials,  which we denote by
\begin{displaymath}
  H_2, \  H_4, \  \dots, \ H_{2n}, \dots
\end{displaymath}

 defined by $H_{2i} = { 1 \over 2i} \  { \rm Tr} \ L^{2i}  $.
 The degrees of the first $n-1$ (independent) polynomials are $2,4, \dots,
2n-2$. We also
define
\begin{displaymath}
P_n=\sqrt{ {\rm det}\,L}  \ .
\end{displaymath}
The degree of $P_n$ is $n$.  The set
$ \{ H_2, \  H_4, \  \dots, \ H_{2n-2}, P_n \}$ corresponds to the Chevalley invariants for a Lie
group of type $D_n$. The exponents of the Lie group is the set
$\{ 1,3,5, \dots, 2n-3, n-1 \} $ which is obtained by subtracting 1 from the degrees of the invariant
polynomials. A conjecture of Flaschka states that the degrees of the Poisson brackets is in one--to--one
correspondence with the exponents of the corresponding Lie group.

Taking $H_2={1\over 2}  {\rm Tr} \ L^2$
as the Hamiltonian we have that
\bd
\pi_1 \nabla H_2
\ed
gives precisely equations (\ref{31}). In this section, we find a bracket $\pi_{-1}$ which satisfies
\bd
\pi_1 \nabla H_2=\pi_{-1} \nabla H_4 \ .
\ed

\smallskip

First, we define a bracket $\pi_3$ which satisfies
\begin{equation}
 \pi_3 \ \nabla \ H_2 = \pi_1 \ \nabla  \ H_4  \label{300} \ ,
\end{equation}

and whose non-zero terms are

\begin{equation}
\begin{array}{lcl}
\{ a_i, a_{i+1} \}& = &a_i a_{i+1} b_{i+1}   \ \ \ \ \ \ \ \ \ \ i=1,2, \dots, n-2 \\
 \{a_{n-2}, a_n \} &=& a_{n-2}a_n b_{n-1} \\
\{ a_{n-1}, a_n \} &=& 2 a_{n-1}a_n b_n \\
 \{ a_i, b_i \}& =&-a_i( b_i^2 +a_i^2)   \ \ \ \ \ \ \ \ \ \ i=1,2, \dots , n-2 \\
\{a_{n-1}, b_{n-1} \}&=& -a_{n-1}(a_{n-1}^2+3a_n^2 +b_{n-1}^2) \\
 \{a_n,b_n \}& = &-a_n (a_n^2+b_n^2-a_{n-1}^2)  \\
\{a_i, b_{i+1} \} &=& a_i(a_i^2+b_{i+1}^2)  \ \ \ \ \ \ \ \ \ \ i=1,2, \dots n-2 \\
\{a_{n-1},b_n \} &=& a_{n-1}(a_{n-1}^2+b_n^2-a_n^2)  \\
 \{a_i, b_{i+2} \}& = &a_ia_{i+1}^2  \ \ \ \ \ \ \ \ \ \  i=1,2, \dots, n-3 \\
\{a_{n-2}, b_n \} &=& a_{n-2} (a_{n-1}^2-a_n^2)  \\
 \{ a_i, b_{i-1} \} &=& -a_{i-1}^2 a_i  \ \ \ \ \ \ \ \ \ \  i=2,3, \dots, n-1 \\
\{a_n, b_{n-2} \} &=& -a_{n-2}^2 a_n \\
 \{a_n, b_{n-1}  \} &=& -a_n (3 a_{n-1}^2+a_n^2+b_{n-1}^2)  \\
\{ b_i, b_{i+1} \}&= &2 a_i^2 (b_i +b_{i+1}) \ \ \ \ \ \ \ \ \ \ i=1,2, \dots,  n-2 \\
\{ b_{n-1}, b_n \} &=& 2 a_{n-1}^2 (b_{n-1}+b_n)+ 2a_n^2( b_n-b_{n-1})   \ . \label{34}
\end{array}
\end{equation}
\smallskip
This bracket appeared recently in \cite{damianou4}.
We summarize the properties of this new bracket in the following:

\begin{theorem}  The bracket $\pi_3$ satisfies

\smallskip
\noindent 1. $\pi_3 $ is Poisson.

\smallskip
\noindent 2. $\pi_3$ is compatible with $\pi_1$.

\smallskip
Define ${\cal R}=\pi_3 \pi_1^{-1}$. Then ${\cal R}$ is a recursion
operator.
 We obtain a hierarchy
$$\pi_1, \pi_3, \pi_5, \dots $$ consisting of compatible Poisson
brackets of odd degree in which   the constants of motion are in
involution.

\smallskip
\noindent 3. All the  $H_{2i}$ and $P_n$ are in involution with respect to all the
brackets $\pi_1, \pi_3, \pi_5, \dots $.

\smallskip
\noindent 4. $\pi_{j+2} \  \nabla    \  H_{2i} = \pi_j \  \nabla  \  H_{2i+2} \ \ \ \ \ \forall \ i, \ j \   \ . $

\end{theorem}

\smallskip

The proof of {\it 1.} is a  straightforward verification of the Jacobi identity.
 We will see later, in the next  subsection, that  $\pi
_{3}$ is the Lie derivative of $\pi _{1}$ in the direction of a master

symmetry and this fact makes  $\pi _{1}$, $\pi _{3}$  compatible.
{\it  4.} follows from properties of the recursion operator.  {\it 3.} is a consequence of {\it 4}
(see for example Proposition 3 for a method of proof). The only part which is not obvious is
the involution of $P_n$ with $H_n$ which will be proved at the end of next subsection using master
 symmetries.

\subsection{Master symmetries}

We would like to make some observations concerning master symmetries.

Due to the presence of a recursion operator, we will use the approach of Oevel. We define
$Z_0$ to be the Euler vector field

\begin{displaymath}
Z_0=\sum_{i=1}^n a_i { \partial \, \over \partial a_i}+ b_i { \partial \, \over \partial b_i} \ .
\end{displaymath}
We define the master symmetries $Z_i$ by:
\begin{displaymath}
Z_i = {\cal R}^i Z_0  \ .
\end{displaymath}

For obvious reasons we use the notations
\[
X_{2i}=Z_{i}\ ,h_{i}=H_{2i+2} \ ,\Pi _{i}=\pi _{2i+1} \ ,
\psi _{i}=\chi _{2i+2} \ ,i=0,1,2,\ldots
\]%
where $\chi _{2i}$ denotes the Hamiltonian vector field generated by $H_{2i},
$ with respect to  $\pi _{1}$.
This notation is convenient since $X_2$ is a master symmetry which raises the degrees of invariants
and Poisson tensors by 2 each time.
 One calculates easily that
\bd
{\cal L }_{Z_{0}}\Pi _{0}=-\Pi _{0} \ ,\ \ \ {\cal L}_{Z_{0}}\Pi
_{1}=\Pi _{1} \ , \ \ \ \  {\cal L}_{Z_{0}}h_{0}=2h_{0} \ .
\ed

Therefore $Z_{0}$ is a conformal symmetry for $\Pi _{0}$, $\Pi _{1}$ and $h_{0}$. The
constants appearing in Oevel's Theorem are $\lambda =-1$ , $\mu =1$ and $\nu
=2.$ Therefore we obtain%

\begin{equation}
\left[ Z_{i},\psi _{j}\right] =\left( 1+2j\right) \psi
_{i+j}\Longleftrightarrow \left[ X_{2i},\chi _{2j+2}\right] =\left(
1+2j\right) \chi _{2\left( i+j+1 \right) } \label{100}
\end{equation}%
\begin{equation}
\left[ Z_{i},Z_{j}\right] =2\left( j-i\right) Z_{i+j}\Longleftrightarrow %
\left[ X_{2i},X_{2j}\right] =2\left( j-i\right) X_{2(i+j)} \label{101}
\end{equation}%
\begin{equation}
{\cal L}_{Z_{i}}\left( \Pi _{j}\right) =\left( 2j-2i-1\right) \Pi
_{i+j}\Longleftrightarrow {\cal L}_{X_{2i}}\left( \pi _{2j+1}\right)
=\left( 2j-2i-1\right) \pi _{2(i+j)+1} \label{102}
\end{equation}
\begin{equation}
Z_{i}\left( h_{j}\right) =\left( 2+2i+2j\right) h_{i+j}\Longleftrightarrow
X_{2i}\left( H_{2j}\right) =2\left( i+j\right) H_{2(i+j)} \label{103}  \ .
\end{equation}

{\it Remark 1:} The relation (\ref{102})  implies that ${\cal L}_{X_{2}}(\pi _{1})=-3\pi _{3}$
and therefore $\pi _{3}$ is Lie-derivative of $\pi _{1}$ in the direction of
a master symmetry. This makes $\pi _{1}$ compatible with $\pi _{3}$
  (see Lemma 1).

{\it Remark 2:} The relation (\ref{100})  gives a procedure for generating almost all the exponents. As we
mentioned in the introduction, the last exponent is generated by the application of the
conformal  symmetry on the Hamiltonian vector field corresponding to the Pfaffian of the Jacobi
matrix.

\bigskip
It is interesting to note that one can obtain the master symmetry $X_{2}$ by
using the matrix equation%
\be
\dot{L}=\left[ B,L\right] +L^{3} \label{200} \ ,
\ee

where $L$ is the Lax matrix (\ref{32}) and $B$ is the skew-symmetric matrix defined as follows:

\[
B=\left(
\begin{array}{cccccccccccc}
0 & x_{1} & y_{1} & 0 &  &  &  &  &  &  &  &  \\
-x_{1} & 0 & x_{2} & y_{2} & \ddots  &  &  &  &  &  &  &  \\
-y_{1} & -x_{2} & 0 & \ddots  & \ddots  & 0 &  &  &  & 0 &  &  \\
0 & -y_{2} & \ddots  & \ddots  & x_{n-2} & y_{n-2} & y_{n-1} &  &  &  &  &  \\
& \ddots  & \ddots  & -x_{n-2} & 0 & x_{n-1} & x_n & 0 &  &  &  &  \\
&  & 0 & -y_{n-2} & -x_{n-1} & 0 & 0 & -x_n & -y_{n-1} &  &  &  \\
&  &  & -y_{n-1} & -x_n & 0 & 0 & -x_{n-1} & -y_{n-2} & 0 &  &  \\
&  &  &  & 0 & x_n & x_{n-1} & 0 & -x_{n-2} & \ddots  & \ddots  &  \\
&  &  &  &  & y_{n-1} & y_{n-2} & x_{n-2} & \ddots  & \ddots  & -y_{2} & 0 \\
&  & 0 &  &  &  & 0 & \ddots  & \ddots  & 0 & -x_{2} & -y_{1} \\
&  &  &  &  &  &  & \ddots  & y_{2} & x_{2} & 0 & -x_{1} \\
&  &  &  &  &  &  &  & 0 & y_{1} & x_{1} & 0%
\end{array}%
\right)
\]%  \ ,

where%
\begin{eqnarray*}
x_{i} &=&a_{i}\left\{ \sum_{j=1}^{i-1}b_{j}+\left( i+1-n\right) \left(
b_{i}+b_{i+1}\right) \right\} , \ i=1,2,\ldots ,n-1 \\
x_n &=&-a_{n}\sum_{j=1}^{n-2}b_{j}  \\
y_{i} &=&\left( i+1-n\right) a_{i}a_{i+1}, \ \ \ \ \ \ \ i=1,2,\ldots ,n-2 \\
 y_{n-1}&=&a_{n-2}a_{n}.
\end{eqnarray*}

 It is interesting to note that the $y_i$ is a constant times a product of $a_ja_k$ where the product is
determined from the Dynkin diagram of a Lie algebra of type $D_n$.

The matrix $B$ was chosen in such a way that both sides of
(\ref{200}) have the same form. The components of the  vector field $X_2$ are defined
by the right hand side of (\ref{200}).

Finally we  note the action of the first master symmetry on $P_n=\sqrt{ {\rm det}\, L}$:
\begin{displaymath}
X_2(P_n)= P_n H_2 \ .
\end{displaymath}

\bigskip
{\it Remark:} This last result should be expected since the eigenvalues of $L$ satisfy $\dot \lambda = \lambda^3$ under
(\ref{200}). Therefore,

\begin{eqnarray*}
X_2(P_n)&=&X_2(\sqrt{ {\rm det}L })\\
&=& X_2 \left( \sqrt{ \lambda_1 \dots \lambda_n} \right) \\
&=& { 1\over 2} \left( \lambda_1 \dots \lambda_n \right)^{ -{1 \over 2}}
 \left( \dot{ \lambda}_1 \lambda_2 \dots \lambda_n
+  \lambda_1 \dot{ \lambda}_2 \dots \lambda_n + \dots + \lambda_1 \dots \dot{ \lambda}_n \right)\\
&=& { 1 \over 2 \sqrt{ {\rm det}L} }
 \left(  \lambda_1^3 \lambda_2 \dots \lambda_n
+  \lambda_1  \lambda_2^3 \dots \lambda_n + \dots + \lambda_1 \dots  \lambda_n^3 \right)\\
&=&{ {\rm det} L \over 2 \sqrt{ {\rm det}L} }   \left( \lambda_1^2+\lambda_2^2+ \dots + \lambda_n^2 \right) \\
&= &\sqrt{ {\rm det}L } \ \  { \left( \lambda_1^2+\lambda_2^2+ \dots + \lambda_n^2 \right)  \over 2} \\
&=& P_n H_2 \ .
\end{eqnarray*}
\bigskip

We conclude this subsection by proving the involution of $H_i$ with $P_n$. It is clearly enough to show
the involution of the eigenvalues of $L$ since $P_n$ and $H_i$ are both functions of the eigenvalues.
It is well--known  that the eigenvalues are in involution with respect to the symplectic bracket $\pi_1$. We
will give here a proof based on the Lenard relations (\ref{300}). Let $\lambda$ and $\mu$ be two distinct
eigenvalues and let $U$, $V$ be the gradients of $\lambda$ and $\mu$ respectively. We use
the notation $\{\ , \ \}$ to denote the bracket $\pi_1$ and $\left<\ , \ \right>$ the standard inner product.
 The Lenard relations
(\ref{300})  translate into $\pi_3\,  U= \lambda^2 \,  \pi_1\, U$ and $\pi_3\, V= \mu^2 \, \pi_1 \, V$. Therefore,

\begin{eqnarray*}
\{\lambda, \mu \}&=& \left< \pi_1 U, V \right> \\
                 &=& { 1 \over \lambda^2} \left< \pi_3 U , V \right> \\
                 &=& -{ 1 \over \lambda^2} \left< U, \pi_3 V \right> \\
                 &=& -{ 1 \over \lambda^2} \left< U,  \mu^2 \pi_1 V \right> \\
                 &=& -{ \mu^2 \over \lambda^2} \left< U, \pi_1 V \right> \\
                 &=& { \mu^2 \over \lambda^2} \left< \pi_1 U,  V \right> \\
                 &=& { \mu^2 \over \lambda^2} \{\lambda, \mu \}  \ .
\end{eqnarray*}
Therefore, $\{\lambda, \mu \}=0$.
To show the involution with respect to all brackets $\pi_{2j+1}$  and in view of (\ref{102}) it is
enough to show the following: Let $f_1$, $f_2$ be  two functions  in involution with respect to the Poisson bracket
 $\pi$, let $X$ be a vector field such that $X(f_i)=f_i^3$ for $i=1,2$. Define a Poisson bracket $w$ by
$w={\cal L}_X \pi $.  Then the functions $f_1$, $f_2$ remain in involution with respect to the bracket
$w$. The proof follows trivially if we write  $w={\cal L}_X \pi $
in Poisson form
\bd
\{f_1, f_2 \}_w=X \{ f_1, f_2\}_{ \pi} - \{ f_1, X(f_2) \}_{\pi} - \{ X(f_1), f_2 \}_{ \pi}  \ .
\ed

\bigskip
{\it Remark:} We have to point out that unlike the case of $B_n$ and $C_n$ the cubic bracket (\ref{34})
was discovered not by manipulating the left hand side of (\ref{300})  but through the use of the
master symmetry $X_2$. In other words, we construct the master symmetry $X_2$ using (\ref{200})
and then compute $\pi_3=-{ 1\over 3} {\cal L}_{X_2} \pi_1$.

\subsection{  A recursion operator  for $D_n$ Toda systems
in natural $(q,p)$ coordinates }

We now define a bi-Hamiltonian formulation for $D_n$ Bogoyavlensky-Toda
systems in natural $(q_i ,p_i)$ coordinates. This bracket is simply the pull-back of
$\pi_3$ under the Flaschka transformation (\ref{30}). After some tedious calculation, we  obtain the
following bracket in $(q_i ,p_i )$ coordinates:

\begin{equation}
\begin{array}{lcl}
\{ q_i, q_j \} &=& -2 p_j\ , \ \ \ \ i<j    \\
\{q_i ,p_i  \} &=& p_i^2+2 e^{q_i-q_{i+1}}   \ \ \ \ \ \ \ \ \ \  i=1,2, \dots, n-2  \\

\{q_{n-1} ,p_{n-1}  \} &=& p_{n-1}^2+2 e^{q_{n-1}-q_n} +2 e^{q_{n-1}+q_n}      \\
\{q_n ,p_n  \} &=&p_n^2   \\
\{  q_i , p_{i-1} \} &=& e^{ q_{i-1}-q_i}    \ \ \ \ \ \ \ \ \ \  i=2,3, \dots, n-1  \\
\{ q_n ,p_{n-1}  \} &=&e^{q_{n-1}-q_n} -e^{q_{n-1}+q_n}   \\
\{ q_i ,p_{i+1}  \} &=& -e^{q_i-q_{i+1}} +2 e^{q_{i+1}-q_{i+2}}   \ \ \ \ \ \ \ \ \ \  i=1,2, \dots, n-3  \\
\{q_{n-2} ,p_{n-1}  \} &=& -e^{q_{n-2}-q_{n-1}}+2 e^{q_{n-1}-q_n} +2 e^{q_{n-1}+q_n}    \\
\{q_{n-1} ,p_n  \} &=& -e^{q_{n-1}-q_n}+e^{q_{n-1}+q_n}   \\
\{q_i ,p_j  \} &=&  -2e^{q_{j-1}-q_j}+2 e^{q_j-q_{j+1}}   \ \ \ \ \ \ \ \ \ \  1\le i<j-1\le n-3  \\
\{q_i , p_{n-1} \} &=& -2e^{q_{n-2}-q_{n-1}} +2e^{q_{n-1}-q_n}+2e^{q_{n-1}+q_n}   \ \ \ \ \ \ \ \ \ \  i=1,2, \dots, n-3 \\
\{ q_i,  p_n\} &=&  -2e^{q_{n-1}-q_n} +2 e^{q_{n-1}+q_n}   \ \ \ \ \ \ \ \ \ \  i=1,2, \dots, n-2  \\
 \{ p_i , p_{i+1} \} & = & -e^{q_i -q_{i+1}} (p_i + p_{i+1}) \ \ \ \ \ \ \ \ \ \ i=1,2,\dots, n-2  \\
\{p_{n-1} ,p_n  \} &=&-(p_{n-1}+p_n)e^{q_{n-1}-q_n} +(p_{n-1}-p_n)e^{q_{n-1}+q_n}  \ ;
  \label{35}
\end{array}
\end{equation}
and all other brackets are zero.

Denote this Poisson tensor by $J_1$ and let $J_0$ be the standard symplectic bracket. A simple computation leads to
the following:

\begin{theorem}
 The bracket $J_1$ satisfies

\smallskip
\noindent 1. $J_1 $ is Poisson.

\smallskip
\noindent 2. $J_1$ is compatible with $J_0$.

\smallskip
\noindent 3. The mapping $F$ given by (\ref{30}) is a Poisson
mapping  between $J_1$ and the cubic bracket $\pi_3$.
\end{theorem}

Thus, in $(q,p)$ coordinates we also have a non-degenerate pair
$(J_0 ,J_1 )$ for $D_n$ Bogoyavlensky-Toda and therefore  we may define a recursion
operator ${\cal R} = J_1 J_0 ^{-1}$. We obtain  a hierarchy of
mutually  compatible Poisson tensors defined by $J_i = {\cal
R}^{i}J_0$.

The vector field

\begin{equation}
 Z_0= \sum_{i=1}^{n}  p_i \frac{\partial}{\partial p_i}+
 \sum_{i=1}^{n}2(n-i) \frac {\partial}{\partial q_i}, \label{36}
\end{equation}

is a conformal symmetry for the Poisson tensors $J_0$ and $J_1$
and for the Hamiltonian

\begin{equation}
 h_0 =        { 1 \over 2} \sum_1^n p_j^2 + e^{ q_1- q_2} + \cdots + e^{ q_{n-1}-q_n}
   + e^{q_{n-1} +q_n}           \ . \label{37}
\end{equation}

We compute
\begin{equation}
{\cal L}_{Z_{0}}J_{0}=-J_{0} \ , \ \ \ \ \  {\cal L}_{Z_{0}}J_{1}=J_{1} \ , \ \ \ \
{\cal L}_{Z_{0}}h_{0}=2h_{0}\ .
\end{equation}
So Oevel's Theorem applies. With $Z_{i}={\cal R}^{i}Z_{0}$ ,  $\chi_{0}=\chi _{h_{0}}$
 and $\chi_{i}={\cal R}^{i}\chi_{0}$ one
calculates easily that%
\begin{eqnarray}
\left( a\right)  \ \ \left[ Z_{i}, \chi_{j}\right]  &=&\left(
1+2j\right) \chi_{i+j} \nonumber \\
\left( b\right) \ \ \ \left[ Z_{i},Z_{j}\right]  &=&2\left( j-i\right)
Z_{i+j}  \nonumber \\
\left( c\right)  \ {\cal L}_{Z_{i}}\left( J_{j}\right)  &=&\left(
2j-2i-1\right) J_{i+j}\ .  \nonumber
\end{eqnarray}

Note that (a)  gives the exponents (except one) for a Lie group of
type $D_n$.

The action of the first master symmetry on $P_n$ is the same as in Flaschka coordinates:
\begin{equation}
Z_{1}\left( P_{n}\right) =h_{0} \, P_{n} \label{38}\ .
\end{equation}

Finally, we calculate that
\begin{equation}
\left[ Z_{0},\chi_{P_{n}}\right] =\left( n-1\right)\chi_{P_{n}} \label{39} \ ,
\end{equation}%
producing the last exponent.

\subsection{Bi--Hamiltonian formulation of Bogoyavlensky--Toda systems of type $D_n$}

In order to show that the $D_n$ Toda systems are bi--Hamiltonian we use the same procedure as in the
previous two cases. The tensors $\pi_1$ and $\pi_3$ are both invertible and
we define $\pi_{-1}=\pi_1 \pi_3^{-1} \pi_1$.
  This
is the second bracket required to obtain a bi--Hamiltonian pair. We illustrate with a small dimensional
example, namely $D_4$. The explicit formula for $\pi_{-1}$ is the following:

Let $A$ be  a skew--symmetric $8\times 8$  matrix   given by the following terms:

\noindent
\begin{eqnarray*}
a_{12}=-a_1a_2(a_2^2a_3^2b_4+b_2b_3^2b_4^2 +2b_4b_3b_2a_4^2-2b_4b_3b_2a_3^2+b_2a_3^4 +a_4^4b_2
-2b_2a_4^2a_3^2 +a_2^2b_2b_4^2 \\
-b_2^2b_3b_4^2+b_4b_2^2a_3^2-b_4b_2^2a_4^2+b_1^2b_3b_4^2+b_1^2a_4^2b_4-b_1^2a_3^2b_4-a_2^2b_3b_4^2-a_2^2a_4^2b_4)
\end{eqnarray*}

\noindent
\begin{eqnarray*}
a_{13}=-a_1a_3(b_2^2b_3b_4^2-b_4b_2^2a_3^2-a_2^2b_2b_4^2+b_4b_2^2a_4^2-b_1^2b_3b_4^2-b_1^2a_4^2b_4
+b_1^2a_3^2b_4-a_2^4b_4 +2a_2^2b_2b_3b_4 \\
-a_1^2a_2^2b_4+a_2^2b_2a_4^2-a_2^2a_3^2b_2-b_2^2b_3^2b_4
+a_3^2b_2^2b_3-b_3b_2^2a_4^2+b_1^2a_2^2b_4+b_1^2b_3^2b_4-b_1^2a_3^2b_3+b_1^2b_3a_4^2)
\end{eqnarray*}

\begin{eqnarray*}
a_{14}=a_1a_4(a_2^2b_2b_4^2-b_2^2b_3b_4^2+b_4b_2^2a_3^2-b_4b_2^2a_4^2+b_1^2b_3b_4^2+b_1^2a_4^2b_4-b_1^2a_3^2b_4-a_2^4b_4
+2a_2^2b_2b_3b_4-a_1^2a_2^2b_4 \\
+a_2^2b_2a_4^2-a_2^2a_3^2b_2-b_2^2b_3^2b_4+a_3^2b_2^2b_3-b_3b_2^2a_4^2+b_1^2a_2^2b_4+b_1^2b_3^2b_4-b_1^2a_3^2b_3
+b_1^2b_3a_4^2)
\end{eqnarray*}

\begin{eqnarray*}
a_{15}=-a_1 (2a_4^2b_3b_2^2b_4+2b_4a_2^2a_3^2b_2+b_2^2a_4^4+a_2^4b_4^2-2b_3b_2a_2^2b_4^2+b_2^2a_3^4-2b_3b_4b_2^2a_3^2-2b_2^2a_4^2a_3^2-2b_4a_2^2a_4^2b_2 \\
+b_3^2b_4^2b_2^2+a_1^2b_3^2b_4^2-2a_1^2b_3b_4a_3^2+2a_1^2b_3b_4a_4^2+a_1^2a_4^4-2a_1^2a_4^2a_3^2+a_1^2a_3^4)
\end{eqnarray*}

\begin{eqnarray*}
a_{16}=a_1(2a_1^2b_3b_4a_4^2+2b_1^2b_3b_4a_4^2-2a_1^2b_3b_4a_3^2-2b_4a_2^2a_4^2b_2+2b_4a_2^2a_3^2b_2+2a_2^4b_4^2
-2b_3b_2a_2^2b_4^2\\
+a_1^2a_4^4+a_1^2a_3^4+b_1^2a_4^4+b_1^2a_3^4-2a_1^2a_4^2a_3^2-2b_1^2a_3^2a_4^2 +a_1^2b_3^2b_4^2-2b_1^2a_3^2b_3b_4
 +b_1^2b_3^2b_4^2)
\end{eqnarray*}

\begin{eqnarray*}
a_{17}=a_1(2b_2^2a_4^4-2b_1^2b_3b_4a_4^2+2b_2^2a_3^4+2a_4^2b_3b_2^2b_4-2a_2^4b_4^2-4b_2^2a_4^2a_3^2+2b_3b_2a_2^2b_4^2-
2b_3b_4b_2^2a_3^2\\
-2b_1^2a_4^4-2b_1^2a_3^4
+4b_1^2a_3^2a_4^2+2b_1^2a_3^2b_3b_4+b_1^2a_2^2b_4^2-a_1^2a_2^2b_4^2)
\end{eqnarray*}

\begin{eqnarray*}
a_{18}=a_1(a_1^2a_2^2a_4^2-a_1^2a_2^2a_3^2-2b_1^2b_3b_4a_4^2+2b_2^2a_4^4-2b_4a_2^2a_4^2b_2-2b_2^2a_3^4+2a_4^2b_3b_2^2b_4
-2b_4a_2^2a_3^2b_2 \\
+2b_3b_4b_2^2a_3^2-2b_1^2a_4^4+2b_1^2a_3^4-2b_1^2a_3^2b_3b_4+b_1^2a_2^2a_3^2-b_1^2a_2^2a_4^2)
\end{eqnarray*}

\begin{eqnarray*}
a_{23}=a_2a_3(b_1^2a_3^2b_4-b_1^2b_3b_4^2-b_1^2a_4^2b_4-b_3a_1^2b_4^2-a_1^2a_4^2b_4+a_1^2a_3^2b_4
+b_1^2b_3^2b_4-b_1^2a_3^2b_3+b_1^2b_3a_4^2\\
+b_3^2b_4a_1^2+a_4^2b_3a_1^2-b_3a_1^2a_3^2-b_4b_1^2b_2^2+2b_4b_1a_1^2b_2-a_1^2a_2^2b_4-b_4a_1^4)
\end{eqnarray*}

\begin{eqnarray*}
a_{24}=-a_2a_4(b_1^2b_3b_4^2+b_1^2a_4^2b_4-b_1^2a_3^2b_4+b_3a_1^2b_4^2+a_1^2a_4^2b_4-a_1^2a_3^2b_4
+b_1^2b_3^2b_4-b_1^2a_3^2b_3+b_1^2b_3a_4^2 \\
+b_3^2b_4a_1^2+a_4^2b_3a_1^2-b_3a_1^2a_3^2-b_4b_1^2b_2^2+2b_4b_1a_1^2b_2-a_1^2a_2^2b_4-b_4 a_1^4)
\end{eqnarray*}

\begin{eqnarray*}
a_{25}=-a_1^2a_2(2b_4b_3a_3^2-b_3^2b_4^2-2b_4a4^2b_3-a_4^4+2a_4^2a_3^2-a_3^4+2b_1b_3b_4^2-2b_4b_1a_3^2
 +2b_4a_4^2b_1 -a_2^2b_4^2 \\
+2b_3b_2b_4^2
+2b_4b_2a_4^2-2b_4a_3^2b_2)
\end{eqnarray*}

\begin{eqnarray*}
a_{26}=-a_2(4a_1^2b_3b_4a_4^2+2b_1^2b_3b_4a_4^2-2a_1^2b_4b_2a_4^2-2a_1^2b_3b_2b_4^2+2a_1^2b_4a_3^2b_2-2b_1a_1^2a_4^2b_4+
2b_1a_1^2a_3^2b_4\\
-2b_1b_3a_1^2b_4^2 -4a_1^2b_3b_4a_3^2  +2a_1^2a_4^4+2a_1^2a_3^4+b_1^2a_4^4+b_1^2a_3^4-4a_1^2a_4^2a_3^2-2b_1^2a_3^2a_4^2+
2a_1^2b_3^2b_4^2- \\
2b_1^2a_3^2b_3b_4+b_1^2a_2^2b_4^2+2a_1^2a_2^2b_4^2+b_1^2b_3^2b_4^2)
\end{eqnarray*}

\begin{eqnarray*}
a_{27}=a_2(a_1^4b_4^2+2a_1^2b_3b_4 a_4^2+2b_1^2b_3b_4a_4^2-2a_1^2 b_3b_4a_3^2+2a_1^2a_4^4+2a_1^2a_3^4+2b_1^2a_4^4
+2b_1^2a_3^4-4a_1^2a_4^2a_3^2\\
-4b_1^2 a_3^2a_4^2-2b_1^2a_3^2b_3b_4
+b_1^2a_2^2b_4^2+2a_1^2a_2^2b_4^2-2b_1a_1^2b_2b_4^2+b_1^2b_2^2b_4^2)
\end{eqnarray*}

\begin{eqnarray*}
a_{28}=-a_2(a_4^2b_1^2b_2^2-2a_1^2b_3b_4a_4^2-2b_1^2b_3b_4a_4^2-2a_1^2b_3b_4a_3^2-2a_1^2a_4^4+2a_1^2a_3^4
-2b_1^2a_4^4+2 b_1^2a_3^4 \\
-2b_1^2a_3^2b_3b_4-2b_1a_4^2a_1^2b_2-b_1^2a_3^2b_2^2
+b_1^2a_2^2a_3^2-b_1^2a_2^2a_4^2-a_1^4a_3^2+a_1^4a_4^2+2b_1a_1^2a_3^2b_2)
\end{eqnarray*}

\begin{eqnarray*}
a_{34}=-2a_3a_4b_4 (b_1^2a_2^2+b_1^2b_2^2-2b_1a_1^2b_2+a_1^2a_2^2+a_1^4) \\
\end{eqnarray*}

\begin{eqnarray*}
a_{35}=-a_3a_1^2(2b_4b_1a_3^2-2b_1b_3b_4^2-2b_4a_4^2b_1+a_2^2b_4^2-2b_3b_2b_4^2-2b_4b_2a_4^2+2b_4a_3^2b_2
+2b_1b_4a_2^2+ \\
2b_1b_3^2b_4-2b_1a_3^2b_3
+2a_4^2b_1b_3-2a_2^2b_3b_4+a_2^2a_3^2-a_2^2a_4^2
+2b_2b_3^2b_4+2a_4^2b_2b_3-2a_3^2b_2b_3)
\end{eqnarray*}

\begin{eqnarray*}
a_{36}=-a_3(2b_1^2b_4a_2^2b_2-4b_1a_1^2a_2^2b_4+2a_1^2a_2^2a_4^2-2a_1^2a_2^2a_3^2+2a_1^2b_4b_2a_4^2
+2a_1^2b_3b_2b_4^2-2a_1^2a_4^2b_2b_3 \\
  +2a_1^2a_3^2b_2b_3 +2a_2^2b_1^2b_3b_4+2a_1^2b_1a_3^2b_3+4a_1^2a_2^2b_3b_4-2a_1^2b_2b_3^2b_4-
2a_1^2a_4^2b_1b_3-2a_1^2b_1b_3^2b_4-2a_1^2b_4a_3^2b_2 \\
+2b_1a_1^2a_4^2b_4-2b_1a_1^2a_3^2b_4+
2b_1b_3a_1^2b_4^2-b_1^2a_2^2b_4^2-2a_1^2a_2^2b_4^2-b_1^2a_2^2a_3^2+b_1^2a_2^2a_4^2)
\end{eqnarray*}

\begin{eqnarray*}
a_{37}=a_3(2b_1^2b_4a_2^2b_2-2b_1a_1^2a_2^2b_4+2a_1^2a_2^2a_4^2-a_1^4b_4^2-2a_1^2a_2^2a_3^2+2a_2^2b_1^2b_3b_4+
2a_1^2a_2^2b_3b_4+a_4^2b_1^2b_2^2 \\
-2b_1^2a_2^2b_4^2-2a_1^2a_2^2b_4^2+2b_1a_1^2b_2b_4^2-2b_1a_4^2a_1^2b_2
-b_1^2a_3^2b_2^2-2b_1^2a_2^2a_3^2+2b_1^2a_2^2a_4^2-b_1^2b_2^2b_4^2-a_1^4a_3^2 \\
+a_1^4a_4^2+2b_1a_1^2a_3^2b_2)
\end{eqnarray*}

\begin{eqnarray*}
a_{38}=a_3(2b_1a_1^2a_2^2b_3+2a_1^2a_2^2a_4^2+2a_1^2a_2^2a_3^2+a_1^4b_3^2+b_1^2b_2^2b_3^2+3a_4^2b_1^2b_2^2
-6b_1a_4^2a_1^2b_2+b_1^2a_3^2b_2^2+b_1^2a_2^4 \\
+2b_1^2a_2^2a_3^2+2b_1^2a_2^2a_4^2
-2b_1^2a_2^2b_3b_2-2b_1a_1^2b_2b_3^2+a_1^4a_3^2+3a_1^4a_4^2-2b_1a_1^2a_3^2b_2)
\end{eqnarray*}

\begin{eqnarray*}
a_{45}=a_4a_1^2(2b_1b_3b_4^2-2b_4b_1a_3^2+2b_4a_4^2b_1-a_2^2b_4^2+2b_3b_2b_4^2+2b_4b_2a_4^2-2b_4a_3^2b_2
+2b_1b_4a_2^2+2b_1b_3^2b_4- \\
2b_1a_3^2b_3
+2a_4^2b_1b_3-2a_2^2b_3b_4+a_2^2a_3^2-a_2^2a_4^2
+2b_2b_3^2b_4+2a_4^2b_2b_3-2a_3^2b_2b_3)
\end{eqnarray*}

\begin{eqnarray*}
a_{46}=a_4(2b_1^2b_4a_2^2b_2-4b_1a_1^2a_2^2b_4+2a_1^2a_2^a4^2-2a_1^2a_2^2a_3^2-2a_1^2b_4b_2a_4^2-
2a_1^2b_3b_2b_4^2-2a_1^2a_4^2b_2b_3+2a_1^2a_3^2b_2b_3 \\
+2a_2^2b_1^2b_3b_4+ 2a_1^2b_1a_3^2b_3+4a_1^2a_2^2b_3b_4-2a_1^2b_2b_3^2b_4
-2a_1^2a_4^2b_1b_3-2a_1^2b_1b_3^2b_4+2a_1^2b_4a_3^2b_2 \\
-2b_1a_1^2a_4^2b_4+2b_1a_1^2a_3^2b_4-2b_1b_3a_1^2b_4^2+b_1^2a_2^2b_4^2+2a_1^2a_2^2b_4^2-b_1^2a_2^2a_3^2+b_1^2a_2^2a_4^2)
\end{eqnarray*}

\begin{eqnarray*}
a_{47}=-a_4(2b_1^2b_4a_2^2b_2-2b_1a_1^2a_2^2b_4+2a_1^2a_2^2a_4^2+a_1^4b_4^2-2a_1^2a_2^2a_3^2+2a_2^2b_1^2b_3b_4
+2a_1^2a_2^2b_3b_4+a_4^2b_1^2b_2^2\\
+2b_1^2a_2^2b_4^2
+2a_1^2a_2^2b_4^2-2b_1a_1^2b_2b_4^2 -2b_1a_4^2a_1^2b_2-b_1^2a_3^2b_2^2-2b_1^2a_2^2a_3^2+2b_1^2a_2^2a_4^2
+b_1^2b_2^2b_4^2\\
-a_1^4a_3^2 +a_1^4a_4^2+2b_1a_1^2a_3^2b_2)
\end{eqnarray*}

\begin{eqnarray*}
a_{48}=-a_4(2b_1a_1^2a_2^2b_3+2a_1^2a_2^2a_4^2+2a_1^2a_2^2a_3^2+a_1^4b_3^2+b_1^2b_2^2b_3^2+a_4^2b_1^2b_2^2-
2b_1a_4^2a_1^2b_2+3b_1^2a_3^2b_2^2 \\
+b_1^2 a_2^4+2b_1^2a_2^2a_3^2+
2b_1^2a_2^2a_4^2-2b_1^2a_2^2b_3b_2-2b_1a_1^2b_2b_3^2+3a_1^4a_3^2+a_1^4a_4^2-6b_1a_1^2a_3^2b_2)
\end{eqnarray*}

\begin{eqnarray*}
a_{56}=-2a_1^2(b_1b_3^2b_4^2-2b_1b_4a_3^2b_3+2b_1b_4a_4^2b_3+b_1a_4^4-2b_1a_3^2a_4^2+b_1a_3^4+b_2b_3^2b_4^2
+2b_4b_3b_2a_4^2-2b_4b_3b_2a_3^2\\
+a_4^4b_2-2b_2a_4^2a_3^2+b_2a_3^4)
\end{eqnarray*}

\begin{eqnarray*}
a_{57}=-2a_1^2(4b_2a_4^2a_3^2-2b_1a_4^4-2b_1a_3^4-2a_4^4b_2-a_2^2b_3b_4^2+2b_4b_3b_2a_3^2-2b_4b_3b_2a_4^2-
2b_2a_3^4- \\
2b_1b_4a_4^2b_3+ 4b_1a_3^2a_4^2 +2b_1b_4a_3^2b_3+b_1a_2^2b_4^2)
\end{eqnarray*}

\begin{eqnarray*}
a_{58}=-2a_1^2(2b_1a_3^4-2b_1a_4^4-2a_4^4b_2-2b_4b_3b_2a_3^2-2b_4b_3b_2a_4^2+2b_2a_3^4+a_2^2a_3^2b_4
-2b_1b_4a_4^2b_3- \\2b_1b_4a_3^2b_3-
b_1a_2^2a_4^2 +a_2^2a_4^2b_4+b_1a_2^2a_3^2)
\end{eqnarray*}

\begin{eqnarray*}
a_{67}=-4a_1^2b_2a_4^4-4a_1^2a_3^4b_2-4b_3a_1^2a_2^2b_4^2+8a_1^2a_3^2b_2a_4^2-4a_1^2b_1a_4^4-4a_1^2b_1a_3^4
+4a_1^2b_1a_2^2b_4^2- 2a_2^2b_1^2b_3b_4^2 \\
-2a_2^2b_1^2b_4^2b_2
+8a_1^2b_1a_3^2a_4^2+4a_1^2b_1b_4a_3^2b_3-4a_1^2b_1b_4a_4^2b_3-4b_3b_4a_1^2b_2a_4^2+4b_3b_4a_1^2a_3^2b_2
\end{eqnarray*}

\begin{eqnarray*}
a_{68}=-4b_3b_4a_1^2a_3^2b_2-4b_3b_4a_1^2b_2a_4^2-4a_1^2b_1b_4a_3^2b_3-4a_1^2b_1b_4a_4^2b_3-4a_1^2b_2a_4^4+
4a_1^2a_3^4b_2-4a_1^2b_1a_4^4 \\
 +4a_1^2b_1a_3^4
+4a_1^2a_2^2a_3^2b_4 +4a_1^2a_2^2a_4^2b_4+4b_1a_1^2a_2^2a_3^2-4b_1a_1^2a_2^2a_4^2+2b_1^2a_2^2a_4^2b_4\\
+2b_1^2a_2^2a_3^2b_4+2b_1^2a_2^2b_2a_4^2
-2b_1^2a_2^2b_2a_3^2
\end{eqnarray*}

\begin{eqnarray*}
a_{78}=2a_4^2b_3a_1^4-2a_1^4a_4^2b_4-4a_1^2a_2^2a_3^2b_4-4a_1^2a_2^2a_4^2b_4+4b_1a_1^2a_3^2b_2b_3
-4b_1a_1^2a_4^2b_2b_3+ 4b_1b_4a_1^2a_3^2b_2  \\
-4b_1a_1^2a_2^2a_3^2+ 4b_1a_1^2a_2^2a_4^2-2b_1^2a_3^2b_3b_2^2-2b_1^2b_4b_2^2a_3^2-2b_1^2b_4b_2^2a_4^2+
4b_1b_4a_4^2a_1^2b_2-
4b_1^2a_2^2a_4^2b_4 \\
+2b_1^2b_3b_2^2a_4^2-4b_1^2a_2^2a_3^2b_4-4b_1^2a_2^2b_2a_4^2+4b_1^2a_2^2b_2a_3^2-
2a_3^2b_3a_1^4-2a_1^4a_3^2b_4  \ .
\end{eqnarray*}

The Poisson tensor of $\pi_{-1}$ is given  by the formula
\bd
\pi_{-1}= { 1 \over 2  {\rm det}\, L} A \ ,
\ed

\bd
{\rm det}\, L=P_4^2=(a_2^2 b_1b_4+ a_3^2 b_1b_2-a_4^2 b_1b_2-b_1 b_2 b_3b_4-a_1^2a_3^2+a_1^2a_4^2+a_1^2 b_3 b_4)^2 \ .
\ed

We note that
\bd
{\rm det}\, \pi_3= {\rm det}\, \pi_1 ({\rm det}\, L)^2= P_4^4 \ ,
\ed
and therefore
\bd
{\rm det}\, {\cal R}=  ({\rm det}\, L)^2 \ .
\ed
This formula (as well as the formulas in the previous two cases) indicates that the eigenvalues of the recursion
operator should be  the squares of the non--zero eigenvalues  of the Jacobi matrix. This is known to hold in  the case
of the classical $A_n$ Toda lattice, see \cite{falqui}. For a general recursion operator, the relation between its eigenvalues and the
eigenvalues of the Lax matrix has not been fully investigated yet.

As in the case of $B_2$ and $C_3$  we have

\bd
\pi_1 \nabla H_2 = \pi_{-1} \nabla H_4  \ .
\ed

\section{Conclusion}

\subsection{Summary of results}

The classical, finite, non--periodic Toda lattice is known to be bi--Hamiltonian. Moreover,
(\ref{a98}) is a multi--Hamiltonian formulation of the system. We have indicated how to
obtain similar results for the other classical Lie algebras and we have illustrated with some
small dimensional examples. These examples  may be generalized:

\begin{theorem}
The $B_n$, $C_n$ and $D_n$ Toda systems are  bi--Hamiltonian. In fact, they are  multi--Hamiltonian.
  In each case we define
\bd
{\cal N}= \pi_1 \pi_3^{-1} \ ,
\ed
where $\pi_1$ is the Lie--Poisson bracket,  $\pi_{3}$ is the cubic Poisson bracket
and
\bd
\pi_{ -(2i-1)} ={\cal N}^{i} \pi_1 \ \ \ i=1,2, \dots  \ .
\ed
Then all the brackets    are mutually compatible, Poisson  and satisfy
\be
\pi_1 \nabla H_2 = \pi_{-1} \nabla H_4 =\pi_{-3} \nabla H_6 =\dots  \label{a17} \ .
\ee

\end{theorem}

\smallskip
{\bf Proof:}
The proof of (\ref{a17}) is trivial. They are just the Lenard relations for the negative hierarchy.
The brackets $\pi_{-1}, \pi_{-3}, \pi_{-5}, \dots $ are all Poisson since they are generated by the
negative recursion operator, ${\cal N}$,  applied to the initial Poisson bracket $\pi_1$.
To prove compatibility of all Poisson brackets appearing in (\ref{a17}) we
take  two brackets $\pi_t$ and $\pi_s$ where  $t$, $s$ are  odd integers,
with  $t<s\le 1$.
   Using condition (b) of Oevel's theorem (for the negative operator) we can express $\pi_{t}$ as the Lie derivative
 of $\pi_s$ in
the direction of a master symmetry.
It is  therefore enough to prove the following simple general  result: If $\pi$ and $\sigma$ are both Poisson tensors and
$\sigma={\cal L}_X \pi $ for some vector field $X$, then $\pi$ and $\sigma$ are compatible. The one line
proof uses the super--Jacobi identity for the Scouten bracket:

\bd
[\pi, \sigma]=[\pi , [\pi , X]]= -{1 \over 2} [ X, [\pi , \pi  ]  ]=0 \ .
\ed
\hfill \rule{5pt}{5pt}

We remark that Oevel's theorem applies in all three cases (and for both hierarchies)
since the Euler vector field $X_0$ (\ref{f2})  is a conformal
symmetry for $\pi_1$, $\pi_3$ and $H_2$. Furthermore, the compatibility condition holds for all brackets in
both hierarchies.
If $\pi_t$ and $\pi_s$ are both in the positive hierarchy
 then the argument of the theorem still works
using the positive recursion operator.
The remaining case, when one of the tensors has negative index and the other one positive, can also be proved
in a straightforward manner using
 similar arguments,
i.e., properties of the Schouten bracket and the fact that the formulas in  Oevel's theorem hold for any integer
value of the
index  (Theorem 8).
The tensor in the positive hierarchy is the Lie derivative of the tensor in the negative hierarchy in the
direction of a suitable master symmetry  and
 the argument of the theorem shows that  they are compatible.  Therefore,  we
have a more general result: Any two brackets in either the positive or negative hierarchy are compatible.

{\it Remark }
The compatibility condition  follows also  from a general result of bi--Hamiltonian geometry: If $\pi$ and $\sigma$
are two compatible Poisson tensors and $\pi$ is invertible, then ${\cal N}=\sigma \pi^{-1}$ is a recursion
 operator (i.e. its torsion vanishes) and  all the tensors ${\cal N}^i \pi$, with $i\ge 0$, are Poisson
and compatible. Using this result one can
prove compatibility of all brackets in both hierarchies. For example, to show that $\pi_5$ is compatible with $\pi_{-3}$
we use the fact that ${\cal R}=\pi_3 \pi_{1}^{-1}=\pi_{-1} \left(\pi_{-3}\right)^{-1}$ to obtain $\pi_5={\cal R}^{4}
\pi_{-3}$. Since $\pi_{-1}$ and $\pi_{-3}$ are compatible and Poisson, then ${\cal R}$ generates a chain of compatible
Poisson tensors. In particular $\pi_5$ and $\pi_{-3}$ are compatible.

\subsection{Open problems}

We conclude  with some open problems and  some possible directions of research  for systems related to the Toda
lattice.

\begin{itemize}

\item

\noindent
{\bf Exceptional Toda lattices}

The case of exceptional simple Lie groups is still an open problem. It is also a much more difficult problem.
The only case that is reasonable to complete  is the Toda system of type  $G_2$. In that case the second Poisson bracket
should be a  homogeneous bracket of degree 5 (a conjecture of Flaschka states  that the degrees of the independent
Poisson tensors  coincide with  the exponents of the corresponding Lie group).
 The other exceptional cases are even more complicated. It is
 a nontrivial task  even to write  down   an explicit   Lax pair  for the systems
 and therefore the methods of this  paper will be difficult  to apply.

\item

\noindent
{\bf Full--Kostant Toda}

One has a tri--Hamiltonian formulation of the $A_n$ system but no hierarchy. In the case of generalized full--Kostant
Toda lattice (associated to simple Lie groups) one could seek to find  similar structures
 as in the present paper. So far, nothing
is known.  The interesting
feature of these systems is the presence of  the rational integrals that are necessary  to prove integrability.
The Lie--algebraic background of the systems will certainly play a prominent role.

\item
\noindent
{\bf  The Volterra or KM--system}
The multi--Hamiltonian structure of this system was  first obtained in \cite{damianou11}. However, there is a symplectic
realization of the system, due to Volterra, and it would be interesting to find a recursion operator in that symplectic
space that projects onto the known hierarchy (as in section 5.1).

\item
\noindent
{\bf Bogoyavlensky--Volterra lattices}

There is also an interesting connection with the corresponding generalized Volterra
systems also defined by Bogoyavlensky \cite{bogo2} in 1988. It seems that the multiple
Hamiltonian structures of  the Volterra and Toda lattices are in one--to--one
correspondence through a procedure of Moser.
Multiple Hamiltonian structures  for the generalized Volterra lattices,
were constructed recently by Kouzaris \cite{Kou},  at least  for the classical Lie algebras.
The relation between the Volterra systems of type $B_{n}$ and  $C_{n} $ and the corresponding Toda $B_{n}$, $ C_{n} $
systems was   demonstrated in  \cite{damianou12}.  The connection between Volterra $D_{n}$
 and Toda $D_{n}$ is still an open problem.

\item
\noindent
{\bf Independence of Poisson structures}
Equations (\ref{a98}) and (\ref{a17}) are  remarkable; one Hamiltonian system, infinite formulations.
On the other hand the systems are finite dimensional and after some point some dependencies should occur. Indeed, the
integrals $H_k$ are not all independent. It is natural to ask a similar question about the infinite sequence of
Poisson structures. Do they become dependent after a certain point? Unfortunately, there exists no widely accepted
definition of independence  for Poisson tensors.

\item
\noindent
{\bf Is the Toda lattice super--integrable?}
This conjecture should be true. A number of well--known systems are super--integrable, i.e. the free particle, the
harmonic oscillator and the Calogero--Moser systems.
In the case of the open Toda lattice,  asymptotically the   particles become free as  time
goes to infinity with asymptotic momenta being the eigenvalues  of the Lax matrix. Therefore, the system behaves
asympotically  like a system of  free particles which is super--integrable.
\end{itemize}

\smallskip
\noindent
{\bf Acknowledgments}
I would like to thank H. Flaschka for introducing me to  the problems
 of the present paper  and
for his considerable imput in this long project.  I thank also a number of people who had discussed the subject with me
 and given me some useful ideas:  J. M. Costa Nunes, R. Fernades,  M. Gekhtman,  S. Kouzaris, F. Magri, I. Marshall,
 W. Oevel, T. Ratiu, C. Sophocleous.


\newpage


\begin{thebibliography}{99}


\bibitem{bogo}  O. I. Bogoyavlensky,  Comm. Math. Phys.  {\bf 51}  (1976), 201.


\bibitem{kostant} B. Kostant,  Adv. Math.  {\bf 34} (1979), 195.

\bibitem{olshanetsky}  M. A. Olshanetsky and A. M. Perelomov, Invent. Math.  {\bf 54} (1979), 261.

\bibitem{avm}   M.  Adler and  P. van Moerbeke,  Adv. in Math. {\bf 38}  (1980), 267.

\bibitem{kozlov} V. V. Kozlov and D. V. Treshchev,  Math. USSR-Izv. {\bf 34}  (1990), 555.

\bibitem{ranada2} M. F. Ranada,  J. Math. Phys.  {\bf 36} (1995), 6846.

\bibitem{anna}  A. Annamalai and K. M. Tamizhmani, J.  Math. Phys. {\bf 34} (1993),  1876.



\bibitem{perelomov} A. M. Perelomov,  Integrable systems of classical mechanics and Lie algebras, Vol. I,
Birkhauser Verlag, Basel, 1990.

\bibitem{damianou1}   P. A.  Damianou,  Lett.   Math.  Phys. {\bf  20} (1990), 101.


\bibitem{damianou2}  P. A.  Damianou, J. Math. Phys. {\bf 35} (1994), 5511.


\bibitem{joana} J. M. Nunes da Costa  and  P. A. Damianou, Bull. Sci. math. {\bf 125}   (2001), 49.

\bibitem{damianou3}   P. A.  Damianou,  Regul. Chaotic dyn. {\bf 5}  (2000), 17.


\bibitem{damianou4} P. A.  Damianou and S. P. Kouzaris, J. Phys. A. {\bf 36}  (2003), 1385.

\bibitem{damianou9} P. A. Damianou, Nonlinearity  {\bf 17}   (2004), 397.


\bibitem{damianou5} P. A.  Damianou,  J.  Geom.  Phys. {\bf 45 } (2003), 184.

\bibitem{lich} A. Lichnerowicz,  J. Diff. Geom.  {\bf 12} (1977), 253.

\bibitem{ratiu} J. E. Marsden  and  T. S. Ratiu, Introduction to mechanics and symmetry, A basic exposition
 of classical mechanical systems,  Springer-Verlag, New York, 1999.

\bibitem{vaisman} I. Vaisman,  Lectures on the geometry of Poisson manifolds,
 Progress in mathematics, {\bf  118}, Birkh\'auser, Basel, 1994.

\bibitem{weinstein} A. Weinstein, J. Diff. Geom.  {\bf 18} (1983), 523.

\bibitem{marmo} J. Grabowski, G. Marmo and A. M.  Perelomov,  Modern Phys. Lett. A {\bf 8} (1993), 1719.

\bibitem{cushman} R. Cushman and M. Roberts,  Bull. Sci. math. {\bf 126} (2002), 525.

\bibitem{damianou10} P. A. Damianou, Bull. Sci. math. {\bf 120} (1996), 195.


\bibitem{chevalley} C. Chevalley  and  S. Eilenberg,    Trans. of Amer. Math. Soc. {\bf 63}  (1948), 85.

\bibitem{koszul} J. L. Koszul,  Soc. Math. France Asterisque hors serie (1985), 257.

\bibitem{magri}  F. Magri, J. Math. Phys.  {\bf 19} (1978), 1156.

\bibitem{falqui} G. Falqui, F. Magri  and  M.Pedroni, J. Nonlinear Math. Phys. {\bf 8} (2001), 118.

\bibitem{gelfand}  I. M. Gelfand and  I. Zakharevich,  Selecta Math.  {\bf  6}  (2000), 131.

\bibitem{smirnov}  R. G. Smirnov, C. R. Math. Rep. Acad. Sci. Canada {\bf 17} (1995), 225.

\bibitem{suris} Y. B.  Suris,  Phys. Lett. A {\bf 180}  (1993), 419.


\bibitem{fuchssteiner} B. Fuchssteiner,  Progr. Theor. Phys.  {\bf 70}  (1983), 1508.

\bibitem{fokas1} A. S.  Fokas  and  B. Fuchssteiner,  Phys. Lett. A {\bf 86}  (1981), 341.

\bibitem{oevel2} W.  Oevel, {\it Topics in Soliton Theory and Exactly Solvable non-linear Equations},
  World Scientific Publ.,  Singapore, 1987.

 \bibitem{flaschka1} H.  Flaschka H, Phys. Rev. B  {\bf 9} (1974), 1924.


\bibitem{flaschka2} H.  Flaschka,  Progr. Theor. Phys. {\bf 51} (1974), 703.

\bibitem{henon} M. Henon,  Phys. Rev. B {\bf 9} (1974), 1921.

\bibitem{manakov}  S.  Manakov, Zh. Exp. Teor. Fiz.  {\bf 67} (1974), 543.


\bibitem{moser} J. Moser, Lect.  Notes  Phys.  {\bf 38} (1976), 97.

\bibitem{moser2} J. Moser,  Adv. Math.  {\bf 16} (1975), 197.

\bibitem{toda} M. Toda,  J. Phys. Soc. Japan  {\bf 22}  (1967), 431.

\bibitem{adler}  M.  Adler, Invent. Math. {\bf 50}  (1979), 219.

\bibitem{kup}  B.  Kupershmidt, Asterisque {\bf 123}  (1985), 1.

\bibitem{das}  A. Das  and S. Okubo,  Ann. Phys.  {\bf 190} (1989), 215.

\bibitem{fernandes} R. L.  Fernandes,  J.  Phys. A  {\bf 26}  (1993), 3797.


\bibitem{morosi}  C. Morosi and  G. Tondo,  Inv. Probl.  {\bf 6} (1990), 557.

\bibitem{olver1}  P. J. Olver, J. Math. Phys.  {\bf 18}  (1977), 1212.


\bibitem{fayb}   L.  Faybusovich and M.  Gekhtman, Phys. Lett. A {\bf 272} (2000), 236.

\bibitem{atiyah} M. F. Atiyah and N. Hitchin, The geometry and dynamics of magnetic monopoles. M. B.
Porter Lectures, Princeton University Press, Princeton, 1988.

\bibitem{moser3} J. Moser, Finitely many mass points on the line under the influence of an exponential
potential. Batelles Recontres, Springer Notes in Physics, 417--497 (1974).

\bibitem{vaninsky} K. L. Vaninsky, J. Geom. Phys. {\bf 46} (2003), 283.

\bibitem{petalidou} F. Petalidou,  Bull. Sci. math. {\bf 124} (2000), 255.

\bibitem{magri2}  Y. Kosmann-Schwarzbach and F.  Magri,  J. Math. Phys. {\bf 37}  (1996), 6173.

\bibitem{olver2} P. J.  Olver,  Applications of Lie groups to Differential
 Equations. GTM, {\bf 107}, Springer-Verlag, New York 1986.

\bibitem{bluman}  G. W. Bluman  and S. Kumei,  Symmetries and Differential
 Equations,  Springer-Verlag, New York 1989.

\bibitem{ovs} L. V. Ovsiannikov, Group Analysis of Differential Equations,  Academic Press, New York 1982.

\bibitem{ib} N. H. Ibragimov, Elementary Lie group analysis and ordinary differential equations,
 Wiley Series in Mathematical Methods in Practice,  4,  John Wiley and Sons, Ltd.,  1999.

\bibitem{costa}   J. M. Nunes da Costa  and  C. M. Marle,  J. Phys. A  {\bf 30} (1997), 7551.


\bibitem{damianou7} P. A. Damianou and C.  Sophocleous, J. Math. Phys. {\bf 40} (1999), 210.


\bibitem{damianou6} P. A. Damianou, J. Phys. A {\bf 26} (1993), 3791.

\bibitem{ranada} M. F. Ranada,  J. Math. Phys. {\bf 40} (1999), 236.

\bibitem{sopho}  C. Sophocleous, S. Moyo, P.G.L. Leach, P. A. Damianou, Noether Symmetries and Integrals
in One, Two and Three dimensions, TR/16/2000, Department of Mathematics and Statistics, University of Cyprus.

\bibitem{damianou8} P. A. Damianou, C. Sophocleous, Master and Noether symmetries for the Toda lattice, Proceedings
of the 16th International Symposium on Nonlinear Acoustics, {\bf  1}, 2003, pp. 618--622.

\bibitem{collin} D. H. Collingwood and  W. M. McGovern,  Nilpotent orbits in semisimple Lie algebras,
Van Nostrand Reinhold Co., New York, 1993.

\bibitem{humphreys} J. E. Humphreys, Introduction to Lie algebras and
representation theory  GTM {\bf 9}, Springer--Verlag 1972.

\bibitem{damianou11}  P. A. Damianou,   Phys.  Lett.  A  {\bf 155} (1991), 126.

\bibitem{bogo2} O. I. Bogoyavlensky,  Phys. Lett. A. {\bf 134} (1988),  34.

\bibitem{Kou}  S. P. Kouzaris,  J.  Nonlinear Math. Phys. {\bf 10} (2003), 431.

\bibitem{damianou12} P. A. Damianou and  R. Fernandes,  Rep. Math. Phys.  {\bf 50} (2002), 361.



\end{thebibliography}
\end{document}